\documentclass{report}
\usepackage{Thesis,epsfig}

\title{Stabilizer Codes and Quantum Error Correction}
\author{Daniel Gottesman}
\date{May 21, 1997}

\newcommand{\G}{{\cal G}}
\newcommand{\ket}[1]{|{#1}\rangle}
\newcommand{\bra}[1]{\langle{#1}|}
\newcommand{\X}{\sigma_x}
\newcommand{\Z}{\sigma_z}
\newcommand{\Y}{\sigma_y}
\newcommand{\Xs}[1]{\sigma_{x{#1}}}
\newcommand{\Ys}[1]{\sigma_{y{#1}}}
\newcommand{\Zs}[1]{\sigma_{z{#1}}}
\newcommand{\Xbar}{\overline{X}}
\newcommand{\Ybar}{\overline{Y}}
\newcommand{\Zbar}{\overline{Z}}
\newcommand{\A}{{\cal A}}
\newcommand{\ptj}[1]{p_{Tof}^{({#1})}}
\newcommand{\pgj}[1]{p_g^{({#1})}}
\newcommand{\psj}[1]{p_{stor}^{({#1})}}
\newcommand{\ppj}[1]{p_{prep}^{({#1})}}

\newcommand{\ttj}[1]{t_{Tof}^{({#1})}}
\newcommand{\tpj}[1]{t_{prep}^{({#1})}}
\newcommand{\tmj}[1]{t_{meas}^{({#1})}}
\newcommand{\Sum}{\sum}
\newcommand{\low}[1]{\raisebox{-0.3ex}{{#1}}}

\begin{document}

\pagenumbering{roman}
\maketitle

\makecopyright

\begin{acknowledgements}
I would like to thank my advisor John Preskill for his guidance, and
the members of the QUIC collaboration, particularly David Beckman, John
Cortese, Jarah Evslin, Chris Fuchs, Sham Kakade, Andrew Landahl, and
Hideo Mabuchi, for many stimulating conversations.  My graduate career
was supported by a National Science Foundation Graduate Fellowship, by the 
U.~S. Department of Energy under Grant No. DE-FG03-92-ER40701, and by
DARPA under Grant No. DAAH04-96-1-0386 administered by the Army Research
Office.
\end{acknowledgements}

\begin{abstract}
Controlling operational errors and decoherence is one of the major 
challenges facing the field of quantum computation and other attempts
to create specified many-particle entangled states.  The field of quantum
error correction has developed to meet this challenge.  A group-theoretical 
structure and associated subclass of quantum codes, the stabilizer codes,
has proved particularly fruitful in producing codes and in understanding the 
structure of both specific codes and classes of codes.  I will give an 
overview of the field of quantum error correction and the formalism of 
stabilizer codes.  In the context of stabilizer codes, I will discuss a 
number of known codes, the capacity of a quantum channel, bounds on 
quantum codes, and fault-tolerant quantum computation.
\end{abstract}

\tableofcontents
\listoftables
\listoffigures

\chapter{Introduction and Preliminary Material}
\label{chap-intro}
\pagenumbering{arabic}

\section{Quantum Computers}

Computers have changed the world in many ways.  They are ubiquitous, 
running air-traffic control and manufacturing plants, providing movie 
special effects and video games, and serving as a substrate for electronic 
mail and the World Wide Web.  While computers allow us to solve many 
problems that were simply impossible before their advent, a number of 
problems require too much computation to be practical even for relatively 
simple inputs, and using the most powerful computers.

The field of classical complexity theory has developed to classify problems
by their difficulty.  A class of problems is generally considered tractable 
if an algorithm exists to solve it with resources (such as time and memory) 
polynomial in the size of the input.  Two well-known classically intractable 
problems are factoring an $n$-bit number and the Traveling Salesman problem (finding the minimum cyclic path connecting $n$ cities with specified distances 
between them).  Both of these problems are in the complexity class NP (for 
``non-deterministic polynomial''):\footnote{Strictly speaking, it is the 
associated decision problems that are in NP.} given a black box that solves 
the problem (an {\em oracle}), we can check in polynomial time that the 
solution is correct.  The Traveling Salesman problem is an NP-complete problem; 
that is, any problem in NP can be transformed into an instance of the Traveling 
Salesman problem in polynomial time.  If we can solve the Traveling 
Salesman problem in polynomial time, we can solve any NP problem in polynomial
time.  Factoring may or may not be NP-complete, but so much work has been done 
attempting to solve it that the consensus is that it is classically 
intractable, and RSA public-key cryptography, which is used, for instance, to 
send credit-card numbers in Web browsing software, depends on the difficulty of 
factoring large numbers.

As computer hardware develops over time, the underlying technology continually 
changes to become faster, smaller, and generally better.  What was 
impossible on yesterday's computers may be quite possible today.  A 
problem that was intractable on the earlier hardware might become 
tractable with the new technology.  However, the strong Church-Turing Thesis 
\cite{church-turing} states that this is not the case, and that every physical 
implementation of universal computation can simulate any other 
implementation with only a polynomial slowdown.\footnote{The original
Church-Turing thesis only states that any universal computer can simulate
any other computer, but the requirement of polynomial resources is a
useful strengthening.}  In this way, the Church-Turing 
Thesis protects complexity theory from obsolescence as computer technology 
improves.  While a new computer may be able to factor larger numbers, the 
difficulty of factoring numbers will still scale the same way with the size 
of the input on the new hardware as on the old hardware.

Another problem that has proven to be classically intractable is simulating 
quantum systems.  A single spin-1/2 particle, such as an electron trapped 
in a quantum dot, has a two-dimensional space of states, which can be 
considered to describe the direction of its spin.  A similar classical particle 
such as a Heisenberg spin would also have a two-dimensional space of states.  
However, $n$ quantum particles have a $2^n$-dimensional state space, while 
$n$ classical Heisenberg spins would only have a $2n$-dimensional space of 
states.  The extra states in the quantum system come from the presence of 
entangled states between many different particles.  Note that while an $n$-bit
classical digital computer has $2^n$ possible states, they only form an 
$n$-dimensional state space, since a state can be described by an $n$-component
binary vector.  To describe a state in a quantum computer with $n$ qubits 
requires a complex vector with $2^n$ components.  I give a basic introduction 
to quantum mechanics in section~\ref{sec-QM}.  Quantum systems are difficult 
to simulate classically because they generically utilize the full 
$2^n$-dimensional Hilbert space as they evolve, requiring exponential 
classical resources.

This fact led Feynman to conjecture that a quantum computer which used 
quantum mechanics intrinsically might be more powerful than a computer 
mired in the classical world~\cite{feynman}.  While this seems a sensible 
suggestion when just looking at quantum mechanics, it is in fact quite 
revolutionary in that it suggests that the strong Church-Turing Thesis is 
wrong!\footnote{A classical computer can simulate a quantum computer, but
only with exponential resources, so the weak Church-Turing Thesis does still
hold.}  This opens up the possibility that classical complexity classes might 
not apply for quantum computers, and that some classically intractable problems 
might become tractable.  The most spectacular instance of this is Shor's 
discovery of an algorithm to factor numbers on a quantum computer in a 
polynomial time in the number of digits~\cite{shor-factoring}.  Another 
impressive algorithm is Grover's algorithm~\cite{grover}, which can find a 
single object in an unsorted database of $N$ objects in $O(\sqrt{N})$ time 
on a quantum computer, while the same task would require an exhaustive 
search on a classical computer, taking $O(N)$ time.  It has been shown
that $O(\sqrt{N})$ time is the best possible speed for this task
\cite{bennett-strengths}, which tends to suggest that NP-complete problems
are still intractable on a quantum computer, although this has not been
shown (note that a proof of this would also show ${\rm P} \neq {\rm NP}$ for a 
classical computer).

However, declaring by theoretical fiat the basic properties of a quantum 
computer is a far cry from actually building one and using it to factor 
large numbers.  Nevertheless, the first steps in building a quantum 
computer have been taken.  Any quantum computer requires a system with
long-lived quantum states and a way to interact them.  Typically, we
consider systems comprised of a number of two-state subsystems, which
are called {\em qubits} (for ``quantum bits'').  There are many proposals 
for how to build a quantum computer.  Some possible physical realizations
of qubits are: 
\begin{itemize}
\item the ground and excited states of ions stored in a linear ion trap, with 
interactions between ions provided through a joint vibrational 
mode~\cite{cirac-zoller,wineland}.
\item photons in either polarization, with interactions via cavity 
QED~\cite{kimble}.
\item nuclear spin states in polymers, with interactions provided by 
nuclear magnetic resonance techniques~\cite{gershenfeld}.
\end{itemize}
While these implementations are seemingly very different, it is 
possible to simulate the computational process of one system on 
any of the others, providing a quantum analogue to the Church-Turing 
Thesis (although there are difficult technical or theoretical problems
with scaling up the size of these implementations).

These suggested implementations of quantum computers all share a 
much higher susceptibility to errors than modern classical computers.  
While further development may reduce the size of errors by orders of 
magnitude, it is unlikely that quantum computers will ever reach the 
incredible reliability of classical computers.  Modern classical computers 
guard against error largely by being digital instead of analog --- instead of 
allowing each bit of the computer to vary continuously between 0 and 1, at 
each time step the hardware kicks the bit back to the nearer of 0 and 1.  
This prevents small errors from building up into large errors, which are 
therefore drastically reduced.  The same technique cannot be used in a 
quantum computer, because continually measuring each qubit would destroy 
the entangled states that distinguish a quantum computer from a classical 
computer.

Entangled states are in general very delicate, and making a measurement on 
one will typically collapse it into a less entangled state.  Small interactions 
with the environment provide a sort of continuous measurement of a system, 
and as the system grows in size, these become harder and harder to ignore.  
The system will {\em decohere} and begin to look like a classical system.  
Decoherence is why the world looks classical at a human scale.  Reducing 
interactions with the environment can reduce the effects of decoherence, 
but not eliminate them entirely.

Even if the basal error rate in a quantum computer can be reduced to some 
small value $\epsilon$ per unit time, after $N$ time steps, the probability 
of surviving without an error is only $(1-\epsilon)^N$, which decreases 
exponentially with $N$.  Even if an algorithm runs in polynomial time on an 
error-free computer, it will require exponentially many runs on a real 
computer unless something can be done to control the errors.

The same problem occurs for classical computers.  There, the problem can 
be solved in principle by the use of error-correcting codes.  In practice, 
they are not usually necessary for normal computer operation, but they 
are essential to overcome noise in communications channels.  I give a basic 
introduction to the theory of classical error-correcting codes in 
section~\ref{sec-classical}.

Classical error-correction techniques cannot be directly carried over to 
quantum computers for two reasons.  First of all, the classical techniques 
assume we can measure all of the bits in the computer.  For a quantum 
computer, this would destroy any entanglement between qubits.  More 
importantly, a classical computer only needs to preserve the bit 
values of 0 and 1.  A quantum computer also needs to keep phase 
information in entangled states.  Thus, while quantum error-correcting 
codes are related to classical codes, they require a somewhat new 
approach.

The first quantum error-correcting codes were discovered by 
Shor~\cite{shor-9qubit} and Steane~\cite{steane-7qubit}.  I discuss Shor's 
original code and some basics of quantum error-correcting codes in 
chapter~\ref{chap-basics}.  I then go on to describe the formalism of 
stabilizer codes in chapter~\ref{chap-stabilizers}, along with some 
simple examples and methods for creating new codes from old ones.  
Chapter~\ref{chap-encoding} describes how to build networks to encode 
and decode stabilizer codes.  Because we will want to use these codes in 
the operation of quantum computers, in chapter~\ref{chap-fault-tolerant}, 
I will discuss how to perform operations on states encoded using a 
quantum error-correcting code without losing the protection against errors. 
Chapter~\ref{chap-concatenation} describes how to use concatenated codes
to do arbitrarily long calculations as long as the basic error rate is
below some threshhold value, and presents a rough calculation of that 
threshhold. Chapter~\ref{chap-bounds} discusses known upper and lower bounds on 
the existence of stabilizer codes and the channel capacity.  Finally, in 
chapter~\ref{chap-examples}, I will give a partial list of known 
quantum error-correcting codes and their properties.  Appendix~\ref{app-gates}
contains a brief discussion of quantum gates and a list of symbols for
them used in figures.  Appendix~\ref{app-glossary} contains a glossary
of useful terms for discussing quantum error-correcting codes.

Since the promise of quantum computation has attracted scientists from a 
number of fields, including computer science, mathematics, and physics, 
some of the background one group takes for granted may be alien to 
others.  Therefore, in the following two sections, I have provided basic 
introductions to quantum mechanics and classical coding theory.  People 
familiar with one or both fields should skip the appropriate section(s).  For 
a more complete treatment of quantum mechanics, 
see~\cite{cohen-tannoudji}.  For a more complete treatment of classical 
error-correcting codes, see~\cite{macwilliams-sloane}.

\section{Introduction to Quantum Mechanics}
\label{sec-QM}

The state of a classical computer is a string of 0s and 1s, which is a vector 
over the finite field ${\bf Z}_2$.  The state of a quantum computer (or any 
quantum system) is instead a vector over the complex numbers $\bf{C}$.
Actually, a quantum state lies in a Hilbert space, since there is an inner
product (which I will define later).  The state is usually written 
$\ket{\psi}$, which is called a {\em ket}.  A classical computer with $n$ 
bits has $2^n$ possible states, but this is only an $n$-dimensional vector 
space over ${\bf Z}_2$.  A quantum computer with $n$ qubits is a state in 
a $2^n$-dimensional complex vector space.  For a single qubit, the standard 
basis vectors are written as $\ket{0}$ and $\ket{1}$.  An arbitrary 
single-qubit state is then
\begin{equation}
\alpha \ket{0} + \beta \ket{1}.
\label{eq-1qubit}
\end{equation}
$\alpha$ and $\beta$ are complex numbers, with $|\alpha|^2 + |\beta|^2 = 
1$.  This is a {\em normalized} state.  With multiple qubits, we can have 
states that cannot be written as the product of single-qubit states.  For 
instance,
\begin{equation}
\frac{1}{\sqrt{2}} \left(\ket{00} + \ket{11} \right)
\label{eq-2qubit}
\end{equation}
cannot be decomposed in this way.  Such a state is said to be {\em 
entangled}.  Entangled states are what provide a quantum computer with 
its power.  They will also play a major role in quantum error correction.
The particular state (\ref{eq-2qubit}) is called an Einstein-Podalsky-Rosen
pair (or EPR) pair, and serves as a useful basic unit of entanglement in
many applications.

If we make a measurement on the qubit in equation (\ref{eq-1qubit}), we get 
a classical number corresponding to one of the basis states.  The measurement 
disturbs the original state, which collapses into the basis state corresponding 
to the measurement outcome.  If we measure the state~(\ref{eq-1qubit}), the 
outcome will be 0 with probability $|\alpha|^2$, and it will be 1 with 
probability $|\beta|^2$.  The normalization ensures that the probability of 
getting some result is exactly 1.  Through most of this thesis, I will instead 
write down unnormalized states.  These states will stand for the 
corresponding normalized states, which are formed by multiplying the 
unnormalized states by an appropriate constant.  The overall phase of a
state vector has no physical significance.

The measurement we made implements one of two projection operators, 
the projections on the basis $\ket{0},\ \ket{1}$.  This is not the only 
measurement we can make on a single qubit.  In fact, we can project on 
any basis for the Hilbert space of the qubit.  If we have multiple qubits, we 
can measure a number of different qubits independently, or we can 
measure some joint property of the qubits, which corresponds to projecting 
on some entangled basis of the system.  Note that the projection on the
basis $\ket{0},\ \ket{1}$ for either qubit destroys the entanglement of
the state (\ref{eq-2qubit}), leaving it in a tensor product state.

A particularly fruitful way to understand a quantum system is to look at 
the behavior of various operators acting on the states of the system.  For 
instance, a nice set of operators to consider for a single qubit is the 
set of Pauli spin matrices
\begin{equation}
\sigma_x = \pmatrix{ 0 & 1 \cr 1 & 0},\ 
\sigma_y = \pmatrix{ 0 & -i \cr i & \ 0},\ {\rm and}\ 
\sigma_z = \pmatrix{ 1 & \ 0 \cr 0 & -1}.
\end{equation}
The original measurement I described corresponds to measuring the 
eigenvalue of $\sigma_z$.  The corresponding projection operators are 
$\frac{1}{2}(I \pm \sigma_z)$.  If we have a spin-$1/2$ particle, this 
measurement is performed by measuring the spin of the particle along the 
$z$ axis.  We could also measure along the $x$ or $y$ axis, which 
corresponds to measuring the eigenvalue of $\sigma_x$ or $\sigma_y$.  
The projections are $\frac{1}{2}(I \pm \sigma_x)$ and $\frac{1}{2}(I \pm 
\sigma_y)$.

We can also make measurements of more general operators, provided they 
have real eigenvalues.  A matrix $A$ has real eigenvalues iff it is 
Hermitian: $A^\dagger = A$, where $A^\dagger$ is the {\em Hermitian 
adjoint} (or just {\em adjoint}), equal to the complex conjugate transpose.  
Note that all of the Pauli spin matrices are Hermitian.

The Pauli matrices also satisfy an important algebraic property --- they 
{\em anticommute} with each other.  That is,
\begin{equation}
\{\sigma_i, \sigma_j \} = \sigma_i \sigma_j + \sigma_j \sigma_i = 0
\end{equation}
whenever $i \neq j$ (with $i, j \in \{ x, y, z \}$).  Another possible 
relationship between two operators $A$ and $B$ is for them to {\em 
commute}.  That is,
\begin{equation}
[A, B] = A B - B A = 0.
\end{equation}
It is possible for two matrices to neither commute nor anticommute, and, 
in fact, this is the generic case.  Two commuting matrices can be 
simultaneously diagonalized.  This means that we can measure the 
eigenvalue of one of them without disturbing the eigenvectors of the other.  
Conversely, if two operators do not commute, measuring one will disturb 
the eigenvectors of the other, so we cannot simultaneously measure 
non-commuting operators.

There is a natural complex inner product on quantum states.  Given an
orthonormal basis $\ket{\psi_i}$, the inner product between $\ket{\alpha} 
= \Sum c_i \ket{\psi_i}$ and $\ket{\beta} = \Sum d_i \ket{\psi_i}$ is
\begin{equation}
\langle \alpha \ket{\beta} = \Sum c_i^* d_j \langle \psi_i \ket{\psi_j}
= \Sum c_i^* d_i.
\end{equation}
Each ket $\ket{\psi}$ corresponds to a {\em bra} $\bra{\psi}$ and the 
Hermitian adjoint is the adjoint with respect to this inner product, so
$U \ket{\psi}$ corresponds to $\bra{\psi} U^\dagger$.  The operator  
$\Sum \ket{\psi} \bra{\phi}$ acts on the Hilbert space as follows:
\begin{equation}
\left( \Sum \ket{\psi} \bra{\phi} \right) \ket{\alpha} =
\Sum \langle \phi \ket{\alpha} \ \ket{\psi}.
\end{equation}
The inner product can reveal a great deal of information about the
structure of a set of states.  For instance, $\langle \psi \ket{\phi} = 1$
if and only if $\ket{\psi} = \ket{\phi}$.

Eigenvectors of a Hermitian operator $A$ with different eigenvalues are 
automatically orthogonal:
\begin{eqnarray}
\bra{\psi} A \ket{\phi} & = & \bra{\psi} \big( A \ket{\phi} \big) = 
\lambda_{\phi} \langle \psi \ket{\phi} \\
& = & \big( \bra{\psi} A \big) \ket{\phi} = \lambda_{\psi}^* \langle \psi 
\ket{\phi}.
\end{eqnarray}
Since the eigenvalues of $A$ are real, it follows that 
$\langle \psi \ket{\phi} = 0$ whenever $\lambda_{\phi} \neq \lambda_{\psi}$.  
Conversely, if $\langle \psi \ket{\phi} = 0$, there exists a Hermitian operator 
for which $\ket{\psi}$ and $\ket{\phi}$ are eigenvectors with different 
eigenvalues.

We often want to consider a subsystem ${\cal A}$ of a quantum system 
${\cal B}$.  Since ${\cal A}$ may be entangled with the rest of the system, it 
is not meaningful to speak of the ``state'' of ${\cal A}$.  If we write the state of ${\cal B}$ as $\Sum \ket{\psi_i} \ket{\phi_i}$, where $\ket{\psi_i}$ 
is an orthonormal basis for ${\cal B} - {\cal A}$, and $\ket{\phi_i}$ are 
possible states for ${\cal A}$, then to an observer who only interacts with 
the subsystem ${\cal A}$, the subsystem appears to be in just one of the states 
$\ket{\phi_i}$ with some probability.  ${\cal A}$ is said to be in a 
{\em mixed state} as opposed to the {\em pure state} of a closed system in 
a definite state.

We can extend the formalism to cover mixed states by introducing the 
{\em density matrix} $\rho$.  For a pure system in the state $\ket{\psi}$, 
the density matrix is $\ket{\psi} \bra{\psi}$.  The density matrix for the 
subsystem for the entangled state above is $\Sum \ket{\phi_i} 
\bra{\phi_i}$.  Density matrices are always positive and have 
${\rm tr}\:\rho = 1$.  To find the density matrix of a subsystem given the 
density matrix of the full system, simply trace over the degrees of freedom of 
the rest of the system.

Given a closed quantum system, time evolution preserves the inner 
product, so the time evolution operator $U$ must be unitary.  That is, 
$U^\dagger U = U U^\dagger = I$.  An open system can be described as a 
subsystem of a larger closed system, so the evolution of the open system 
descends from the global evolution of the full system.  Time evolution of 
the subsystem is described by some {\em superoperator} acting 
on the density matrix of the subsystem.

One fact about quantum states that has profound implications for quantum
computation is that it is impossible to make a copy of an arbitrary
unknown quantum state.  This is known as the ``No Cloning Theorem,''
\cite{no-cloning} and is a consequence of the linearity of quantum mechanics.
The proof is straightforward:  Suppose we wish to have an operation that
maps an arbitrary state 
\begin{equation}
\ket{\psi} \rightarrow \ket{\psi} \otimes \ket{\psi}.
\end{equation}
Then arbitrary $\ket{\phi}$ is mapped by
\begin{equation}
\ket{\phi} \rightarrow \ket{\phi} \otimes \ket{\phi}
\end{equation}
as well.  Because the transformation must be linear, it follows that
\begin{equation}
\ket{\psi} + \ket{\phi} \rightarrow \ket{\psi} \otimes \ket{\psi} +
\ket{\phi} \otimes \ket{\phi}.
\end{equation}
However,
\begin{equation}
\ket{\psi} \otimes \ket{\psi} + \ket{\phi} \otimes \ket{\phi} \neq
(\ket{\psi} + \ket{\phi}) \otimes (\ket{\psi} + \ket{\phi}),
\end{equation}
so we have failed to copy $\ket{\psi} + \ket{\phi}$.  In general, if we
pick an orthonormal basis, we can copy the basis states, but we will
not have correctly copied superpositions of those basis states.  We will
instead have either measured the original system and therefore destroyed the
superposition, or we will have produced a state that is entangled between
the original and the ``copy.''  This means that to perform quantum error
correction, we cannot simply make backup copies of the quantum state to
be preserved.  Instead, we must protect the original from any likely error.

\section{Introduction to Classical Coding Theory}
\label{sec-classical}

Classical coding theory tends to concentrate on {\em linear codes}, a 
subclass of all possible codes with a particular relation between codewords.  
Suppose we wish to encode $k$ bits using $n$ bits.  The data can be 
represented as a $k$-dimensional binary vector $v$.  Because we are 
dealing with binary vectors, all the arithmetic is mod two.  For a linear code, 
the encoded data is then $G v$ for some $n \times k$ matrix $G$ (with 
entries from ${\bf Z}_2$), which is independent of $v$.  $G$ is called the 
{\em generator matrix} for the code.  Its columns form a basis for the 
$k$-dimensional coding subspace of the $n$-dimensional binary vector space, 
and represent basis codewords.  The most general possible codeword is an
arbitrary linear combination of the basis codewords; thus the name ``linear
code.''

Given a generator matrix $G$, we can calculate the dual matrix $P$, which 
is an $(n-k) \times n$ matrix of 0s and 1s of maximal rank $n-k$ with $P G = 
0$.  Since any codeword $s$ has the form $G v$, $P s = P G v = 0 v = 0$, and 
$P$ annihilates any codeword.  Conversely, suppose $P s = 0$.  Since $P$ 
has rank $n-k$, it only annihilates a $k$-dimensional space spanned by 
the columns of $G$, and $s$ must be a linear combination of these columns.  
Thus, $s = G v$ for some $v$, and $s$ is a valid codeword.  The matrix $P$ 
is called the {\em parity check matrix} for the code.  It can be used to test 
if a given vector is a valid codeword, since $P s = 0$ iff $s$ is a codeword.   
The {\em dual code} is defined to be the code with generator matrix $P^T$ 
and parity matrix $G^T$.

In order to consider the error-correcting properties of a code, it is useful to 
look at the {\em Hamming distance} between codewords.  The Hamming 
distance between two vectors is the minimum number of bits that must be 
flipped to convert one vector to the other.  The distance between $a$ and 
$b$ is equal to the {\em weight} (the number of 1s in the vector) of $a+b$.  
For a code to correct $t$ single-bit errors, it must have distance at least 
$2t+1$ between any two codewords.  A $t$ bit error will take a codeword exactly 
distance $t$ away from its original value, so when the distance between 
codewords is at least $2t+1$, we can distinguish errors on different 
codewords and correct them to the proper codewords.  A code to encode 
$k$ bits in $n$ bits with minimum distance $d$ is said to be an $[n, k, d]$ 
code.

Now suppose we consider a $t$ bit error.  We can write down a vector $e$ 
to describe this vector by putting ones in the places where bits are flipped
and zeros elsewhere.  Then if the original codeword is $s$, after the error it 
is $s' = s + e$.  If we apply the parity check matrix, we get
\begin{equation}
P s' = P (s + e) = P s + P e = 0 + P e = P e,
\end{equation}
so the value of $P s'$ does not depend on the value of $s$, only on $e$.  If 
$P e$ is different for all possible errors $e$, we will be able to determine 
precisely what error occurred and fix it.  $P e$ is called the {\em error 
syndrome}, since it tells us what the error is.  Since $P e = P f$ iff $P (e-f) 
= 0$, to have a code of distance $d$, we need $P e \neq 0$ for all vectors $e$ 
of weight $d-1$ or less.  Equivalently, any $d-1$ columns of $P$ must be 
linearly independent.

We can place upper and lower bounds on the existence of linear codes to 
correct $t$ errors.  Each of the $2^k$ codewords has a {\em Hamming 
sphere} of radius $t$.  All the words inside the Hamming sphere come from 
errors acting on the same codeword.  For a code on $n$ bits, there are $n$ 
one-bit errors, \mbox{\tiny $\pmatrix{n \cr 2}$} two-bit errors, and in 
general \mbox{\tiny $\pmatrix{n \cr j}$} $j$-bit errors.  The Hamming 
spheres cannot overlap, but they must all fit inside the vector space, which 
only has $2^n$ elements.  Thus,
\begin{equation}
\Sum_{j=0}^{t} \pmatrix{n \cr j} 2^k \leq 2^n.
\end{equation}
This is called the {\em Hamming bound} on $[n, k, 2t+1]$ codes.  As $n$, 
$k$, and $t$ get large, this bound approaches the asymptotic form
\begin{equation}
\frac{k}{n} \leq 1 - H \left( \frac{t}{n} \right),
\label{eq-Hamming}
\end{equation}
where $H(x)$ is the {\em Hamming entropy}
\begin{equation}
H(x) = - x \log_2 x - (1 - x) \log_2 (1-x). 
\label{eq-H(x)}
\end{equation}

We can set a lower bound on the existence of $[n, k, 2t+1]$ linear codes as 
well, called the {\em Gilbert-Varshamov} bound.  Suppose we have such a 
code (if necessary with $k=0$) with
\begin{equation}
\Sum_{j=0}^{2t} \pmatrix{n \cr j} 2^k < 2^n.
\end{equation}
Then the spheres of distance $2t$ around each codeword do not fill the 
space, so there is some vector $v$ that is at least distance $2t +1$ from 
each of the other codewords.  In addition, $v + s$ (for any codeword $s$) is 
at least distance $2t +1$ from any other codeword $s'$, since the distance 
is just $(v + s) + s' = v + (s + s')$, which is the distance between $v$ and 
the codeword $s + s'$.  This means that we can add $v$ and all the vectors 
$v+s$ to the code without dropping the distance below $2t + 1$.  This 
gives us an $[n, k+1, 2t+1]$ code.  We can continue this process until
\begin{equation}
\Sum_{j=0}^{2t} \pmatrix{n \cr j} 2^k \geq 2^n.
\label{eq-Gilbert-Varshamov}
\end{equation}
Asymptotically, this becomes
\begin{equation}
\frac{k}{n} \geq 1 - H \left( \frac{2t}{n} \right).
\end{equation}

Another case of great interest is the capacity of a classical channel.  This 
is equal to the {\em efficiency} $k/n$ of the most efficient code on an 
asymptotically large block that corrects measure one of the errors occuring.  
For instance, a common channel is the {\em binary symmetric channel}, where an 
error occurs independently on each bit with probability $p$ for both $0$ and 
$1$.  Shannon showed that channel capacity is just equal to one minus the 
entropy introduced by the channel~\cite{shannon}.  For the binary symmetric 
channel, the entropy is just the Hamming entropy $H(p)$, so the capacity is 
$1-H(p)$, coinciding with the Hamming bound for the expected number of errors 
$t = pn$.  Shannon also showed that the capacity of a channel can be achieved 
by choosing codewords at random, then discarding only a few of them (measure 
zero asymptotically).

\chapter{Basics of Quantum Error Correction}
\label{chap-basics}

\section{The Quantum Channel}

Now we turn to the quantum channel.  A noisy quantum channel can be a 
regular communications channel which we expect to preserve at least some 
degree of quantum coherence, or it can be the passage of time as a set of 
qubits sits around, interacting with its environment, or it can be the result 
of operating with a noisy gate on some qubits in a quantum computer.  
In any of these cases, the input of a pure quantum state can produce a 
mixed state as output as the data qubits become entangled with the 
environment.  Even when a pure state comes out, it might not be the same 
state as the one that went in.

At first it appears that trying to correct a mixed state back into the correct
pure state is going to be harder than correcting an erroneous pure state, but 
this is not the case.  The output mixed state can be considered as an ensemble 
of pure states.  If we can correct each of the pure states in the ensemble 
back to the original input state, we have corrected the full mixed state.  
Another way of phrasing this is to say the channel applies a 
superoperator to the input density matrix.  We can diagonalize 
this superoperator and write it as the direct sum of a number of 
different matrices acting directly on the possible input pure states with 
various probabilities.  If the code can correct any of the possible matrices, 
it can correct the full superoperator.  A key point is that the 
individual matrices need not be unitary.  From now on, I will only consider 
the effects of a (possibly non-unitary) matrix acting on a pure state.

\section{A Simple Code}

For the moment, let us consider only channels which cause an error on a 
single qubit at a time.  We wish to protect a single logical qubit against 
error.  We cannot send it through the channel as is, because the one qubit 
that is affected might be the one we want to keep.  Suppose we send through 
nine qubits after encoding the logical qubit as follows:
\begin{eqnarray}
\ket{0} & \rightarrow & \ket{\overline{0}} = (\ket{000} + \ket{111}) 
(\ket{000} + \ket{111}) (\ket{000} + \ket{111}) \\
\ket{1} & \rightarrow & \ket{\overline{1}} = (\ket{000} - \ket{111}) 
(\ket{000} - \ket{111}) (\ket{000} - \ket{111}).
\end{eqnarray}
The data is no longer stored in a single qubit, but instead spread out 
among nine of them.  Note that even if we know the nine qubits are in one 
of these two states, we cannot determine which one without making a 
measurement on at least three qubits.  This code is due to 
Shor~\cite{shor-9qubit}.

Suppose the channel flips a single qubit, say the first one, switching 
$\ket{0}$ and $\ket{1}$.  Then by comparing the first two qubits, we find 
they are different, which is not allowed for any valid codeword.  Therefore 
we know an error occurred, and furthermore, it flipped either the first or 
second qubit.  Note that we do not actually measure the first and second 
qubits, since this would destroy the superposition in the codeword; we just 
measure the difference between them.

Now we compare the first and third qubits.  Since the first qubit was 
flipped, it will disagree with the third; if the second qubit had been 
flipped, the first and third would have agreed.  Therefore, we have 
narrowed down the error to the first qubit and we can fix it simply by 
flipping it back.  To handle possible bit flips on the other blocks of three, 
we do the same comparisons inside the other blocks.

However, this is not the only sort of error that could have occurred.  The 
channel might have left the identity of the 0 and 1 alone, but altered their 
relative phase, introducing, for instance, a relative factor of $-1$ when 
the first qubit is $\ket{1}$.  Then the two basis states become
\begin{eqnarray}
\ket{\overline{0}} & \rightarrow & (\ket{000} - \ket{111}) (\ket{000} + 
\ket{111}) (\ket{000} + \ket{111}) \\
\ket{\overline{1}} & \rightarrow & (\ket{000} + \ket{111}) (\ket{000} - 
\ket{111}) (\ket{000} - \ket{111}).
\end{eqnarray}
By comparing the sign of the first block of three with the second block of 
three, we can see that a sign error has occurred in one of those blocks.  
Then by comparing the signs of the first and third blocks of three, we 
narrow the sign error down to the first block, and flip the sign back to 
what it should be.  Again, we do not want to actually measure the signs, 
only whether they agree.  In this case, measuring the signs would give us 
information about whether the state is $\ket{\overline{0}}$ or 
$\ket{\overline{1}}$, which would destroy any superposition between 
them.

This does not exhaust the list of possible one qubit errors.  For instance, we 
could have both a bit flip and a sign flip on the same qubit.  However, by 
going through both processes described above, we will fix first the bit flip, 
then the sign flip (in fact, this code will correct a bit flip and a sign flip 
even if they are on different qubits).  The original two errors can be 
described as the operation of
\begin{equation}
\X = \pmatrix{0 & 1 \cr 1 & 0} \ {\rm and}\ 
\Z = \pmatrix{1 & \ 0 \cr 0 & -1}.
\end{equation}
The simultaneous bit and sign flip is
\begin{equation}
\Y = i \X \Z = \pmatrix{0 & -i \cr i & \ 0}.
\end{equation}
Sometimes I will write $\Xs{i}$, $\Ys{i}$, or $\Zs{i}$ to represent $\X$, $\Y$,
or $\Z$ acting on the $i$th qubit.

The most general one-qubit error that can occur is some $2 \times 2$ 
matrix; but such a matrix can always be written as the (complex) linear 
combination of $\X$, $\Y$, $\Z$, and the $2 \times 2$ identity matrix $I$.  
Consider what happens to the code when such an error occurs:
\begin{equation}
\ket{\psi} = \alpha \ket{\overline{0}} + \beta \ket{\overline{1}}
\rightarrow a \Xs{i} \ket{\psi} + b \Ys{i} \ket{\psi} + c \Zs{i} \ket{\psi} +
d \ket{\psi}.
\end{equation}
Suppose we perform the process above, comparing bits within a block of 
three, and comparing the signs of blocks of three.  This acts as a 
measurement of which error (or the identity) has occurred, causing the 
state, originally in a superposition, to collapse to $\Xs{i} \ket{\psi}$ with 
probability $|a|^2$, to $\Ys{i} \ket{\psi}$ with probability $|b|^2$, to 
$\Zs{i} \ket{\psi}$ with probability $|c|^2$, and to $\ket{\psi}$ with
probability $|d|^2$.  In any of the four cases, we have determined which error
occurred and we can fix it.

\section{Properties of Any Quantum Code}
\label{sec-general-prop}

Now let us consider properties of more general codes.  A code to encode 
$k$ qubits in $n$ qubits will have $2^k$ basis codewords corresponding to 
the basis of the original states.  Any linear combination of these basis 
codewords is also a valid codeword, corresponding to the same linear 
combination of the unencoded basis states.  The space $T$ of valid 
codewords (the {\em coding space}) is therefore a Hilbert space in its own 
right, a subspace of the full $2^n$-dimensional Hilbert space.  As with 
Shor's nine-qubit code, if we can correct errors $E$ and $F$, we can correct 
$aE + bF$, so we only need to consider whether the code can correct a basis of 
errors.  One convenient basis to use is the set of tensor products of $\X$, 
$\Y$, $\Z$, and $I$.  The {\em weight} of an operator of this form is the 
number of qubits on which it differs from the identity.  The set of all these 
tensor products with a possible overall factor of $-1$ or $\pm i$ forms a 
group $\G$ under multiplication.  $\G$ will play a major role in the 
stabilizer formalism.  Sometimes I will write it $\G_n$ to distinguish the 
groups for different numbers of qubits.  $\G_1$ is just the quaternionic 
group; $\G_n$ is the direct product of $n$ copies of the quaternions 
modulo all but a global phase factor.

In order for the code to correct two errors $E_a$ and $E_b$, we must 
always be able to distinguish error $E_a$ acting on one basis codeword 
$\ket{\psi_i}$ from error $E_b$ acting on a different basis codeword 
$\ket{\psi_j}$.  We can only be sure of doing this if $E_a \ket{\psi_1}$ is 
orthogonal to $E_b \ket{\psi_2}$; otherwise there is some chance of 
confusing them.  Thus,
\begin{equation}
\bra{\psi_i} E_a^\dagger E_b \ket{\psi_j} = 0
\label{eq-cond-orthogonal}
\end{equation}
when $i \neq j$ for correctable errors $E_a$ and $E_b$.  Note that we 
normally include the identity in the set of possible ``errors,'' since we do
not want to confuse an error on one qubit with nothing happening to another.  
If we have a channel in which we are certain {\em some} error occurred, 
we do not need to include the identity as a possible error.  In any case,
the set of correctable errors is unlikely to be a group --- it does not
even need to be closed under multiplication.

However, (\ref{eq-cond-orthogonal}) is insufficient to guarantee a code will 
work as a quantum error-correcting code.  When we make a measurement to find 
out about the error, we must learn nothing about the actual state of the code 
within the coding space.  If we did learn something, we would be disturbing 
superpositions of the basis states, so while we might correct the basis 
states, we would not be correcting an arbitrary valid codeword.  We learn 
information about the error by measuring $\bra{\psi_i} E_a^\dagger E_b 
\ket{\psi_i}$ for all possible errors $E_a$ and $E_b$.  This quantity must 
therefore be the same for all the basis codewords:
\begin{equation}
\bra{\psi_i} E_a^\dagger E_b \ket{\psi_i} = \bra{\psi_j} E_a^\dagger E_b 
\ket{\psi_j}.
\label{eq-cond-structure}
\end{equation}
We can combine equations (\ref{eq-cond-orthogonal}) and 
(\ref{eq-cond-structure}) into a single equation:
\begin{equation}
\bra{\psi_i} E_a^\dagger E_b \ket{\psi_j} = C_{ab} \delta_{ij},
\label{eq-condition}
\end{equation}
where $\ket{\psi_i}$ and $\ket{\psi_j}$ run over all possible basis 
codewords, $E_a$ and $E_b$ run over all possible errors, and $C_{ab}$ is 
independent of $i$ and $j$.  This condition was found by Knill and 
Laflamme~\cite{knill-laflamme-theory} and Bennett {\it et 
al.}~\cite{bennett-tome}.

The above argument shows that (\ref{eq-condition}) is a necessary 
condition for the code to correct the errors $\{E_a\}$.  It is also a 
sufficient condition:  The matrix $C_{ab}$ is Hermitian, so it can be 
diagonalized.  If we do this and rescale the errors $\{E_a\}$ appropriately, we 
get a new basis $\{F_a\}$ for the space of possible errors, with either
\begin{equation}
\bra{\psi_i} F_a^\dagger F_b \ket{\psi_j} = \delta_{ab} \delta_{ij}
\end{equation}
or
\begin{equation}
\bra{\psi_i} F_a^\dagger F_b \ket{\psi_j} = 0,
\end{equation}
depending on $a$.  Note that this basis will not necessarily contain 
operators that are tensor products of one-qubit operators.  Errors of the 
second type actually annihilate any codeword, so the probability of one 
occuring is strictly zero and we need not consider them.  The other errors 
always produce orthogonal states, so we can make some measurement that 
will tell us exactly which error occurred, at which point it is a simple 
matter to correct it.  Therefore, a code satisfies equation 
(\ref{eq-condition}) for all $E_a$ and $E_b$ in some set ${\cal E}$ iff the
code can correct all errors in ${\cal E}$.

Another minor basis change allows us to find a basis where any two errors 
acting on a given codeword either produce orthogonal states or exactly the 
same state.  The errors $F_a$  that annihilate codewords correspond to two 
errors that act the same way on codewords.  For instance, in Shor's 
nine-qubit code, $\Zs{1}$ and $\Zs{2}$ act the same way on the code, so $\Zs{1} 
- \Zs{2}$ will annihilate codewords.  This phenomenon will occur iff $C_{ab}$ 
does not have maximum rank.  A code for which $C_{ab}$ is singular is 
called a {\em degenerate} code, while a code for which it is not is {\em 
nondegenerate}.  Shor's nine-qubit code is degenerate; we will see many 
examples of nondegenerate codes later.  Note that whether a code is 
degenerate or not depends on the set of errors it is intended to correct.  
For instance, a two-error-correcting degenerate code might be 
nondegenerate when considered as a one-error-correcting code.

In equation~(\ref{eq-condition}), $E = E_a^\dagger E_b$ is still in the group 
$\G$ when $E_a$ and $E_b$ are in $\G$.  The weight of the smallest $E$ in 
$\G$ for which (\ref{eq-condition}) does {\em not} hold is called the {\em 
distance} of the code.  A quantum code to correct up to $t$ errors must 
have distance at least $2t+1$.  Every code has distance at least one.  A
distance $d$ code encoding $k$ qubits in $n$ qubits is described as an
$[n, k, d]$ code.  Note that a quantum $[n,k,d]$ code is often written
in the literature as $[[n,k,d]]$ to distinguish it from a classical
$[n,k,d]$ code.  I have chosen the notation $[n,k,d]$ to emphasize the
similarities with the classical theory; when I need to distinguish, I
will do so using the words ``quantum'' and ``classical.''

We can also consider variations of the usual error-correction problem.  
For instance, suppose we only want to detect if an error has occurred, not 
to correct it.  This could, for instance, be used to prevent errors using the 
quantum Zeno effect~\cite{vaidman}.  In this case, we do not need to 
distinguish error $E_a$ from $E_b$, only from the identity.  We can use the 
same argument to find (\ref{eq-condition}), only now $E_b = I$ always.  
This means a code to detect $s$ errors must have distance at least $s+1$.  
Another variation is when we know in which qubit(s) an error has 
occurred, as in the quantum erasure channel~\cite{grassl}.  In this case, we 
only need distinguish $E_a$ from those $E_b$ affecting the same qubits.  
This means that $E_a^\dagger E_b$ has the same weight as $E_a$, and to correct 
$r$ such located errors, we need a code of distance at least $r+1$.  We can 
also imagine combining all of these tasks.  A code to correct $t$ arbitrary 
errors, $r$ additional located errors, and detect a further $s$ errors must 
have distance at least $r + s + 2t + 1$.

\section{Error Models}

In this thesis, I will mostly assume that errors occur independently on 
different qubits, and that when an error occurs on a qubit, it is equally 
likely to be a $\X$, $\Y$, or $\Z$ error.  If the probability $\epsilon$ of 
error per qubit is fairly small, it is often useful to simply ignore the 
possibility of more than $t$ errors, since this only occurs with probability 
$O(\epsilon^{t+1})$.  Thus, I will typically deal with codes that correct up 
to $t$ arbitrary errors.  Such a code will handle any error on up to $t$ qubits 
that leaves the data somewhere in the normal computational space (although 
moving it outside of the space of valid codewords).

In some systems, there will be errors that move the system outside of the
computational space.  For instance, if the data is stored as the ground or
metastable excited state of an ion, the electron might instead end up in a 
different excited state.  If the data is stored in the polarization of a
photon, the photon might escape.  In both of these cases, the normal error 
correction networks will not function properly, since they assume that the
qubit is either in the state $\ket{0}$ or $\ket{1}$.  However, by performing
some measurement that distinguishes between the computational Hilbert space
and other possible states, we can determine not only that this sort of
{\em leakage error} has occurred, but also on which qubit it has occurred.
Then we can cool the atom to the ground state or introduce a new photon with
random polarization, and the error becomes a located error, which was
discussed at the end of the previous section.  One possible network of gates 
to detect a leakage error is given in figure~\ref{fig-leakage} (see appendix
\ref{app-gates} for a description of the symbols used in this and later
figures).
\begin{figure}
\centering
\begin{picture}(120,60)

\put(0,34){\makebox(20,12){$\ket{\psi}$}}
\put(0,14){\makebox(20,12){$\ket{0}$}}

\put(20,20){\line(1,0){100}}
\put(20,40){\line(1,0){100}}

\put(40,40){\circle*{4}}
\put(40,40){\line(0,-1){24}}
\put(40,20){\circle{8}}

\put(60,40){\circle{8}}
\put(60,36){\line(0,1){8}}

\put(80,40){\circle*{4}}
\put(80,40){\line(0,-1){24}}
\put(80,20){\circle{8}}

\put(100,40){\circle{8}}
\put(100,36){\line(0,1){8}}

\end{picture}
\caption{Network to detect leakage errors.}
\label{fig-leakage}
\end{figure}
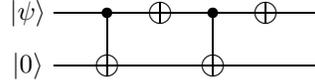
This network asssumes that states outside the normal computational space do
not interact at all with other qubits.  If the data state $\ket{\psi}$ is
either $\ket{0}$ or $\ket{1}$, the ancilla qubit will flip and become $\ket{1}$.
If the data state is neither $\ket{0}$ nor $\ket{1}$, the ancilla will remain
$\ket{0}$, thus signalling a leakage error on this data qubit.

Another possible difficulty arises when correlated errors on multiple qubits
can occur.  While this can in principle be a severe problem, it can be handled 
without a change in formalism as long as the chance of a correlated error drops
rapidly enough with the size of the blocks of errors.  Since a $t$-qubit
error will occur with probability $O(\epsilon^t)$ when the probability of
uncorrelated single-qubit errors is $\epsilon$, as long as the probability
of a $t$-qubit correlated error is $O(\epsilon^t)$, the correlated errors
cause no additional problems.

In real systems, the assumption that errors are equally likely to be $\X$,
$\Y$, and $\Z$ errors is a poor one.  In practice, some linear combinations
of $\X$, $\Y$, and $\Z$ are going to be more likely than others.  For instance,
when the qubits are ground or excited states of an ion, a likely source of
errors is spontaneous emission.  After some amount of time, the excited
state will either decay to the ground state, producing the error $\X + i\Y$
with probability $\epsilon$, or it will not, which changes the relative 
amplitudes of $\ket{0}$ and $\ket{1}$, resulting in the error $I - \Z$ with
probability $O (\epsilon^2)$.  A channel that performs this sort of time
evolution is known as an {\em amplitude damping} channel.  Since the only 
$O(1)$ effect of time evolution is the identity, this sort of error can be 
protected against to lowest order by a code to correct an arbitrary single 
error.  However, codes that take account of the restricted possibilities
for errors can be more efficient than codes that must correct a general
error~\cite{leung}, and understanding the physically likely sources of
error will certainly be an important part of engineering quantum computers.

\chapter{Stabilizer Coding}
\label{chap-stabilizers}

\section{The Nine-Qubit Code Revisited}

Let us look more closely at the procedure we used to correct errors for the 
nine-qubit code.  To detect a bit flip error on one of the first three qubits, 
we compared the first two qubits and the first and third qubits.  This is 
equivalent to measuring the eigenvalues of $\Zs{1} \Zs{2}$ and $\Zs{1} \Zs{3}$.
If the first two qubits are the same, the eigenvalue of $\Zs{1} \Zs{2}$ is 
$+1$; if they are different, the eigenvalue is $-1$.  Similarly, to detect a 
sign error, we compare the signs of the first and second blocks of three and 
the first and third blocks of three.  This is equivalent to measuring the 
eigenvalues of $\Xs{1} \Xs{2} \Xs{3} \Xs{4} \Xs{5} \Xs{6}$ and $\Xs{1} \Xs{2}
\Xs{3} \Xs{7} \Xs{8} \Xs{9}$.  Again, if the signs agree, the eigenvalues will 
be $+1$; if they disagree, the eigenvalues will be $-1$.  In order to totally 
correct the code, we must measure the eigenvalues of a total of eight operators.
They are listed in table~\ref{table-9qubit}.
\begin{table}
\centering
\begin{tabular}{c|ccccccccc}
$M_1$ & $\Z$ & $\Z$ & $I$ & $I$ & $I$ & $I$ & $I$ & $I$ & $I$ \\
$M_2$ & $\Z$ & $I$ & $\Z$ & $I$ & $I$ & $I$ & $I$ & $I$ & $I$ \\
$M_3$ & $I$ & $I$ & $I$ & $\Z$ & $\Z$ & $I$ & $I$ & $I$ & $I$ \\
$M_4$ & $I$ & $I$ & $I$ & $\Z$ & $I$ & $\Z$ & $I$ & $I$ & $I$ \\
$M_5$ & $I$ & $I$ & $I$ & $I$ & $I$ & $I$ & $\Z$ & $\Z$ & $I$ \\
$M_6$ & $I$ & $I$ & $I$ & $I$ & $I$ & $I$ & $\Z$ & $I$ & $\Z$ \\
$M_7$ & $\X$ & $\X$ & $\X$ & $\X$ & $\X$ & $\X$ & $I$ & $I$ & $I$ \\
$M_8$ & $\X$ & $\X$ & $\X$ & $I$ & $I$ & $I$ & $\X$ & $\X$ & $\X$
\end{tabular}
\caption{The stabilizer for Shor's nine-qubit code}
\label{table-9qubit}
\end{table}

The two valid codewords $\ket{\overline{0}}$ and $\ket{\overline{1}}$ in 
Shor's code are eigenvectors of all eight of these operators with eigenvalue 
$+1$.  All the operators in $\G$ that fix both $\ket{\overline{0}}$ and 
$\ket{\overline{1}}$ can be written as the product of these eight operators.  
The set of operators that fix $\ket{\overline{0}}$ and $\ket{\overline{1}}$ 
form a group $S$, called the {\em stabilizer} of the code, and $M_1$ 
through $M_8$ are the generators of this group.

When we measure the eigenvalue of $M_1$, we determine if a bit flip 
error has occurred on qubit one or two, i.e., if $\Xs{1}$ or $\Xs{2}$ has 
occurred.  Note that both of these errors anticommute with $M_1$, while 
$\Xs{3}$ through $\Xs{9}$, which cannot be detected by just $M_1$, commute 
with it.  Similarly, $M_2$ detects $\Xs{1}$ or $\Xs{3}$, which anticommute 
with it, and $M_7$ detects $\Zs{1}$ through $\Zs{6}$.  In general, if 
$M \in S$, $\{M, E\} = 0$, and $\ket{\psi} \in T$, then
\begin{equation}
M E \ket{\psi} = - E M \ket{\psi} = - E \ket{\psi},
\end{equation}
so $E \ket{\psi}$ is an eigenvector of $M$ with eigenvalue $-1$ instead of 
$+1$ and to detect $E$ we need only measure $M$.

The distance of this code is in fact three.  Even a cursory perusal reveals 
that any single-qubit operator $\Xs{i}$, $\Ys{i}$, or $\Zs{i}$ will anticommute 
with one or more of $M_1$ through $M_8$.  Since states with different 
eigenvalues are orthogonal, condition~(\ref{eq-condition}) is satisfied when 
$E_a$ has weight one and $E_b = I$.  We can also check that every two-qubit 
operator $E$ anticommutes with some element of $S$, except for those of 
the form $\Zs{a} \Zs{b}$ where $a$ and $b$ are in the same block of three.  
However, the operators of this form are actually in the stabilizer.  This means 
that $\Zs{a} \Zs{b} \ket{\psi} = \ket{\psi}$ for any codeword $\ket{\psi}$, and 
$\bra{\psi} \Zs{a} \Zs{b} \ket{\psi} = \langle \psi \ket{\psi} = 1$ for all 
codewords $\ket{\psi}$, and these operators also satisfy 
equation~(\ref{eq-condition}).  Since $\Zs{a} \Zs{b}$ is in the stabilizer,
both $\Zs{a}$ and $\Zs{b}$ act the same way on the codewords, and there is no 
need to distinguish them.  When we get to operators of weight three, we do 
find some for which (\ref{eq-condition}) fails.  For instance, $\Xs{1} \Xs{2} 
\Xs{3}$ commutes with everything in $S$, but
\begin{eqnarray}
\bra{\overline{0}} \Xs{1} \Xs{2} \Xs{3} \ket{\overline{0}} & = & +1 \\
\bra{\overline{1}} \Xs{1} \Xs{2} \Xs{3} \ket{\overline{1}} & = & -1.
\end{eqnarray}

\section{The General Stabilizer Code}

The stabilizer construction applies to many more codes than just the 
nine-qubit one~\cite{gottesman-stab,calderbank-stab}.  In general, the 
stabilizer $S$ is some Abelian subgroup of $\G$ and the coding space $T$ is 
the space of vectors fixed by $S$.  Since $\Y$ has imaginary components, 
while $\X$ and $\Z$ are real, with an even number of $\Y$'s in each element
of the stabilizer, all the coefficients in the basis codewords can be chosen
to be real; if there are an odd number of $\Y$'s, they may be imaginary.  
However, Rains has shown that whenever a (possibly complex) code exists, a 
real code exists with the same parameters~\cite{rains-shadow}.  Therefore, I 
will largely restrict my attention to real codes.

For a code to encode $k$ qubits in $n$, $T$ has $2^k$ dimensions and $S$ 
has $2^{n-k}$ elements.  $S$ must be an Abelian group, since only 
commuting operators can have simultaneous eigenvectors, but provided it 
is Abelian and neither $i$ nor $-1$ is in $S$, the space $T = \{ \ket{\psi}\ 
{\rm s.t.} \ M \ket{\psi} = \ket{\psi} \ \forall M \in S \}$ does have 
dimension $2^k$.  At this point it will be helpful to note a few properties of 
$\G$.  Since $\X^2 = \Y^2 = \Z^2 = +1$, every element in $\G$ squares to 
$\pm 1$.  Also, $\X$, $\Y$, and $\Z$ on the same qubit anticommute, while 
they commute on different qubits.  Therefore, any two elements of $\G$ 
either commute or they anticommute.  $\X$, $\Y$, and $\Z$ are all 
Hermitian, but of course $(iI)^\dagger = -i I$, so elements of $\G$ can be 
either Hermitian or anti-Hermitian.  In either case, if $A \in \G$, 
$A^\dagger \in G$ also.  Similarly, $\X$, $\Y$, and $\Z$ are all unitary, so 
every element of $\G$ is unitary.

As before, if $M \in S$, $\ket{\psi_i} \in T$, and $\{M, E \} = 0$, then
$M E \ket{\psi_i} =- E \ket{\psi_i}$, so
\begin{equation}
\bra{\psi_i} E \ket{\psi_j} = \bra{\psi_i} M E \ket{\psi_j} = - \bra{\psi_i} E 
\ket{\psi_j} = 0.
\end{equation}
Therefore the code satisfies~(\ref{eq-cond-orthogonal}) whenever $E = 
E_a^\dagger E_b = \pm E_a E_b$ anticommutes with $M$ for some $M \in 
S$.  In fact, in such a case it also satisfies~(\ref{eq-cond-structure}), since 
$\bra{\psi_i} E \ket{\psi_i} = \bra{\psi_j} E \ket{\psi_j} = 0$.  Therefore, if 
$E_a^\dagger E_b$ anticommutes with some element of $S$ for all errors 
$E_a$ and $E_b$ in some set, the code will correct that set of errors.

Of course, strictly speaking, this is unlikely to occur.  Generally, $I$ will 
be an allowed error, and $E = I^\dagger I$ commutes with everything.  
However, $S$ is a group, so $I \in S$.  In general, if $E \in S$,
\begin{equation}
\bra{\psi_i} E \ket{\psi_j} = \langle \psi_i \ket{\psi_j} = \delta_{ij}.
\end{equation}
This will satisfy equation~(\ref{eq-condition}) also.

Now, there generally are many elements of $\G$ that commute with 
everything in $S$ but are not actually in $S$.  The set of elements in $\G$ 
that commute with all of $S$ is defined as the centralizer $C(S)$ of $S$ in 
$\G$.  Because of the properties of $S$ and $\G$, the centralizer is actually 
equal to the normalizer $N(S)$ of $S$ in $\G$, which is defined as the set of 
elements of $\G$ that fix $S$ under conjugation.  To see this, note that for
any $A \in \G$, $M \in S$, 
\begin{equation}
A^\dagger M A = \pm A^\dagger A M = \pm M.
\end{equation}
Since $-1 \notin S$, $A \in N(S)$ iff $A \in C(S)$, so $N(S) = C(S)$.  Note 
that $S \subseteq N(S)$.  In fact, $S$ is a normal subgroup of $N(S)$.  
$N(S)$ contains $4 \cdot 2^{n+k}$ elements.  The factor of four is for the 
overall phase factor.  Since an overall phase has no effect on the physical
quantum state, often, when considering $N(S)$, I will only really 
consider $N(S)$ without this global phase factor.

If $E \in N(S)-S$, then $E$ rearranges elements of $T$ but does not take 
them out of $T$: if $M \in S$ and $\ket{\psi} \in T$, then
\begin{equation}
M E \ket{\psi} = EM \ket{\psi} = E \ket{\psi},
\end{equation}
so $E \ket{\psi} \in T$ also.  Since $E \notin S$, there is some state in $T$ 
that is not fixed by $E$.  Unless it differs from an element of $S$ by an 
overall phase, $E$ will therefore be undetectable by this code.

Putting these considerations together, we can say that a quantum code 
with stabilizer $S$ will detect all errors $E$ that are either in $S$ or 
anticommute with some element of $S$.  In other words, $E \in S \cup (\G - 
N(S))$.  This code will correct any set of errors $\{ E_i \}$ iff $E_a E_b \in 
S \cup (\G - N(S)) \ \forall E_a, E_b$ (note that $E_a^\dagger E_b$ commutes 
with $M \in \G$ iff $E_a E_b = \pm E_a^\dagger E_b$ does).  For instance, the 
code will have distance $d$ iff $N(S) - S$ contains no elements of weight less 
than $d$.  If $S$ has elements of weight less than $d$ (except the identity), 
it is a degenerate code; otherwise it is a nondegenerate code.  For instance, 
the nine-qubit code is degenerate, since it has distance three and $\Zs{1} 
\Zs{2} \in S$.  A nondegenerate stabilizer code satisfies
\begin{equation}
\bra{\psi_i} E_a^\dagger E_b \ket{\psi_j} = \delta_{ab} \delta_{ij}.
\end{equation}
By convention, an $[n, 0, d]$ code must be nondegenerate.  When $E_a E_b \in S$,
we say that the errors $E_a$ and $E_b$ are degenerate.  We cannot distinguish
between $E_a$ and $E_b$, but there is no need to, since they have the same
effect on the codewords.

It is sometimes useful to define the {\em error syndrome} for a stabilizer 
code.  Let $f_M : \G \rightarrow {\bf Z}_2$, 
\begin{equation}
f_M (E) = \left\{ \begin{array}{ll} 0 & \mbox{if $[M, E] = 0$} \\ 1 & 
\mbox{if $\{M, E\} = 0$} \end{array} \right.
\end{equation}
and $f (E) = (f_{M_1} (E), \ldots, f_{M_{n-k}} (E) )$, where $M_1, \ldots, 
M_{n-k}$ are the generators of $S$.  Then $f(E)$ is some $(n-k)$-bit binary 
number which is $0$ iff $E \in N(S)$.  $f(E_a) = f(E_b)$ iff $f(E_a E_b) = 0$, 
so for a nondegenerate code, $f(E)$ is different for each correctable error 
$E$.

In order to perform the error-correction operation for a stabilizer code, all 
we need to do is measure the eigenvalue of each generator of the 
stabilizer.  The eigenvalue of $M_i$ will be $(-1)^{f_{M_i} (E)}$, so this 
process will give us the error syndrome.  The error syndrome in turn tells 
us exactly what error occurred (for a nondegenerate code) or what set of 
degenerate errors occurred (for a degenerate code).  The error will always be 
in $\G$ since the code uses that error basis, and every operator in $\G$ is 
unitary, and therefore invertible.  Then we just apply the error operator (or 
one equivalent to it by multiplication by $S$) to fix the state.  Note that
even if the original error that occurred is a nontrivial linear combination of 
errors in $\G$, the process of syndrome measurement will project onto one
of the basis errors.  If the resulting error is not in the correctable set, 
we will end up in the wrong encoded state, but otherwise, we are in the 
correct state.  In chapter~\ref{chap-fault-tolerant}, I describe a few ways 
of measuring the error syndrome that are tolerant of imperfect component gates.

Since the elements of $N(S)$ move codewords around within $T$, they 
have a natural interpretation as encoded operations on the codewords.  
Since $S$ fixes $T$, actually only $N(S) / S$ will act on $T$ nontrivially.  If 
we pick a basis for $T$ consisting of eigenvectors of $n$ commuting elements 
of $N(S)$, we get an automorphism $N(S) / S \rightarrow \G_k$.  $N(S)/S$ can 
therefore be generated by $i$ (which we will by and large ignore) and $2k$ 
equivalence classes, which I will write $\Xbar_i$ and $\Zbar_i$ ($i=1 
\ldots k$), where $\Xbar_i$ maps to $\Xs{i}$ in $\G_k$ and $\Zbar_i$ 
maps to $\Zs{i}$ in $\G_k$.  They are encoded $\X$ and $\Z$ operators for 
the code.  If $k=1$, I will write $\Xbar_1 = \Xbar$ and $\Zbar_1 = \Zbar$.  
The $\Xbar$ and $\Zbar$ operators satisfy
\begin{eqnarray}
[\Xbar_i, \Xbar_j] & = & 0 \\
{[}\Zbar_i, \Zbar_j] & = & 0 \\
{[}\Xbar_i, \Zbar_j] & = & 0\ (i \neq j) \\
\{\Xbar_i, \Zbar_i \} & = & 0.
\end{eqnarray}

\section{Some Examples}
\label{sec-stab-examples}

I shall now present a few short codes to use as examples.  The first 
encodes one qubit in five qubits~\cite{bennett-tome,laflamme-5qubit} and is 
given in table~\ref{table-5qubit}.
\begin{table}
\centering
\begin{tabular}{c|ccccc}
$M_1$ & $\X$ & $\Z$ & $\Z$ & $\X$ & $I$ \\
$M_2$ & $I$ & $\X$ & $\Z$ & $\Z$ & $\X$ \\
$M_3$ & $\X$ & $I$ & $\X$ & $\Z$ & $\Z$ \\
$M_4$ & $\Z$ & $\X$ & $I$ & $\X$ & $\Z$ \\
\hline
\low{$\Xbar$} & \low{$\X$} & \low{$\X$} & \low{$\X$} & \low{$\X$} & \low{$\X$} 
\\
\low{$\Zbar$} & \low{$\Z$} & \low{$\Z$} & \low{$\Z$} & \low{$\Z$} & \low{$\Z$}
\end{tabular}
\caption{The stabilizer for the five-qubit code.}
\label{table-5qubit}
\end{table}
I have also included $\Xbar$ and $\Zbar$, which, along with $M_1$ 
through $M_4$, generate $N(S)$.  Note that this code is {\em cyclic} (i.e., 
the stabilizer and codewords are invariant under cyclic permutations of
the qubits).  It has distance three (for instance, $\Ys{1} \Zs{2} \Ys{3} \in 
N(S)-S$) and is nondegenerate.  We can take the basis codewords for this code 
to be
\begin{equation}
\ket{\overline{0}} = \Sum_{M \in S} M \ \ket{00000}
\end{equation}
and
\begin{equation}
\ket{\overline{1}} = \Xbar \ket{\overline{0}}.
\end{equation}
That is,
\begin{eqnarray}
\ket{\overline{0}} & = & \ket{00000} + M_1 \ket{00000} + M_2 \ket{00000} +
M_3 \ket{00000} + M_4 \ket{00000} \nonumber \\
& & \quad \mbox{} + M_1 M_2 \ket{00000} + M_1 M_3 \ket{00000} +
M_1 M_4 \ket{00000} \nonumber \\
& & \quad \mbox{} + M_2 M_3 \ket{00000} + M_2 M_4 \ket{00000} + 
M_3 M_4 \ket{00000} \\
& & \quad \mbox{} + M_1 M_2 M_3 \ket{00000} + M_1 M_2 M_4 \ket{00000} + 
M_1 M_3 M_4 \ket{00000} \nonumber \\
& & \quad \mbox{} + M_2 M_3 M_4 \ket{00000} + M_1 M_2 M_3 M_4 \ket{00000} 
\nonumber \\
& = & \ket{00000} + \ket{10010} + \ket{01001} + \ket{10100} \nonumber \\
& & \mbox{} + \ket{01010} - \ket{11011} - \ket{00110} - \ket{11000} \nonumber \\
& & \mbox{} - \ket{11101} - \ket{00011} - \ket{11110} - \ket{01111} \\
& & \mbox{} - \ket{10001} - \ket{01100} - \ket{10111} + \ket{00101}, \nonumber
\end{eqnarray}
and
\begin{eqnarray}
\ket{\overline{1}} & = & \Xbar \ket{\overline{0}} \nonumber \\
& = & \ket{11111} + \ket{01101} + \ket{10110} + \ket{01011} \nonumber \\
& & \mbox{} + \ket{10101} - \ket{00100} - \ket{11001} - \ket{00111} \nonumber \\
& & \mbox{} - \ket{00010} - \ket{11100} - \ket{00001} - \ket{10000} \\
& & \mbox{} - \ket{01110} - \ket{10011} - \ket{01000} + \ket{11010}. \nonumber
\end{eqnarray}
Since multiplying by an element of the stabilizer merely rearranges the 
sum $\Sum M$, these two states are in $T$.  When these are the encoded 
$0$ and $1$, $\Xbar$ is the encoded bit flip operator $\X$ and $\Zbar$ is 
the encoded $\Z$.  This code also has the property that every possible 
error syndrome is used by the single-qubit errors.  It is therefore a {\em 
perfect} code.  There are a number of other perfect 
codes~\cite{gottesman-pasting,calderbank-GF4}, which will be discussed in 
chapter~\ref{chap-examples}.

A code encoding three qubits in eight 
qubits~\cite{gottesman-stab,calderbank-stab,steane-8qubit} appears in 
table~\ref{table-8qubit}.
\begin{table}
\centering
\begin{tabular}{c|cccccccc}
$M_1$ & $\X$ & $\X$ & $\X$ & $\X$ & $\X$ & $\X$ & $\X$ & $\X$ \\
$M_2$ & $\Z$ & $\Z$ & $\Z$ & $\Z$ & $\Z$ & $\Z$ & $\Z$ & $\Z$ \\
$M_3$ & $I$ & $\X$ & $I$ & $\X$ & $\Y$ & $\Z$ & $\Y$ & $\Z$ \\
$M_4$ & $I$ & $\X$ & $\Z$ & $\Y$ & $I$ & $\X$ & $\Z$ & $\Y$ \\
$M_5$ & $I$ & $\Y$ & $\X$ & $\Z$ & $\X$ & $\Z$ & $I$ & $\Y$ \\
\hline
\low{$\Xbar_1$} & \low{$\X$} & \low{$\X$} & \low{$I$} & \low{$I$} & \low{$I$} & 
\low{$\Z$} & \low{$I$} & \low{$\Z$} \\
\low{$\Xbar_2$} & \low{$\X$} & \low{$I$} & \low{$\X$} & \low{$\Z$} & \low{$I$} 
& \low{$I$} & \low{$\Z$} & \low{$I$} \\
\low{$\Xbar_3$} & \low{$\X$} & \low{$I$} & \low{$I$} & \low{$\Z$} & \low{$\X$} 
& \low{$\Z$} & \low{$I$} & \low{$I$} \\
\low{$\Zbar_1$} & \low{$I$} & \low{$\Z$} & \low{$I$} & \low{$\Z$} & \low{$I$} & 
\low{$\Z$} & \low{$I$} & \low{$\Z$} \\
\low{$\Zbar_2$} & \low{$I$} & \low{$I$} & \low{$\Z$} & \low{$\Z$} & \low{$I$} & \low{$I$} & \low{$\Z$} & \low{$\Z$} \\
\low{$\Zbar_3$} & \low{$I$} & \low{$I$} & \low{$I$} & \low{$I$} & \low{$\Z$} & 
\low{$\Z$} & \low{$\Z$} & \low{$\Z$}
\end{tabular}
\caption{The stabilizer for the eight-qubit code.}
\label{table-8qubit}
\end{table}
Again, $M_1$ through $M_5$ generate the stabilizer, and generate $N(S)$ with 
$\Xbar_i$ and $\Zbar_i$.  This is also a nondegenerate distance three 
code.  The codewords are
\begin{equation}
\ket{\overline{c_1 c_2 c_3}} = \Xbar_1^{c_1} \Xbar_2^{c_2} \Xbar_3^{c_3} 
\Sum_{M \in S} M \ket{00000000}.
\end{equation}
The operators $\Xbar_i$ and $\Zbar_i$ are the encoded $\X$ and $\Z$ on 
the $i$th encoded qubit.  This code is one of an infinite family of 
codes~\cite{gottesman-stab,steane-RM}, which I present in 
chapter~\ref{chap-examples}.

A particularly useful class of codes with simple stabilizers is the 
Calderbank-Shor-Steane (or {\em CSS}) class of 
codes~\cite{calderbank-CSS,steane-CSS}.  Suppose we have a classical code 
with parity check matrix $P$.  We can make a quantum code to correct just 
$\X$ errors using a stabilizer with elements corresponding to the rows of 
$P$, with a $\Z$ wherever $P$ has a $1$ and $I$'s elsewhere.  The error 
syndrome $f(E)$ for a product of $\X$ errors $E$ is then equal to the 
classical error syndrome for the same set of classical bit flip errors.  Now 
add in stabilizer generators corresponding to the parity check matrix $Q$ 
of a second classical code, only now with $\X$'s instead of $\Z$'s.  These 
generators will identify $\Z$ errors.  Together, they can also identify $\Y$ 
errors, which will have a nontrivial error syndrome for both parts.  In
general, a code formed this way will correct as many $\X$ errors as the code 
for $P$ can correct, and as many $\Z$ errors as the code for $Q$ can correct; 
a $\Y$ error counts as one of each.

We can only combine $P$ and $Q$ into a single stabilizer in the CSS form if 
the generators derived from the two codes commute.  This will be true iff the 
rows of $P$ and $Q$ are orthogonal using the binary dot product.  This 
means that the dual code of each code must be a subset of the other code.  
The minimum distance of the quantum code will be the minimum of the 
distances of $P$ and $Q$.  An example of a code of this sort is given in 
table~\ref{table-7qubit}.  It is based on the classical $[7,4,3]$ Hamming 
code, which is self-dual.
\begin{table}
\centering
\begin{tabular}{c|ccccccc}
$M_1$ & $\X$ & $\X$ & $\X$ & $\X$ & $I$ & $I$ & $I$ \\
$M_2$ & $\X$ & $\X$ & $I$ & $I$ & $\X$ & $\X$ & $I$ \\
$M_3$ & $\X$ & $I$ & $\X$ & $I$ & $\X$ & $I$ & $\X$ \\
$M_4$ & $\Z$ & $\Z$ & $\Z$ & $\Z$ & $I$ & $I$ & $I$ \\
$M_5$ & $\Z$ & $\Z$ & $I$ & $I$ & $\Z$ & $\Z$ & $I$ \\
$M_6$ & $\Z$ & $I$ & $\Z$ & $I$ & $\Z$ & $I$ & $\Z$ \\
\hline
\low{$\Xbar$} & \low{$I$} & \low{$I$} & \low{$I$} & \low{$I$} & \low{$\X$} & 
\low{$\X$} & \low{$\X$} \\
\low{$\Zbar$} & \low{$I$} & \low{$I$} & \low{$I$} & \low{$I$} & \low{$\Z$} & 
\low{$\Z$} & \low{$\Z$}
\end{tabular}
\caption{The seven-qubit CSS code.}
\label{table-7qubit}
\end{table}
For this code, the codewords are
\begin{eqnarray}
\ket{\overline{0}} & = & \ket{0000000} + \ket{1111000} + \ket{1100110} +
\ket{1010101} \nonumber \\
& & \mbox{} + \ket{0011110} + \ket{0101101} + \ket{0110011} + \ket{1001011}
\end{eqnarray}
and
\begin{eqnarray}
\ket{\overline{1}} & = & \ket{0000111} + \ket{1111111} + \ket{1100001} +
\ket{1010010} \nonumber \\
& & \mbox{} + \ket{0011001} + \ket{0101010} + \ket{0110100} + \ket{1001100}.
\end{eqnarray}
The encoded $\ket{0}$ state is the superposition of the even codewords in
the Hamming code and the encoded $\ket{1}$ state is the superposition of
the odd codewords in the Hamming code.  This behavior is characteristic
of CSS codes; in general, the various quantum codewords are superpositions
of the words in subcodes of one of the classical codes.

CSS codes are not as efficient as the most general quantum 
code, but they are easy to derive from known classical codes and their 
simple form often makes them ideal for other purposes.  For instance, the 
seven-qubit code is particularly well suited for fault-tolerant computation 
(as I will discuss in chapter~\ref{chap-fault-tolerant}).

\section{Alternate Languages for Stabilizers}
\label{sec-alternate}

There are number of possible ways of describing the stabilizer of a 
quantum code.   They each have advantages and are useful in different 
circumstances.  The description I have used so far uses the language of 
finite group theory and is particularly useful for making contact with the 
usual language of quantum mechanics.  This is the form presented in 
\cite{gottesman-stab}.

We can instead write the stabilizer using binary vector spaces, as in 
\cite{calderbank-stab}, which emphasizes connections with the classical 
theory of error-correcting codes.  To do this, we write the stabilizer as a 
pair of $(n-k) \times n$ binary matrices (or often one $(n-k) \times 2n$ 
matrix with a line separating the two halves).  The rows correspond to the 
different generators of the stabilizer and the columns correspond to 
different qubits.  One matrix has a $1$ whenever the generator has a $\X$ 
or a $\Y$ in the appropriate place, the other has a $1$ whenever the 
generator has a $\Y$ or $\Z$.  Overall phase factors get dropped.  For 
instance, the five-qubit code in this form becomes
\begin{equation}
\left( \begin{array}{ccccc|ccccc}
1 & 0 & 0 & 1 & 0 & 0 & 1 & 1 & 0 & 0 \\
0 & 1 & 0 & 0 & 1 & 0 & 0 & 1 & 1 & 0 \\
1 & 0 & 1 & 0 & 0 & 0 & 0 & 0 & 1 & 1 \\
0 & 1 & 0 & 1 & 0 & 1 & 0 & 0 & 0 & 1
\end{array} \right).
\end{equation}
Other elements of $\G$ get converted to two $n$-dimensional vectors in the 
same way.  We can convert back to the group theory formalism by writing 
down operators with a $\X$ if the left vector or matrix has a $1$, a $\Z$ if 
the right vector or matrix has a $1$, and a $\Y$ if they are both $1$.  The 
generators formed this way will never have overall phase factors, although 
other elements of the group might.  Multiplication of group elements
corresponds to addition of the corresponding binary vectors.

In the binary formalism, the condition that two operators commute with 
each other becomes the condition that the following inner product is 0:
\begin{equation}
Q(a|b, c|d) = \Sum_{i=1}^n (a_{i} d_{i} + b_{i} c_{i}) = 0,
\label{eq-commute-bin}
\end{equation}
using binary arithmetic as usual.  $a_i$, $b_i$, $c_i$, and $d_i$ are the 
$i$th components of the corresponding vectors.    Therefore the condition 
that the stabilizer be Abelian converts to the condition that the stabilizer 
matrix $(A|B)$ satisfy
\begin{equation}
\Sum_{l=1}^n (A_{il} B_{jl} + B_{il} A_{jl}) = 0.
\end{equation}
We determine the vectors in $N(S)$ by evaluating the inner product 
(\ref{eq-commute-bin}) with the rows of $(A|B)$.  To get a real code (with 
an even number of $\Y$'s), the code should also satisfy
\begin{equation}
\Sum_{l=1}^n A_{il} B_{il} = 0.
\end{equation}

Another formalism highlights connections with the classical theory of codes 
over the field GF(4) \cite{calderbank-GF4}.  This is a field of characteristic 
two containing four elements, which can be written 
$\{0, 1, \omega, \omega^2\}$.  Since the field has characteristic two, 
\begin{equation}
1 + 1 = \omega + \omega = \omega^2 + \omega^2 = 0.
\end{equation}
Also, $\omega^3 = 1$ and $1 + \omega = \omega^2$.  We can rewrite the 
generators as an $n$-dimensional ``vector'' over GF(4) by substituting $1$ 
for $\X$, $\omega$ for $\Z$, and $\omega^2$ for $\Y$.  The multiplicative 
structure of $\G$ becomes the additive structure of GF(4).  I put vector in 
quotes because the code need not have the structure of a vector space over 
GF(4).  If it does (that is, the stabilizer is closed under multiplication by 
$\omega$), the code is a {\em linear} code, which is essentially a classical 
code over GF(4).  The most general quantum code is sometimes called an {\em 
additive} code, because the stabilizer is only closed under sums of its elements.  In this formalism, the five-qubit code appears as
\begin{equation}
\left( \begin{array}{ccccc}
1 & \omega & \omega & 1 & 0 \\
0 & 1 & \omega & \omega & 1 \\
1 & 0 & 1 & \omega & \omega \\
\omega & 1 & 0 & 1 & \omega
\end{array} \right).
\end{equation}
Note that the five-qubit code is a linear quantum code.

Again, there is an additional condition for a quantum code.  Define the 
``trace'' operator by ${\rm Tr}\ \omega = {\rm Tr}\ \omega^2 = 1$, ${\rm 
Tr}\ 1 = {\rm Tr}\ 0 = 0$.  Two operators in $\G$ commute iff their images, 
the vectors $u$ and $v$ over GF(4), satisfy
\begin{equation}
{\rm Tr}\ u \cdot \overline{v} = {\rm Tr}\left( \Sum_{j=1}^n u_j 
\overline{v}_j \right) = 0,
\end{equation}
where $\overline{v}_j$ is conjugation on the $j$th component of $v$, 
switching $\omega$ and $\omega^2$, and leaving $0$ and $1$ alone.

\section{Making New Codes From Old Codes}
\label{sec-construction}

Using old codes to find new ones can simplify the task of finding codes, 
which can otherwise be quite a difficult problem.  There are a number of simple 
modifications we can make to existing codes to produce new codes with 
different parameters~\cite{gottesman-pasting,calderbank-GF4}.

One trivial change is to perform a permutation of $\X$, $\Y$, and $\Z$ on 
each qubit.  This leaves the distance and size of the code the same, 
although it may be useful for codes that can correct different numbers of 
$\X$, $\Y$, and $\Z$ errors.  A slightly less trivial manipulation is to add a 
new qubit and a new generator which is $\X$ for the new qubit.  The other 
generators are tensored with the identity on the new qubit to form the
generators of the new code.  This makes an $[n, k, d]$ code (degenerate or 
nondegenerate) into an $[n+1, k, d]$ degenerate code:  Any operator acting as 
$\Y$ or $\Z$ on the new qubit will anticommute with the new generator, and any 
operator with the form $M \otimes \Xs{(n+1)}$ will be equivalent to the 
operator $M \otimes I$.  Therefore, an operator must have at least weight $d$ 
when restricted to the first $n$ qubits to be in $N(S)-S$.

A less trivial manipulation is to remove the last qubit, converting an $[n, k, 
d]$ code into an $[n-1, k+1, d-1]$ code.  To do this, we choose the $n-k$ 
generators of $S$ so that $M_1$ ends $\X$, $M_2$ ends $\Z$, and $M_3$ 
through $M_{n-k}$ end $I$.  We can always do this when $d>1$ by picking 
the first two and then multiplying by combinations of them to make the 
others end appropriately.\footnote{If the code has been formed by adding 
a single $\X$ (or $\Y$ or $\Z$) generator, as above, we may not be able to 
do this for a given qubit, but there will always be at least one qubit for 
which we can.}  Then the new code has a stabilizer formed from the last 
$n-k-2$ generators, dropping $M_1$ and $M_2$.  Suppose we have an 
operator $A$ on the first $n-1$ qubits of weight $w$ that commutes with 
$M_3$ through $M_{n-k}$.  There are four possibilities, all of which lead to 
an operator of weight at most $w+1$ that commutes with the original 
stabilizer:
\begin{enumerate}
\item $A$ commutes with both $M_1$ and $M_2$.  
\item $A$ commutes with $M_1$, but not $M_2$.  Then $A \otimes \Xs{n}$ 
commutes with $M_1$ and $M_2$.
\item $A$ commutes with $M_2$, but not $M_1$.  Then $A \otimes \Zs{n}$ 
commutes with $M_1$ and $M_2$.
\item $A$ anticommutes with both $M_1$ and $M_2$.  Then $A \otimes 
\Ys{n}$ commutes with $M_1$ and $M_2$.
\end{enumerate}
Since the original code had distance $d$, $w$ must be at least $d-1$, 
which is therefore the distance of the new code.  The stabilizer has $n-k-2$ 
generators, so the code encodes $(n-1)-(n-k-2) = k+1$ qubits.  The new $\Xbar$
and $\Zbar$ operators are $M_1$ and $M_2$ (in either order), restricted to
the first $n-1$ qubits.  An example of this construction is to remove the
last qubit from the $[5,1,3]$ code of figure~\ref{fig-5qubit} to produce
a $[4,2,2]$ code: the generators of the new code are $M_1$ and $M_3 M_4$,
both without the last qubit.  The new stabilizer is given in figure
\ref{fig-droplast}.  Note that the $\Zbar_1$ operator is equal to
$M_3 \Zbar$ for the five-qubit code.  I have multiplied by $M_3$ so that 
$\Zbar_1$ anticommutes with $\Xbar_1$.
\begin{table}
\centering
\begin{tabular}{c|cccc}
$M_1'$ & $\X$ & $\Z$ & $\Z$ & $\X$ \\
$M_2'$ & $\Y$ & $\X$ & $\X$ & $\Y$ \\
\hline
\low{$\Xbar_1$} & \low{$\X$} & \low{$\X$} & \low{$\X$} & \low{$\X$} \\
\low{$\Xbar_2$} & \low{$\X$} & \low{$I$} & \low{$\X$} & \low{$\Z$} \\
\low{$\Zbar_1$} & \low{$\Y$} & \low{$\Z$} & \low{$\Y$} & \low{$I$} \\
\low{$\Zbar_2$} & \low{$I$} & \low{$\X$} & \low{$\Z$} & \low{$\Z$}
\end{tabular}
\caption{A $[4,2,2]$ code derived from the $[5,1,3]$ code.}
\label{fig-droplast}
\end{table}

Another way to make new codes is by {\em pasting} together old codes.  Suppose 
we have four stabilizers $R_1$, $R_2$, $S_1$, and $S_2$, with $R_1 \subset 
S_1$ and $R_2 \subset S_2$.  Let $R_1$ define an $[n_1, l_1, c_1]$ code, 
$R_2$ be an $[n_2, l_2, c_2]$ code, $S_1$ be an $[n_1, k_1, d_1]$ code, and 
$S_2$ be an $[n_2, k_2, d_2]$ code.  Then $k_i < l_i$ and $c_i \leq d_i$.  We 
require $l_1-k_1 = l_2-k_2$ and for $S_1$ and $S_2$ to be 
nondegenerate.\footnote{We can actually allow $S_1$ and $S_2$ to be degenerate, 
as long as all the degenerate operators are confined to $R_1$ and $R_2$}  Let 
generators of $R_1$ be $\{M_1, \ldots, M_{n_1 - l_1}\}$, the generators of 
$S_1$ be $\{M_1, \ldots, M_{n_1-k_1}\}$, the generators of $R_2$ be $\{N_1, 
\ldots, N_{n_2-l_2}\}$, and the generators of $S_2$ be $\{N_1, \ldots, 
N_{n_2-k_2}\}$.  We form a new stabilizer $S$ on $n_1 + n_2$ qubits generated 
by 
\begin{eqnarray}
& & \{M_1 \otimes I, \ldots, M_{n_1-l_1} \otimes I, I \otimes N_1, \ldots, 
I \otimes N_{n_2-l_2}, \nonumber \\
& & \quad M_{n_1-l_1+1} \otimes N_{n_2-l_2+1}, \ldots,
M_{n_1-k_1} \otimes N_{n_2-k_2} \}.
\end{eqnarray}
The code has $(n_1-l_1) + (n_2-l_2) + (l_i-k_i)$ generators, and therefore 
encodes $l_1+k_2 = l_2+k_1$ qubits.  For instance, if $S_1$ is the eight-qubit 
code and $S_2$ is the five-qubit code, with $R_1$ generated by $\X \X \X \X 
\X \X \X \X$ and $\Z \Z \Z \Z \Z \Z \Z \Z$ and $R_2$ generated by $\X \Z 
\Z \X I$, we can make the $[13,7,3]$ code given in 
table~\ref{table-13qubit}.
\begin{table}
\centering
\begin{tabular}{c|cccccccc|ccccc}
$M_1$ & $\X$ & $\X$ & $\X$ & $\X$ & $\X$ & $\X$ & $\X$ & $\X$ & $I$ 
& $I$ & $I$ & $I$ & $I$ \\
$M_2$ & $\Z$ & $\Z$ & $\Z$ & $\Z$ & $\Z$ & $\Z$ & $\Z$ & $\Z$ & $I$ & 
$I$ & $I$ & $I$ & $I$ \\
\hline
$M_3$ & $I$ & $I$ & $I$ & $I$ & $I$ & $I$ & $I$ & $I$ & $\X$ & $\Z$ & 
$\Z$ & $\X$ & $I$ \\
\hline
$M_4$ & $I$ & $\X$ & $I$ & $\X$ & $\Y$ & $\Z$ & $\Y$ & $\Z$ & $I$ & 
$\X$ & $\Z$ & $\Z$ & $\X$ \\
$M_5$ & $I$ & $\X$ & $\Z$ & $\Y$ & $I$ & $\X$ & $\Z$ & $\Y$ & $\X$ & 
$I$ & $\X$ & $\Z$ & $\Z$ \\
$M_6$ & $I$ & $\Y$ & $\X$ & $\Z$ & $\X$ & $\Z$ & $I$ & $\Y$ & $\Z$ & 
$\X$ & $I$ & $\X$ & $\Z$
\end{tabular}
\caption{The thirteen-qubit code formed by pasting together the five- and 
eight-qubit codes.}
\label{table-13qubit}
\end{table}

In general, the distance of the new code will be ${\rm min}\{d_1, d_2, c_1 
+ c_2 \}$.  This is because an operator acting on just the first $n_1$ qubits 
can only commute with $S$ if it commutes with $S_1$, an operator acting 
on the last $n_2$ qubits can only commute with $S$ if it commutes with 
$S_2$, and an operator acting on both parts must commute with both $R_1 
\otimes I$ and $I \otimes R_2$.

Another very useful way of producing new codes is to {\em concatenate} 
two codes to produce a code of greater total distance.  Suppose we have an 
$[n_1, k, d_1]$ code (stabilizer $S_1$) and we encode each of its $n_1$ 
qubits again using an $[n_2, 1, d_2]$ code (stabilizer $S_2$).  The result is 
an $[n_1 n_2, k, d_1 d_2]$ code.  Its stabilizer $S$ is $n_1$ copies of $S_2$, 
acting on the physical qubits in blocks of size $n_2$, plus an additional 
$n_1 - k$ generators corresponding to the generators of $S_1$.  However, these 
generators are encoded to act on the second code.  That is, a $\X$ acting 
on the first code must be replaced by an $\Xbar$ for the second code.  For 
instance, the code resulting from concatenating the five-qubit code with itself 
has the stabilizer given in table~\ref{table-25qubit}.
\begin{table}
{\setlength{\tabcolsep}{0.1em}
\begin{tabular}{c|ccccc|ccccc|ccccc|ccccc|ccccc}
$M_1$ & $\X$ & $\Z$ & $\Z$ & $\X$ & $I$ & $I$ & $I$ & $I$ & $I$ & $I$ & 
$I$ & $I$ & $I$ & $I$ & $I$ & $I$ & $I$ & $I$ & $I$ & $I$ & $I$ & $I$ & 
$I$ & $I$ & $I$ \\
$M_2$ & $I$ & $\X$ & $\Z$ & $\Z$ & $\X$ & $I$ & $I$ & $I$ & $I$ & $I$ & 
$I$ & $I$ & $I$ & $I$ & $I$ & $I$ & $I$ & $I$ & $I$ & $I$ & $I$ & $I$ & 
$I$ & $I$ & $I$ \\
$M_3$ & $\X$ & $I$ & $\X$ & $\Z$ & $\Z$ & $I$ & $I$ & $I$ & $I$ & $I$ & 
$I$ & $I$ & $I$ & $I$ & $I$ & $I$ & $I$ & $I$ & $I$ & $I$ & $I$ & $I$ & 
$I$ & $I$ & $I$ \\
$M_4$ & $\Z$ & $\X$ & $I$ & $\X$ & $\Z$ & $I$ & $I$ & $I$ & $I$ & $I$ & 
$I$ & $I$ & $I$ & $I$ & $I$ & $I$ & $I$ & $I$ & $I$ & $I$ & $I$ & $I$ & 
$I$ & $I$ & $I$ \\
$M_5$ & $I$ & $I$ & $I$ & $I$ & $I$ & $\X$ & $\Z$ & $\Z$ & $\X$ & $I$ & 
$I$ & $I$ & $I$ & $I$ & $I$ & $I$ & $I$ & $I$ & $I$ & $I$ & $I$ & $I$ & 
$I$ & $I$ & $I$ \\
$M_6$ & $I$ & $I$ & $I$ & $I$ & $I$ & $I$ & $\X$ & $\Z$ & $\Z$ & $\X$ & 
$I$ & $I$ & $I$ & $I$ & $I$ & $I$ & $I$ & $I$ & $I$ & $I$ & $I$ & $I$ & 
$I$ & $I$ & $I$ \\
$M_7$ & $I$ & $I$ & $I$ & $I$ & $I$ & $\X$ & $I$ & $\X$ & $\Z$ & $\Z$ & 
$I$ & $I$ & $I$ & $I$ & $I$ & $I$ & $I$ & $I$ & $I$ & $I$ & $I$ & $I$ & 
$I$ & $I$ & $I$ \\
$M_8$ & $I$ & $I$ & $I$ & $I$ & $I$ & $\Z$ & $\X$ & $I$ & $\X$ & $\Z$ & 
$I$ & $I$ & $I$ & $I$ & $I$ & $I$ & $I$ & $I$ & $I$ & $I$ & $I$ & $I$ & 
$I$ & $I$ & $I$ \\
$M_9$ & $I$ & $I$ & $I$ & $I$ & $I$ & $I$ & $I$ & $I$ & $I$ & $I$ & $\X$ 
& $\Z$ & $\Z$ & $\X$ & $I$ & $I$ & $I$ & $I$ & $I$ & $I$ & $I$ & $I$ & 
$I$ & $I$ & $I$ \\
$M_{10}$ & $I$ & $I$ & $I$ & $I$ & $I$ & $I$ & $I$ & $I$ & $I$ & $I$ & $I$ 
& $\X$ & $\Z$ & $\Z$ & $\X$ & $I$ & $I$ & $I$ & $I$ & $I$ & $I$ & $I$ & 
$I$ & $I$ & $I$ \\
$M_{11}$ & $I$ & $I$ & $I$ & $I$ & $I$ & $I$ & $I$ & $I$ & $I$ & $I$ & 
$\X$ & $I$ & $\X$ & $\Z$ & $\Z$ & $I$ & $I$ & $I$ & $I$ & $I$ & $I$ & 
$I$ & $I$ & $I$ & $I$ \\
$M_{12}$ & $I$ & $I$ & $I$ & $I$ & $I$ & $I$ & $I$ & $I$ & $I$ & $I$ & 
$\Z$ & $\X$ & $I$ & $\X$ & $\Z$ & $I$ & $I$ & $I$ & $I$ & $I$ & $I$ & 
$I$ & $I$ & $I$ & $I$ \\
$M_{13}$ & $I$ & $I$ & $I$ & $I$ & $I$ & $I$ & $I$ & $I$ & $I$ & $I$ & $I$ 
& $I$ & $I$ & $I$ & $I$ & $\X$ & $\Z$ & $\Z$ & $\X$ & $I$ & $I$ & $I$ & 
$I$ & $I$ & $I$ \\
$M_{14}$ & $I$ & $I$ & $I$ & $I$ & $I$ & $I$ & $I$ & $I$ & $I$ & $I$ & $I$ 
& $I$ & $I$ & $I$ & $I$ & $I$ & $\X$ & $\Z$ & $\Z$ & $\X$ & $I$ & $I$ & 
$I$ & $I$ & $I$ \\
$M_{15}$ & $I$ & $I$ & $I$ & $I$ & $I$ & $I$ & $I$ & $I$ & $I$ & $I$ & $I$ 
& $I$ & $I$ & $I$ & $I$ & $\X$ & $I$ & $\X$ & $\Z$ & $\Z$ & $I$ & $I$ & 
$I$ & $I$ & $I$ \\
$M_{16}$ & $I$ & $I$ & $I$ & $I$ & $I$ & $I$ & $I$ & $I$ & $I$ & $I$ & $I$ 
& $I$ & $I$ & $I$ & $I$ & $\Z$ & $\X$ & $I$ & $\X$ & $\Z$ & $I$ & $I$ & 
$I$ & $I$ & $I$ \\
$M_{17}$ & $I$ & $I$ & $I$ & $I$ & $I$ & $I$ & $I$ & $I$ & $I$ & $I$ & $I$ 
& $I$ & $I$ & $I$ & $I$ & $I$ & $I$ & $I$ & $I$ & $I$ & $\X$ & $\Z$ & 
$\Z$ & $\X$ & $I$ \\
$M_{18}$ & $I$ & $I$ & $I$ & $I$ & $I$ & $I$ & $I$ & $I$ & $I$ & $I$ & $I$ 
& $I$ & $I$ & $I$ & $I$ & $I$ & $I$ & $I$ & $I$ & $I$ & $I$ & $\X$ & $\Z$ 
& $\Z$ & $\X$ \\
$M_{19}$ & $I$ & $I$ & $I$ & $I$ & $I$ & $I$ & $I$ & $I$ & $I$ & $I$ & $I$ 
& $I$ & $I$ & $I$ & $I$ & $I$ & $I$ & $I$ & $I$ & $I$ & $\X$ & $I$ & 
$\X$ & $\Z$ & $\Z$ \\
$M_{20}$ & $I$ & $I$ & $I$ & $I$ & $I$ & $I$ & $I$ & $I$ & $I$ & $I$ & $I$ 
& $I$ & $I$ & $I$ & $I$ & $I$ & $I$ & $I$ & $I$ & $I$ & $\Z$ & $\X$ & $I$ 
& $\X$ & $\Z$ \\
$M_{21}$ & $\X$ & $\X$ & $\X$ & $\X$ & $\X$ & $\Z$ & $\Z$ & $\Z$ & $\Z$ 
& $\Z$ & $\Z$ & $\Z$ & $\Z$ & $\Z$ & $\Z$ & $\X$ & $\X$ & $\X$ & $\X$ & 
$\X$ & $I$ & $I$ & $I$ & $I$ & $I$ \\
$M_{22}$ & $I$ & $I$ & $I$ & $I$ & $I$ & $\X$ & $\X$ & $\X$ & $\X$ & 
$\X$ & $\Z$ & $\Z$ & $\Z$ & $\Z$ & $\Z$ & $\Z$ & $\Z$ & $\Z$ & $\Z$ & 
$\Z$ & $\X$ & $\X$ & $\X$ & $\X$ & $\X$ \\
$M_{23}$ & $\X$ & $\X$ & $\X$ & $\X$ & $\X$ & $I$ & $I$ & $I$ & $I$ & 
$I$ & $\X$ & $\X$ & $\X$ & $\X$ & $\X$ & $\Z$ & $\Z$ & $\Z$ & $\Z$ & 
$\Z$ & $\Z$ & $\Z$ & $\Z$ & $\Z$ & $\Z$ \\
$M_{24}$ & $\Z$ & $\Z$ & $\Z$ & $\Z$ & $\Z$ & $\X$ & $\X$ & $\X$ & $\X$ 
& $\X$ & $I$ & $I$ & $I$ & $I$ & $I$ & $\X$ & $\X$ & $\X$ & $\X$ & $\X$ 
& $\Z$ & $\Z$ & $\Z$ & $\Z$ & $\Z$
\end{tabular}
\caption{Result of concatenating the five-qubit code with itself.}
\label{table-25qubit}}
\end{table}
The concatenated code has distance $d_1 d_2$ because operators in $N(S) - 
S$ must have distance at least $d_2$ on at least $d_1$ blocks of $n_2$ 
qubits, so have weight at least $d_1 d_2$.  Note that it is not strictly 
necessary to use the same code to encode each qubit of $S_1$.

There are two possible ways to concatenate when $S_2$ encodes multiple 
qubits.  Suppose $S_1$ is an $[n_1, k_1, d_1]$ code and $S_2$ is an $[n_2, 
k_2, d_2]$ code.  Further, suppose $n_1$ is a multiple of $k_2$.  Then we 
can encode blocks of $S_1$ of size $k_2$ using $S_2$.  This will result in a 
code using $n_1 n_2/k_2$ qubits to encode $k_1$ qubits.  It still takes an 
operator of distance at least $d_2$ to cause an error on an $n_2$-qubit 
block, but such an error can cause up to $k_2$ errors on $S_1$, so the 
resulting code need only have distance $\lceil d_1/k_2 \rceil d_2$.  
However, the $k_2$ errors that result are not a general set of $k_2$ errors, 
so the code may actually be better.  Suppose $S_1$ has distance $d_1'$ 
($d_1' \geq \lceil d_1/k_2 \rceil$) for blocks of $k_2$ errors, i.e., $d_1'$ 
such blocks must have errors before the code fails.  Then the concatenated 
code has distance $d_1' d_2$.

Another way to concatenate codes encoding multiple qubits is to add 
additional blocks of $S_1$ to fill the spaces in $S_2$.  That is, we actually 
encode $k_2$ copies of $S_1$, encoding the $i$th qubit of each copy in the 
same $S_2$ block.  This produces an $[n_1 n_2, k_1 k_2, d_1 d_2]$ code, 
since any failure of an $S_2$ block only produces one error in each $S_1$ 
block.

\section{Higher Dimensional States}
\label{sec-qudits}

So far, we have only considered systems for which the Hilbert space is the
tensor product of two-state systems.  However, it may turn out that a good
physical implementation of quantum computation uses three- or four-level
atoms, or spin-one particles, or some other system where it makes more sense
to consider it as the tensor product of $d$-dimensional systems, where
$d > 2$.  I will call the fundamental unit of such a system a {\em qudit}.
In such a case, we will want to consider error correcting codes where a
single qudit error can occur with reasonable probability.  For these systems,
the stabilizer code formalism needs to be modified to deal with the extra
dimensions.

Fundamental to the success of the stabilizer formalism was the use of the
Pauli spin matrix basis for possible errors.  The algebraic properties of
this basis allowed a straightforward characterization of errors depending
on whether they commuted or anticommuted with elements of an Abelian group.
Knill~\cite{knill-qudit} has codified the properties necessary for this 
construction to generalize to $d$-dimensional spaces.  Suppose we have a
set of $d^2$ unitary operators $E_1, \ldots, E_{n^2}$ (including the 
identity) acting on a single qudit such that the $E_i$'s form a basis for all
possible $d \times d$ complex matrices.  If $E_i E_j = w_{ij} E_{i*j}$ for all 
$i, j$ (where $*$ is some binary group operation), then the $E_i$'s are said to 
form a {\em nice} error basis.  The values $w_{ij}$ will then have modulus one. 
Given a nice error basis, we form the group $\G_n$ for this basis as the tensor
product of $n$ copies of the error basis, with possible overall phases
generated by the $w_{ij}$'s.  Then an Abelian subgroup $S$ of $\G_n$ that does
not contain any nontrivial phase times the identity will have a nontrivial
set $T$ of states in the Hilbert space in the $+1$ eigenspace of every operator
in $S$.  The code $T$ can detect any error $E$ for which $E M = c M E$ for some
$M \in \G_n$ and some $c \neq 1$.

One interesting complication of codes over $d$-dimensional spaces is that
when $S$ has $n-k$ generators, $T$ need not encode $k$ qudits.  This can
only occur when $d$ is composite and the order of a generator of $S$ is
a nontrivial factor of $d$.  It is still true that if $S$ has $r$ elements,
then $T$ will be $(d^n/r)$-dimensional.  If all the generators of $S$ have
order $d$, $T$ does encode $k$ qudits.

One particularly convenient error basis for any $d$ is generated by
$D_\omega$ and $C_n$, where $(D_\omega)_{ij} = \delta_{ij} \omega^i$ and
$(C_n)_{ij} = \delta_{j, (i+1 \bmod n)}$.  $\omega$ is a primitive $n$th
root of unity.  For $d=2$, this just reduces to the usual Pauli basis,
since $C_2 = \X$ and $D_{-1} = \Z$.  For higher $d$, $D_\omega$ maps
$\ket{i} \rightarrow \omega^i \ket{i}$ and $C_n$ adds one modulo $n$.
This is a nice error basis, with
\begin{equation}
C_n D_\omega = \omega D_\omega C_n.
\end{equation}
The elements of the basis can be written $C_n^a D_\omega^b$, and 
\begin{equation}
\left( C_n^a D_\omega^b \right) \left( C_n^c D_\omega^d \right) =
\omega^{ad-bc} \left( C_n^c D_\omega^d \right) \left( C_n^a D_\omega^b \right).
\end{equation}

Codes for higher-dimensional systems have not been as extensively studied as
those for two-dimensional systems, but some constructions are given in
\cite{knill-qudit, chau-d^2, chau-5qudit, aharonov, rains-orthogonal}.

\chapter{Encoding and Decoding Stabilizer Codes}
\label{chap-encoding}

\section{Standard Form for a Stabilizer Code}

To see how to encode a general stabilizer code~\cite{cleve}, it is helpful to 
describe the code in the language of binary vector spaces (see section
\ref{sec-alternate}).  Note that the specific choice of generators is not 
at all unique.  We can always replace a generator $M_i$ with $M_i M_j$ for 
some other generator $M_j$.  The corresponding effect on the binary matrices 
is to add row $j$ to row $i$ in both matrices.  For simplicity, it is also 
helpful to rearrange qubits in the code.  This has the effect of rearranging 
the corresponding columns in both matrices.  Combining these two operations, 
we can perform Gaussian elimination on the first matrix, putting the code in 
this form:
\vspace{2.5ex}
\begin{equation}
\begin{array}{r} r\{ \\ n-k-r\{ \end{array} \!\!\!\!
\left( \begin{array}{cc|cc}
\raisebox{0ex}[1.5ex]{$\overbrace{I}^r$} & 
\raisebox{0ex}[1.5ex]{$\overbrace{A}^{n-r}$} & 
\raisebox{0ex}[1.5ex]{$\overbrace{B}^r$} & 
\raisebox{0ex}[1.5ex]{$\overbrace{C}^{n-r}$} \\
0 & 0 & D & E
\end{array} \right).
\end{equation}
Here, $r$ is the rank of the $\X$ portion of the stabilizer generator matrix.

Then we perform another Gaussian elimination on $E$ to get
\vspace{2.5ex}
\begin{equation}
\begin{array}{r} r\{ \\ n-k-r-s \{ \\ s \{ \end{array} \!\!\!\!
\left( \begin{array}{ccc|ccc}
\raisebox{0ex}[1.5ex]{$\overbrace{I}^r$} & 
\raisebox{0ex}[1.5ex]{$\overbrace{A_1}^{n-k-r-s}$} & 
\raisebox{0ex}[1.5ex]{$\overbrace{A_2}^{k+s}$} & 
\raisebox{0ex}[1.5ex]{$\overbrace{B}^r$} & 
\raisebox{0ex}[1.5ex]{$\overbrace{C_1}^{n-k-r-s}$} & 
\raisebox{0ex}[1.5ex]{$\overbrace{C_2}^{k+s}$} \\
0 & 0 & 0 & D_1 & I & E_2 \\
0 & 0 & 0 & D_2 & 0 & 0
\end{array} \right).
\end{equation}
The rank of $E$ is $n-k-r-s$.  However, the first $r$ generators will not 
commute with the last $s$ generators unless $D_2 = 0$, which really implies 
that $s=0$.  Thus we can always put the code into the standard form
\vspace{2.5ex}
\begin{equation}
\begin{array}{r} r\{ \\ n-k-r\{ \end{array} \!\!\!\!
\left( \begin{array}{ccc|ccc}
\raisebox{0ex}[1.5ex]{$\overbrace{I}^r$} & 
\raisebox{0ex}[1.5ex]{$\overbrace{A_1}^{n-k-r}$} & 
\raisebox{0ex}[1.5ex]{$\overbrace{A_2}^k$} & 
\raisebox{0ex}[1.5ex]{$\overbrace{B}^r$} & 
\raisebox{0ex}[1.5ex]{$\overbrace{C_1}^{n-k-r}$} & 
\raisebox{0ex}[1.5ex]{$\overbrace{C_2}^k$} \\
0 & 0 & 0 & D & I & E
\end{array} \right).
\label{eq-standard-form}
\end{equation}
For instance, the standard form for the five-qubit code of 
table~\ref{table-5qubit} is
\begin{equation}
\left( \begin{array}{ccccc|ccccc}
1 & 0 & 0 & 0 & 1 & 1 & 1 & 0 & 1 & 1 \\
0 & 1 & 0 & 0 & 1 & 0 & 0 & 1 & 1 & 0 \\
0 & 0 & 1 & 0 & 1 & 1 & 1 & 0 & 0 & 0 \\
0 & 0 & 0 & 1 & 1 & 1 & 0 & 1 & 1 & 1
\end{array} \right).
\end{equation}

Suppose we have an $\Xbar$ operator which in this language is written 
$(u|v) = (u_1 u_2 u_3| v_1 v_2 v_3)$, where $u_1$ and $v_1$ are 
$r$-dimensional vectors, $u_2$ and $v_2$ are $(n-k-r)$-dimensional 
vectors, and $u_3$ and $v_3$ are $k$-dimensional vectors.  However, 
elements of $N(S)$ are equivalent up to multiplication by elements of $S$.  
Therefore, we can also perform eliminations on $\Xbar$ to force $u_1 = 0$ 
and $v_2 = 0$.  Then, because $\Xbar$ is in $N(S)$, we must satisfy
(\ref{eq-commute-bin}), so
\begin{eqnarray}
\left( \begin{array}{cccccc}
I & A_1 & A_2 & B & C_1 & C_2 \\
0 & 0 & 0 & D & I & E
\end{array} \right)
\left( \begin{array}{c} v_1^T \\ 0 \\ v_3^T \\ 0 \\ u_2^T \\ u_3^T 
\end{array} \right)
& \!\!\! = & \!\!\! \left( \begin{array}{c}
 v_1^T + A_2 v_3^T + C_1 u_2^T + C_2 u_3^T \\
u_2^T + E u_3^T
\end{array} \right) \nonumber \\
& \!\!\! = & \!\!\! \left( \begin{array}{c} 0 \\ 0 \end{array} \right).
\label{eq-Xbar-commute}
\end{eqnarray}

Suppose we want to choose a complete set of $k$ $\Xbar$ operators.  We 
can combine their vectors into two $k \times n$ matrices $(0 U_2 U_3 | 
V_1 0 V_3)$.  We want them to commute with each other, so $U_3 V_3^T + 
V_3 U_3^T = 0$.  Suppose we pick $U_3 = I$.  Then we can take $V_3 = 0$, and 
by equation (\ref{eq-Xbar-commute}), $U_2 = E^T$ and $V_1 = E^T C_1^T + C_2^T$. 
The rest of the construction will assume that this choice has actually been 
made.  Another choice of $U_3$ and $V_3$ will require us to perform some 
operation on the unencoded data to compensate.  For the five-qubit code, the 
standard form of the $\Xbar$ generator would be $(0 0 0 0 1 | 1 0 0 1 0)$.  
We can see that this is equivalent (mod $S$) to the $\Xbar$ given in 
table~\ref{table-5qubit}.

We can also pick a complete set of $k$ $\Zbar$ operators, which act on the 
code as encoded $\Z$ operators.  They are uniquely defined (up to 
multiplication by $S$, as usual) given the $\Xbar$ operators.  $\Zbar_i$ is 
an operator that commutes with $M \in S$, commutes with $\Xbar_j$ for $i 
\neq j$, and anticommutes with $\Xbar_i$.  We can bring it into the 
standard form $(0 U_2' U_3' | V_1' 0 V_3')$.  Then
\begin{equation}
U_3' V_3^T + V_3' U_3^T = I.
\end{equation}
When $U_3 = I$ and $V_3 = 0$, $V_3' = I$.  Since equation 
(\ref{eq-Xbar-commute}) holds for the $\Zbar$ operators too, $U_2' = U_3' 
= 0$ and $V_1' = A_2^T$.  For instance, for the five-qubit code, the standard 
form of the $\Zbar$ generator is $(0 0 0 0 0 | 1 1 1 1 1)$, which is exactly 
what is given in table~\ref{table-5qubit}.

\section{Network for Encoding}
\label{sec-encode-network}

Given a stabilizer in standard form along with the $\Xbar$ operators in 
standard form, it is straightforward to produce a network to encode the 
corresponding code.  The operation of encoding a stabilizer code can be 
written as
\begin{eqnarray}
\ket{c_1 \ldots c_k} & \rightarrow & \left( \Sum_{M \in S} M \right) 
\Xbar_1^{c_1} \cdots \Xbar_k^{c_k} \ket{0 \ldots 0} \label{eq-encoding}\\
& = & (I + M_1) \cdots (I + M_{n-k}) \Xbar_1^{c_1} \cdots \Xbar_k^{c_k} 
\ket{0\ldots 0}, \label{eq-encoding2}
\end{eqnarray}
where $M_1$ through $M_{n-k}$ generate the stabilizer, and $\Xbar_1$ 
through $\Xbar_k$ are the encoded $\X$ operators for the $k$ encoded 
qubits.  This is true because, in general, for any $N \in S$,
\begin{equation}
N \left( \Sum_{M \in S} M \right) \ket{\psi} = \left( \Sum_{M \in S} NM 
\right) \ket{\psi} = \left( \Sum_{M' \in S} M' \right) \ket{\psi},
\end{equation}
so $\Sum M \ket{\psi}$ is in the coding space $T$ for any state 
$\ket{\psi}$.  If we define the encoded $0$ as
\begin{equation}
\ket{\overline{0}} = \Sum_{M \in S} M \ket{\overbrace{0 \ldots 0}^{n}},
\end{equation}
then by the definition of the $\Xbar$'s, we should encode
\begin{equation}
\ket{c_1 \ldots c_k} \rightarrow \Xbar_1^{c_1} \cdots \Xbar_k^{c_k} \left( 
\Sum_{M \in S} M \right)  \ket{0 \ldots 0}.
\end{equation}
Since $\Xbar_i$ commutes with $M \in S$, this is just (\ref{eq-encoding}).  
Naturally, to encode this, we only need to worry about encoding the basis 
states $\ket{c_1 \ldots c_k}$.

The standard form of $\Xbar_i$ has the form $Z^{(r)} X^{(n-k-r)} 
\Xs{(n-k+i)}$ ($Z^{(r)}$ is the product of $\Z$'s on the first $r$ qubits and 
$X^{(n-k-r)}$ is the product of $\X$'s on the next $n-k-r$ qubits).  Suppose 
we put the $k$th input qubit $\ket{c_k}$ in the $n$th spot, 
following $n-1$ $0$s.  The state $\Xbar_k^{c_k} \ket{0 \ldots 0}$ therefore 
has a $1$ for the $n$th qubit iff $\ket{c_k} = \ket{1}$.  This means we can 
get the state $\Xbar_k^{c_k} \ket{0 \ldots 0}$ by applying $\Xbar_k$ 
(without the final $\Xs{n}$) to the input state conditioned on the $n$th 
qubit.  For instance, for the five-qubit code, $\Xbar = Z \otimes I \otimes I 
\otimes Z \otimes X$.  The corresponding operation is illustrated in 
figure~\ref{fig-5qubit-Xbar}.
\begin{figure}
\centering
\begin{picture}(60,120)

\put(20,20){\line(1,0){40}}
\put(20,40){\line(1,0){14}}
\put(46,40){\line(1,0){14}}
\put(20,60){\line(1,0){40}}
\put(20,80){\line(1,0){40}}
\put(20,100){\line(1,0){14}}
\put(46,100){\line(1,0){14}}

\put(2,14){\makebox(12,12){$c$}}
\put(2,34){\makebox(12,12){$0$}}
\put(2,54){\makebox(12,12){$0$}}
\put(2,74){\makebox(12,12){$0$}}
\put(2,94){\makebox(12,12){$0$}}

\put(40,20){\circle*{4}}
\put(40,20){\line(0,1){14}}
\put(34,34){\framebox(12,12){$\Z$}}
\put(40,46){\line(0,1){48}}
\put(34,94){\framebox(12,12){$\Z$}}

\end{picture}
\caption{Creating the state $\Xbar \ket{00000}$ for the five-qubit code.}
\label{fig-5qubit-Xbar}
\end{figure}
In this case $r=n-k=4$, so there are no bit flips, only controlled $\Z$'s.

In the more general case, we also need to apply $\Xbar_1$ through 
$\Xbar_{k-1}$, depending on $c_1$ through $c_{k-1}$.  Since the form of 
the $\Xbar$'s ensures that each only operates on a single one of the last 
$k$ qubits, we can substitute $\ket{c_i}$ for the $(n-k+i)$th qubit and 
apply $\Xbar_i$ conditioned on it, as with $\ket{c_k}$.  This produces the 
state $\Xbar_1^{c_1} \cdots \Xbar_k^{c_k} \ket{0 \ldots 0}$.

Further, note that the $\Xbar$ operators only act as $\Z$ on the first $r$ 
qubits and as $\X$ on the next $n-k-r$ qubits.  Since $\Z$ acts trivially on 
$\ket{0}$, we can just ignore that part of the $\Xbar$'s when implementing 
this part of the encoder, leaving just the controlled NOTs.  The first $r$ 
qubits automatically remain in the state $\ket{0}$ after this step of 
encoding.  This means that for the five-qubit code, this step of encoding is 
actually trivial, with no operations.  In general, this step is only necessary 
if $r<n-k$.

For the next step of the encoding, we note that the standard form of the 
first $r$ generators only applies a single bit flip in the first $r$ qubits.  
This means that when we apply $I + M_i$, the resulting state will be the 
sum of a state with $\ket{0}$ for the $i$th qubit and a state with $\ket{1}$ 
for the $i$th qubit.  We therefore apply the Hadamard transform
\begin{equation}
R = \frac{1}{\sqrt{2}} \pmatrix{1 & \ 1 \cr 1 & -1}
\label{eq-hadamard}
\end{equation}
to the first $r$ qubits, putting each in the state $\ket{0} + \ket{1}$.  Then 
we apply $M_i$ (for $i = 1, \ldots, r$) conditioned on qubit $i$ (ignoring 
the factor of $\Xs{i}$).  While these operators may perform phase 
operations on the first $r$ qubits, they do not flip them, so there is no risk 
of one operation confusing the performance of another one.  The one 
possible complication is when $M_i$ has a factor of $\Zs{i}$.  In this case, 
$\Zs{i}$ only introduces a minus sign if the qubit is $\ket{1}$ anyway, so we 
do not need to condition it on anything.  Just performing $\Zs{i}$ after the
Hadamard transform is sufficient.  For the five-qubit code, the full network 
for encoding is given in figure~\ref{fig-5qubit}.
\begin{figure}
\centering
\begin{picture}(160,120)

\put(20,20){\line(1,0){54}}
\put(86,20){\line(1,0){48}}
\put(146,20){\line(1,0){14}}

\put(20,40){\line(1,0){14}}
\put(46,40){\line(1,0){8}}
\put(66,40){\line(1,0){8}}
\put(86,40){\line(1,0){8}}
\put(106,40){\line(1,0){54}}

\put(20,60){\line(1,0){14}}
\put(46,60){\line(1,0){48}}
\put(106,60){\line(1,0){28}}
\put(146,60){\line(1,0){14}}

\put(20,80){\line(1,0){14}}
\put(46,80){\line(1,0){28}}
\put(86,80){\line(1,0){28}}
\put(126,80){\line(1,0){34}}

\put(20,100){\line(1,0){14}}
\put(46,100){\line(1,0){8}}
\put(66,100){\line(1,0){48}}
\put(126,100){\line(1,0){8}}
\put(146,100){\line(1,0){14}}

\put(4,14){\makebox(12,12){$c$}}
\put(4,34){\makebox(12,12){$0$}}
\put(4,54){\makebox(12,12){$0$}}
\put(4,74){\makebox(12,12){$0$}}
\put(4,94){\makebox(12,12){$0$}}

\put(34,34){\framebox(12,12){$R$}}
\put(34,54){\framebox(12,12){$R$}}
\put(34,74){\framebox(12,12){$R$}}
\put(34,94){\framebox(12,12){$R$}}

\put(54,34){\framebox(12,12){$\Z$}}
\put(54,94){\framebox(12,12){$\Z$}}

\put(80,100){\circle*{4}}
\put(80,100){\line(0,-1){14}}
\put(74,74){\framebox(12,12){$\Z$}}
\put(80,74){\line(0,-1){28}}
\put(74,34){\framebox(12,12){$\Z$}}
\put(80,34){\line(0,-1){8}}
\put(74,14){\framebox(12,12){$\Y$}}

\put(100,80){\circle*{4}}
\put(100,80){\line(0,-1){14}}
\put(94,54){\framebox(12,12){$\Z$}}
\put(100,54){\line(0,-1){8}}
\put(94,34){\framebox(12,12){$\Z$}}
\put(100,34){\line(0,-1){18}}
\put(100,20){\circle{8}}

\put(120,60){\circle*{4}}
\put(120,60){\line(0,1){14}}
\put(114,74){\framebox(12,12){$\Z$}}
\put(120,86){\line(0,1){8}}
\put(114,94){\framebox(12,12){$\Z$}}
\put(120,60){\line(0,-1){44}}
\put(120,20){\circle{8}}

\put(140,40){\circle*{4}}
\put(140,40){\line(0,1){14}}
\put(134,54){\framebox(12,12){$\Z$}}
\put(140,66){\line(0,1){28}}
\put(134,94){\framebox(12,12){$\Z$}}
\put(140,40){\line(0,-1){14}}
\put(134,14){\framebox(12,12){$\Y$}}

\end{picture}
\caption{Network for encoding the five-qubit code.}
\label{fig-5qubit}
\end{figure}

For more general codes, $r<n-k$, and there are $n-k-r$ generators that are 
formed just of the tensor product of $\Z$'s.  However, we do not need to 
consider such generators to encode.  Let $M$ be such a generator.  Since 
$M$ commutes with all the other generators and every $\Xbar$, we can 
commute $I + M$ through until it acts directly on $\ket{0 \ldots 0}$.  
However, $\Z$ acts trivially on $\ket{0}$, so $I+M$ fixes $\ket{0 \ldots 0}$, 
and in equation~(\ref{eq-encoding2}), we can skip any $M_i$ that is the tensor 
product of $\Z$'s.  The effect of these operators is seen just in 
the form of the $\Xbar$ operators, which must commute with them.

Applying each of the $\Xbar$ operators requires up to $n-k-r$ two-qubit 
operations.  Each of the first $r$ qubits must be prepared with a Hadamard 
transform and possibly a $\Z$, which we can combine with the Hadamard 
transform.  Then applying each of the first $r$ generators requires up to 
$n-1$ two-qubit operations.  The whole encoder therefore requires up to $r$ 
one-qubit operations and at most
\begin{equation}
k(n-k-r) + r(n-1) \leq (k+r)(n-k) \leq n(n-k)
\end{equation} 
two-qubit operations.

\section{Other Methods of Encoding and Decoding}

We can decode a code by performing the above network in reverse.  In 
order to do this, we should first perform an error correction cycle, since 
the network will not necessarily work properly on an encoded state.  Note 
that in principle we can build a decoder that corrects while decoding.  We 
can form a basis for the Hilbert space from the states $A \ket{\psi_i}$, 
where $A \in \G$ and $\ket{\psi_i}$ is a basis state for the coding space 
$T$.  The combined corrector/decoder would map $A \ket{\psi_i}$ to 
$\ket{i} \otimes \ket{f(A)}$, where $f(A)$ is the error syndrome for $A$.  
If $A$ is not a correctable error, $\ket{i}$ will not necessarily be the state 
encoded by $\ket{\psi_i}$, but if $A$ is correctable, it will be.  It is not 
usually worthwhile using a quantum network that does this, since the error 
correction process is usually dealt with more easily using classical 
measurements.  However, some proposed implementations of quantum computation 
cannot be used to measure a single system~\cite{gershenfeld}, so this sort of 
network would be necessary.  The decoding method presented in 
\cite{divincenzo-decoder} can easily be adapted to produce networks that
simultaneously correct and decode.

One good reason not to decode by running the encoder backwards is that 
most of the work in the encoder went into producing the encoded $0$.  
There is no actual information in that state, so we might be able to save 
time decoding if we could remove the information without dealing with the 
structure of the encoded $0$.  We can do this by using the $\Xbar$ and 
$\Zbar$ operators.  If we want to measure the $i$th encoded qubit 
without decoding, we can do this by measuring the eigenvalue of 
$\Zbar_i$.  If the eigenvalue is $+1$, the $i$th encoded qubit is $\ket{0}$; 
if it is $-1$, the $i$th encoded qubit is $\ket{1}$.  In standard form, 
$\Zbar_i$ is the tensor product of $\Z$'s.  That means it will have 
eigenvalue $(-1)^P$, where $P$ is the parity of the qubits acted on by 
$\Zbar_i$.  Therefore, if we apply a controlled-NOT from each of these 
qubits to an ancilla qubit, we have performed a controlled-NOT from the 
$i$th encoded qubit to the ancilla --- we will flip the ancilla iff the $i$th 
encoded qubit is $\ket{1}$.

If the original state of the code is $\ket{\overline{0}} \ket{\psi} + 
\ket{\overline{1}} \ket{\phi}$ (with the first ket representing the $i$th 
{\em logical} qubit) and the ancilla begins in the state $\ket{0}$, after 
applying this CNOT operation, we have
\begin{equation}
\ket{\overline{0}} \ket{\psi} \ket{0} + \ket{\overline{1}} \ket{\phi} 
\ket{1}.
\end{equation}
Now we apply $\Xbar_i$ conditioned on the ancilla qubit.  This will flip the 
$i$th encoded qubit iff the ancilla is $\ket{1}$.  This produces the state
\begin{equation}
\ket{\overline{0}} \ket{\psi} \ket{0} + \ket{\overline{0}} \ket{\phi} 
\ket{1} = \ket{\overline{0}} \left(\ket{\psi} \ket{0} + \ket{\phi} \ket{1} 
\right).
\end{equation}
The $i$th encoded qubit has been set to $0$ and the ancilla holds the state 
that the $i$th encoded qubit used to hold.  The rest of the code has been 
left undisturbed.  We can repeat this operation with each of the encoded 
qubits, transferring them to $k$ ancilla qubits.  Each such operation requires 
at most $2(n-k+1)$ two-qubit operations (since $\Zbar$ requires at most $r + 1$ 
operations and $\Xbar$ could require $n-k+1$ operations).  Therefore, the 
full decoder uses at most $2k(n-k+1)$ operations, which is often less than 
is required to encode.  At the end of the decoding, the original $n$ qubits 
holding the code are left in the encoded $0$ state.

We can run this process backwards to encode, but we need an encoded $0$ 
state to begin with.  This could be a residue from an earlier decoding 
operation, or could be produced separately.  One way to produce it would 
be to use the network of section~\ref{sec-encode-network}, using 
$\ket{0\ldots 0}$ as the input data.  Alternately, we could produce it by 
performing an error correction cycle on a set of $n$ $\ket{0}$'s for the 
stabilizer generated by $M_1, \ldots M_{n-k}, \Zbar_1, \ldots, \Zbar_k$.  
This stabilizer has $n$ generators, so there is only one joint $+1$ 
eigenvector, which is just the encoded $0$ for the original code.

\chapter{Fault-Tolerant Computation}
\label{chap-fault-tolerant}

\section{Encoded Computation and Fault-Tolerance}

I have shown how to encode qubits in blocks to protect them from 
individual errors.  This, by itself, is useful for transmitting quantum data 
down a noisy communications line, for instance --- we can encode the data 
using the code, send it, correct the errors, and decode it.  Then we can 
process the data normally.  However, the framework so far is insufficient 
for performing computations on a realistic quantum computer.  If we need 
to decode the data in order to perform quantum gates on it, it is vulnerable 
to noise during the time it is decoded.  Even if we know how to perform 
gates on the data while it is still encoded, we must be careful to make sure 
that a single error does not cause us to accidentally perform the wrong 
computation.

For instance, suppose a single qubit has been flipped and we apply a 
con\-trolled-NOT from it to another qubit.  Then the second qubit will flip 
exactly when it is supposed to stay the same.  In consequence, now both 
the first and the second qubits have bit flip errors.  If both qubits are part 
of the same block, we now have two errors in the block instead of one.  
Before very much of this occurs, we will have too many errors in the block 
to correct.  If we correct errors often enough, we can salvage the 
situation~\cite{aharonov}, but in the process we lose a lot of the power of 
the error-correcting code.  Therefore, I will define a fault-tolerant 
operation as one for which a single error introduces at most one error per 
block of the code.  In a large computer, we have many encoded blocks of 
data, and a given operation may introduce one error in a number of them.  
However, each block retains its ability to correct that single error.

In the example above, an error propagated forward from the control qubit 
to the target qubit of the CNOT.  In a quantum computer, errors can also 
propagate backwards.  For instance, suppose we have the state
\begin{equation}
(\alpha \ket{0} + \beta \ket{1}) (\ket{0} \pm \ket{1})
\end{equation}
and perform a CNOT from the first qubit to the second.  The resulting state 
is
\begin{equation}
\alpha \ket{0} (\ket{0} \pm \ket{1}) + \beta \ket{1} (\pm 1) (\ket{0} \pm 
\ket{1}) = (\alpha \ket{0} \pm \beta \ket{1}) (\ket{0} \pm \ket{1}).
\end{equation}
Initially flipping the sign on the second qubit will result in a sign flip on 
the first qubit after the CNOT.  In a CNOT, amplitude (bit flip) errors 
propagate forwards, and phase errors propagate backwards.

This means that not only must we make sure not to perform operations 
from one qubit to another within a block, we must also be sure not to 
perform multiple CNOTs from a block onto the same target qubit, even if it 
is a disposable ancilla qubit.  Otherwise, a single phase error in the ancilla
qubit can produce multiple errors within a block.  Operations for which each 
qubit in a block only interacts with the corresponding qubit, either in another 
block or in a specialized ancilla, will be called {\em transversal} operations. 
Any transversal operation is automatically fault-tolerant, although there are 
some fault-tolerant operations which are not transversal.

\section{Measurement and Error Correction}
\label{sec-error-cor}

Suppose we want to measure the operator $\Zs{1} \Zs{2}$, as with Shor's 
nine-qubit code.  The eigenvalue is $+1$ if both qubits are the same and $-1$ 
if they are different.  One natural way to do this is perform a CNOT from 
both qubits to a third ancilla qubit, initially in the state $\ket{0}$.  If 
both qubits are $\ket{0}$, the ancilla is left alone, and if both are 
$\ket{1}$, the ancilla gets flipped twice, returning to the state $\ket{0}$.  
If only one of the two qubits is $\ket{1}$, the ancilla only flips once, ending 
up in the state $\ket{1}$.  Measuring the ancilla will then tell us the 
eigenvalue of $\Zs{1} \Zs{2}$.

However, this procedure is not a transversal operation.  Both qubits 
interact with the same ancilla qubit, and a single phase error on the ancilla 
qubit could produce phase errors in both data qubits, producing two errors 
in the block (actually, this particular example does not have this problem, 
since a phase error on the ancilla qubit is meaningless until after it has 
interacted with the first data qubit; but if we were measuring $\Zs{1} \Zs{2} 
\Zs{3}$ instead, the problem would be a real one).  One possible solution to 
the problem is to use two ancilla qubits, both initially $\ket{0}$, instead of 
one.  Then we perform CNOTs from the first data qubit to the first ancilla 
qubit and from the second data qubit to the second ancilla qubit.  Then we 
measure the ancilla qubits and determine their parity.  This will again tell 
us the eigenvalue of $\Zs{1} \Zs{2}$, and we do not run the risk of introducing 
two phase errors into the data.

However, we have instead done something worse.  By measuring both 
ancilla qubits, we have, in effect, measured the original data qubits, which 
destroys any superposition of the $+1$-eigenstates of $\Zs{1} \Zs{2}$.  To make 
this work, we need to be able to measure the ancilla without finding out 
anything about the data.  Since we are only interested in the parity of the 
data qubits, we could have just as well started the ancilla in the state 
$\ket{11}$ as $\ket{00}$.  If both or neither ancilla qubits are flipped, the 
parity is still even, and if only one is flipped, the parity is odd, as it 
should be.  However, measuring the ancilla still tells us what states the data 
qubits were in.  The state of a data qubit is equal to the reverse of the 
measured state of the corresponding ancilla qubit.

This means if we start the ancilla in the superposition $\ket{00} + 
\ket{11}$ and perform CNOTs from the data qubits to the ancilla qubits, 
measuring the ancilla will again tell us the parity of the data qubits.  
However, we do not know whether the state we measure originally 
corresponded to the ancilla state $\ket{00}$ or $\ket{11}$, which means 
we cannot deduce the state of the data.  The two ancilla states correspond 
to the two possible states of the data qubits with the same parity.  This 
means that measuring the ancilla does not destroy a superposition of these 
two states of the data.  This is what we desired.

Because we interact each data qubit with a separate ancilla qubit, a single 
phase error in the ancilla will only produce a single phase error in the 
data.  Of course, if a single qubit in the ancilla flips so we start in the 
state $\ket{01} + \ket{10}$, we will measure the wrong parity.  We can 
circumvent this problem by simply preparing multiple ancillas in the same 
state, performing the CNOTs to each of them, and measuring each.  If we 
prepare three such ancillas and determine the parity as the majority 
result, the answer will be correct unless two errors have occurred.  If the 
chance of a single error is $\epsilon$, the chance of getting two errors in 
the data or getting the wrong measurement result is $O(\epsilon^2)$.

We can use this trick on products of more than two $\Z$ 
operators~\cite{shor-fault-tol} by preparing the ancilla in a state which is 
the sum of all even parity states.  Such a state can be made by preparing a 
``cat'' state $\ket{0 \ldots 0} + \ket{1 \ldots 1}$ (named after 
Schr\"odinger's cat) and performing a Hadamard transform 
(\ref{eq-hadamard}) on each qubit.  Again, we perform a CNOT from the 
data qubits to corresponding qubits in the ancilla and measure the ancilla.  
The result will have even parity iff the selected data qubits have even 
parity, but the measurement does not destroy superpositions of the 
possible data states with that parity.  Again, a single error in the ancilla 
could give the wrong parity, so we should repeat the measurement.  Also, 
the preparation of the ``cat'' state is not at all fault-tolerant, so we could 
easily have multiple bit flip errors in the ``cat'' state, which will result 
in multiple phase errors in the ancilla state.  Since phase errors will feed 
back into the data, we should carefully verify the ``cat'' state to make sure 
that we do not have multiple amplitude errors.

Suppose we want to measure a more general operator in $\G$, such as 
$M_1 = \X \otimes\Z \otimes \Z \otimes \X \otimes I$, the first generator 
for the five-qubit code.  Note that under the Hadamard transform
\begin{eqnarray}
\ket{0} & \leftrightarrow & \ket{0} + \ket{1} \nonumber \\
\ket{1} & \leftrightarrow & \ket{0} - \ket{1},
\end{eqnarray}
so the eigenvectors of $\Z$ transform to the eigenvectors of $\X$ and 
vice-versa.  This means to measure $M_1$, we should perform the 
Hadamard transform on qubits one and four and instead measure $\Z 
\otimes \Z \otimes \Z \otimes \Z \otimes I$.  We know how to do this from 
the above discussion.  Then we should perform the Hadamard transform 
again to return to the original state (modulo any collapse caused by the 
measurement).  In a similar way, we can rotate $\Y$ into $\Z$ (exactly 
how is discussed in more detail in section~\ref{sec-normalizer}), and 
therefore measure any operator in $\G$.

From the ability to make measurements, we can easily perform error 
correction for any stabilizer code~\cite{divincenzo}.  Recall that to correct 
errors, we measure the eigenvalue of each generator of the stabilizer.  This 
we now know how to do fault-tolerantly.  This tells us the error syndrome, 
which tells us the error (or class of degenerate errors).  This error is some 
operator in $\G$, and to correct it, we just apply the operator to the code.  
Since it is the tensor product of single qubit operators, this is a transversal 
operation, and is therefore fault-tolerant.

Because a full measurement of the error syndrome takes a fair amount of 
time, the possibility of an error in the data while measuring the syndrome 
cannot be ignored.  An error in the data in the middle of the syndrome 
measurement will result in the wrong syndrome, which could correspond 
to a totally different error with nothing in common with the actual error.  
Therefore, we should measure the syndrome multiple times, only stopping 
when we have sufficient confidence that we have determined the correct 
current error syndrome.  Since we are measuring the syndrome multiple 
times, we only need to measure each bit once per overall syndrome 
measurement; repetitions of the syndrome measurement will also protect 
against individual errors in the syndrome bits.  The true error syndrome will 
evolve over the course of repeated measurements.  Eventually, more 
errors will build up in the data than can be corrected by the code, 
producing a real error in the data.  Assuming the basic error rate is low 
enough, this occurance will be very rare, and we can do many error 
correction cycles before it happens.  However, eventually the computation will 
fail.  In chapter~\ref{chap-concatenation}, I will show how to avoid this 
result and do arbitrarily long computations provided the basic error rate is 
sufficiently low.

\section{Transformations of the Stabilizer}
\label{sec-normalizer}

Now I will begin to discuss how to perform actual operations on encoded 
states.  We already know how to perform encoded $\X$, $\Y$, and $\Z$ 
operations on stabilizer codes.  These operations all commute with the
stabilizer and therefore leave the generators of the stabilizer alone.  A 
more general unitary operation $U$ will not necessarily do this.  If $M \in 
S$, then $\ket{\psi} = M \ket{\psi}$ for $\ket{\psi} \in T$, and
\begin{equation}
U \ket{\psi} = U M \ket{\psi} = UMU^\dagger U \ket{\psi},
\end{equation}
so $U M U^\dagger$ fixes $U \ket{\psi}$.  Even if we have an operator $N$
which is not in $S$, $U$ will take the eigenvectors of $N$ to eigenvectors
of $U N U^\dagger$, effectively transforming $N \rightarrow U N U^\dagger$.
Suppose $U M U^\dagger \in \G$.  Then if we want an operation that takes an 
encoded codeword to another valid codeword, we need $U M U^\dagger \in S$.  
If this is true for all $M \in S$, then $U \ket{\psi} \in T$ as well, and $U$ 
is a valid encoded operation.  If it is also transversal, we know it will be 
fault-tolerant as well.

The set of $U$ such that $U A U^\dagger \in \G$ for all $A \in \G$ is the 
normalizer $N(\G)$ of $\G$ in $U(n)$.  It turns out that $N(\G)$ is 
generated by the single qubit operations $R$ (the Hadamard transform) 
and
\begin{equation}
P = \pmatrix{1 & 0 \cr 0 & i},
\end{equation}
and the controlled NOT~\cite{bennett-tome,calderbank-stab}.  The set of 
$U$ such that $U M U^\dagger \in S$ for all $M \in S$ is the normalizer 
$N_{U(n)}(S)$ of $S$ in $U(n)$, which need not be a subset of $N(\G)$.  Any 
transversal operator in $N_{U(n)}(S)$ is a valid fault-tolerant operation.  
However, operators outside of $N(\G)$ are much more difficult to work 
with and analyze.  Therefore, I will restrict my attention to operators in 
the intersection of $N(\G)$ and $N_{U(n)}(S)$.

The operators in $N(\G)$ acting on $\G$ by conjugation permute tensor products 
of $\X$, $\Y$, and $\Z$.  For instance,
\begin{eqnarray}
R \X R^\dagger & = \frac{1}{2} \pmatrix{1 & \ 1 \cr 1 & -1} \pmatrix{0 & 
1 \cr 1 & 0} \pmatrix{1 & \ 1 \cr 1 & -1} = \pmatrix{1 & \ 0 \cr 0 & -1} = 
& \Z \\
R \Z R^\dagger & = \frac{1}{2} \pmatrix{1 & \ 1 \cr 1 & -1} \pmatrix{1 & \ 
0 \cr 0 & -1} \pmatrix{1 & \ 1 \cr 1 & -1} = \pmatrix{0 & 1 \cr 1 & 0} = & 
\X.
\end{eqnarray}
Also, 
\begin{equation}
R \Y R^\dagger = -i R \X \Z R^\dagger = -i R \X R^\dagger R \Z R^\dagger = -
i \Z \X = -\Y.
\end{equation}
$R$ switches $\X$ and $\Z$.  Similarly,
\begin{eqnarray}
P \X P^\dagger & = \pmatrix{1 & 0 \cr 0 & i} \pmatrix{0 & 1 \cr 1 & 0} 
\pmatrix{1 & \ 0 \cr 0 & -i} = \pmatrix{0 & -i \cr i & \ 0} = & \Y \\
P \Z P^\dagger & = \pmatrix{1 & 0 \cr 0 & i} \pmatrix{1 & \ 0 \cr 0 & 
-1} \pmatrix{1 & \ 0 \cr 0 & -i} = \pmatrix{1 & \ 0 \cr 0 & -1} = & \Z.
\end{eqnarray}
$P$ switches $\X$ and $\Y$.  These two operations generate all possible 
permutations of $\X$, $\Y$, and $\Z$.  Operators in $N(\G_1)$ can be
viewed as transformations of the Bloch sphere which permute the coordinate
axes.

The third generator of $N(\G)$ is the controlled NOT.  It acts on two qubits, 
and therefore permutes the elements of $\G_2$.  Its action is as follows:
\begin{eqnarray}
\X \otimes I & \rightarrow & \X \otimes \X \nonumber \\
I \otimes \X & \rightarrow & I \otimes \X \\
\Z \otimes I & \rightarrow & \Z \otimes I \nonumber \\
I \otimes \Z & \rightarrow & \Z \otimes \Z \nonumber.
\end{eqnarray}
Amplitudes are copied forwards and phases are copied backwards, as I 
described before.  In the same way, any element of $N(\G)$ gives a 
permutation of $\G$.  These permutations of $\G$ always preserve the 
group structure of $\G$, so are actually automorphisms of $\G$.

Given an automorphism of $\G$, we can always find an element of $N(\G)$ 
that produces that automorphism~\cite{gottesman-fault-tol}, modulo the 
automorphism $iI \rightarrow -iI$.  We can find the matrix of a given 
transformation $U$ corresponding to some automorphism by determining 
the action of $U$ on basis states.  $\ket{0}$ is an eigenvector of $\Z$, so it 
is mapped to an eigenvector of $U \Z U^\dagger$.  $\ket{1} = \X \ket{0}$, 
so it becomes $(U \X U^\dagger) U \ket{0}$.  For instance, the 
automorphism $T: \X \rightarrow \Y,\ \Z \rightarrow \X$ maps $\ket{0} 
\rightarrow (1 / \sqrt{2})\,(\ket{0} + \ket{1})$ and $\ket{1} 
\rightarrow \Y T \ket{0} = -(i / \sqrt{2})\,(\ket{0} - \ket{1})$.  Thus, 
the matrix of $T$ is
\begin{equation}
T = \frac{1}{\sqrt{2}} \pmatrix{ 1 & -i \cr 1 & \ i }.
\end{equation}

Another useful operation is to swap two qubits in a block.  This is not a 
transversal operation, and it is not fault-tolerant by itself.  An error during 
the swap gate can produce errors in the two qubits to be swapped, 
producing two errors in the same block.  However, we do not need to 
worry about error propagation because the swap gate swaps the errors 
along with the correct states.  Therefore, to get a fault-tolerant swap gate, 
we only need to produce a circuit to swap qubits that does not directly 
interact them.  Such a circuit is given in figure~\ref{fig-swap}.
\begin{figure}
\centering
\begin{picture}(270,80)

\put(0,14){\makebox(20,12){$\ket{\gamma}$}}
\put(0,34){\makebox(20,12){$\ket{\beta}$}}
\put(0,54){\makebox(20,12){$\ket{\alpha}$}}

\put(20,20){\line(1,0){80}}
\put(20,40){\line(1,0){80}}
\put(20,60){\line(1,0){80}}

\put(100,14){\makebox(20,12){$\ket{\gamma}$}}
\put(100,34){\makebox(20,12){$\ket{\alpha}$}}
\put(100,54){\makebox(20,12){$\ket{\beta}$}}

\put(40,16){\line(0,1){48}}
\put(40,20){\circle{8}}
\put(40,60){\circle{8}}

\put(60,36){\line(0,1){28}}
\put(60,40){\circle{8}}
\put(60,60){\circle{8}}

\put(80,16){\line(0,1){28}}
\put(80,20){\circle{8}}
\put(80,40){\circle{8}}

\put(180,30){\line(1,0){20}}
\put(180,50){\line(1,0){20}}
\put(190,30){\circle{8}}
\put(190,50){\circle{8}}
\put(190,26){\line(0,1){28}}
\put(200,34){\makebox(40,12){swap}}

\end{picture}
\caption{Network to swap $\ket{\alpha}$ and $\ket{\beta}$ using ancilla 
$\ket{\gamma}$.}
\label{fig-swap}
\end{figure}

In order to produce a valid fault-tolerant encoded operation, we may 
combine swap operations within a block of an error-correcting code and 
transversal operations on the block to get something that permutes the 
elements of the stabilizer.  The set of such operations is the automorphism 
group $\A (S)$ of $S$.  Codes with a large automorphism group are 
therefore better suited for performing fault-tolerant operations.  For 
instance, the seven-qubit code of table~\ref{table-7qubit} is invariant 
under any single-qubit operation in $N(\G)$ performed bitwise.  There are 
also a number of permutations of its qubits in the automorphism group, 
although they turn out to be unimportant in this case.  The five-qubit code 
of table~\ref{table-5qubit} has fewer automorphisms.  The only 
transversal operations in its automorphism group are
\begin{equation}
T: \X \rightarrow \Y, \ \Z \rightarrow \X
\end{equation}
and $T^2$.  Note that in the language of GF(4) codes, the operation $T$ 
corresponds to multiplication by $\omega^2$.  Therefore it is a valid 
transversal operation for any linear quantum code.  The five-qubit code is 
also invariant under cyclic permutations of the five component qubits, 
although these operations turn out to leave the encoded data unchanged, 
so are not very useful.

Once we have a possible encoded operation $U$, we must discover what it 
actually does to the encoded states.  We can do this by analyzing the 
behavior of $N(S)/S$ under the operation.  Because $U$ is in $N(\G) \cap
N_{U(n)} (S)$, it also has a natural action on $N(S)/S \cong \G_k$.  This 
action on $\G_k$ is equivalent to some operation in $N(\G_k)$.  This is the 
operation that is performed on the $k$ encoded qubits.  For instance, the 
Hadamard transform $R$ applied bitwise to the seven-qubit code switches $\Xbar 
= \Xs{5} \Xs{6} \Xs{7}$ and $\Zbar = \Zs{5} \Zs{6} \Zs{7}$.  This is just $R$ 
applied to the $\G_1$ group for the single encoded qubit.  In the same way, $P$ 
bitwise for the seven-qubit code converts $\Xbar$ into $-\Ybar$ ($\Ybar$ is the 
encoded $\Y$), and thus performs an encoded $P^\dagger$.  The minus sign
for $\Ybar$ occurs because $\Ybar = -i \Xbar \Zbar = -i (i^3) \Ys{5} \Ys{6}
\Ys{7} = - \Ys{5} \Ys{6} \Ys{7}$.

For the five-qubit code, 
$\Xbar = \Xs{1} \Xs{2} \Xs{3} \Xs{4} \Xs{5}$ and $\Zbar = \Zs{1} \Zs{2} \Zs{3} 
\Zs{4} \Zs{5}$, so $T$ bitwise transforms $\Xbar$ to $\Ybar$ and $\Zbar$ to 
$\Xbar$, and therefore acts as an encoded $T$ operation.  For both the 
five- and seven-qubit codes, the qubit permutations in $\A(S)$ produce 
the identity operation on the encoded qubits.  For a block encoding $k$ 
qubits, an operation in the automorphism group might perform any 
multiple-qubit operation in $N(\G_k)$.

We can also do multiple-qubit operations interacting two blocks by 
applying multiple-qubit operations transversally between the blocks.  For 
instance, we can apply a CNOT from the $i$th qubit in the first block to the 
$i$th qubit in the second block.  We can interact $r$ blocks by applying 
transversally any operation in $N(\G_r)$.  We can even apply different 
operations to different qubits within a block.  However, we should not also 
apply swaps within a block unless we can perform error correction 
afterwards, since otherwise errors could spread from one qubit in a block 
to the corresponding qubit in a different block, then back to a different 
qubit in the first block, producing two errors in the first block.

The stabilizer of two blocks of a code is just $S \times S$.  Therefore, the 
operation, to be valid, must permute the elements of this group.  For 
instance, bitwise CNOT applied between two blocks of the seven-qubit code 
is a valid operation, because
\begin{eqnarray}
M_i \otimes I & \rightarrow & M_i \otimes M_i \ (i = 1, 2, 3) \nonumber \\
M_i \otimes I & \rightarrow & M_i \otimes I \ (i = 4, 5, 6) \\
I \otimes M_i & \rightarrow & I \otimes M_i \ (i = 1, 2, 3) \nonumber \\
I \otimes M_i & \rightarrow & M_i \otimes M_i \ (i = 4, 5, 6). \nonumber
\end{eqnarray}
Since this also takes 
\begin{eqnarray}
\Xbar \otimes I & \rightarrow & \Xbar \otimes \Xbar \nonumber \\
I \otimes \Xbar & \rightarrow & I \otimes \Xbar \\
\Zbar \otimes I & \rightarrow & \Zbar \otimes I \nonumber \\
I \otimes \Zbar & \rightarrow & \Zbar \otimes \Zbar, \nonumber
\end{eqnarray}
it acts as a CNOT on the encoded qubits.  On the other hand, bitwise CNOT 
applied to the five-qubit code is not a valid operation, because, for instance, 
$M_1 = \X \otimes \Z \otimes \Z \otimes \X \otimes I$, so $M_1 \otimes I 
\rightarrow M_1 \otimes (\X \otimes I \otimes I \otimes \X \otimes I)$ 
and $\X \otimes I \otimes I \otimes \X \otimes I$ is not in $S$.

The CSS codes are those for which the stabilizer is the direct product of a 
part where the elements are tensor products of $\Xs{i}$'s and a part where 
the elements are tensor products of $\Zs{i}$'s.  We can also pick the $\Xbar$ 
and $\Zbar$ operators to be tensor products of $\Xs{i}$'s and $\Zs{i}$'s, 
respectively.  This means that just as with the seven-qubit code, bitwise CNOT 
will be a valid operation for any CSS codes, and will perform the CNOT between 
corresponding encoded qubits in the two blocks.

Conversely, if bitwise CNOT is a valid operation for a code, that means it is 
a CSS code: Let $M = X Y$ be an arbitrary element of the stabilizer $S$, 
where $X$ is the tensor product of $\Xs{i}$'s and $Z$ is the tensor product of 
$\Zs{i}$'s.  Then, under CNOT, $M \otimes I \rightarrow M \otimes X$ and 
$I \otimes M \rightarrow Z \otimes M$.  Thus, $X$ and $Z$ are themselves 
elements of $S$.  The stabilizer therefore breaks up into a $\X$ part and a 
$\Z$ part, which means it is a CSS code.

\section{The Effects of Measurements}
\label{sec-measurements}

We are not strictly limited to unitary operations in a quantum 
computation.  We can also make measurements, which correspond to 
randomly applying one of a set of complete projection operators, usually 
labeled by eigenvalues of a Hermitian operator.  Based on the classical 
measurement result, we can then apply one of a number of possible 
operators to the resulting quantum state.  This process can be converted 
into a purely quantum process, but in the idealization where classical 
computation is error-free while quantum computation is not, there is a 
distinct advantage in converting as much as possible to classical 
information.  Even in a more realistic situation, classical computation is 
likely to be much more reliable than quantum computation and classical 
error-correction methods are simpler than quantum ones.  In addition, we 
may know how to perform operations conditioned on classical information 
fault-tolerantly even when we do not know how to perform the 
corresponding quantum operations fault-tolerantly.  As we shall see, 
ancilla preparation and measurement are powerful tools for expanding the 
available set of fault-tolerant quantum operations.

Suppose we wish to measure operator $A$, with $A^2 = I$.  Measuring $A$ 
for a state $\ket{\psi}$ will typically give one of two results $\ket{\psi_+}$ 
or $\ket{\psi_-}$, corresponding to the two eigenvalues $\pm 1$ of $A$.  
In order to keep the description of our algorithm under control, we would 
like a way to convert $\ket{\psi_-}$ to $\ket{\psi_+}$ for any possible 
input state $\ket{\psi}$.  This will not be possible unless we know 
something more about the possible states $\ket{\psi}$.  Suppose we know 
that there is a unitary operator $M$, with $M \ket{\psi} = \ket{\psi}$ and 
$\{M, A\} = 0$.  Then
\begin{eqnarray}
M^\dagger \ket{\psi_-} & = & M^\dagger \, \frac{1}{2} (I - A) \ket{\psi} = 
M^\dagger \, \frac{1}{2} (I - A) M \ket{\psi} \nonumber \\
& = & M^\dagger M \, \frac{1}{2} (I + A) \ket{\psi} = 
\frac{1}{2} (I + A) \ket{\psi} \\
& = & \ket{\psi_+}. \nonumber
\end{eqnarray}
If we make the measurement, then apply $M^\dagger$ if the result is $-1$ 
and do nothing if the result is $+1$, then we have applied the nonunitary 
operator $P_+ = \frac{1}{2} (I + A)$.  We can then continue the computation 
with the assurance that the computer is in the state $\ket{\psi_+}$.  In
order to perform this nonunitary operator, we have taken advantage of the
fact that $\ket{\psi}$ is a $+1$-eigenstate of $M$.  This trick cannot
be used if we do not know anything about the state of $\ket{\psi}$.

We know how to measure operators in $\G$ fault-tolerantly.  If we prepare an 
ancilla in a known state and apply a known set of operations in $N(\G)$, 
the resulting state can be at least partially described by a stabilizer $S$.
This stabilizer is not the stabilizer of a quantum error-correcting code,
but simply a way of describing the information we have about the state.  In
many of the applications below, there will be one stabilizer for the 
error-correcting code, and another which describes the restricted state of 
the data due to our preparation of the ancilla in a known state.  We can 
fault-tolerantly measure (fault-tolerant with respect to the error-correcting 
code) an operator $A \in \G$ that anticommutes with some $M \in S$ (the 
stabilizer describing the data) and correct the result as above to 
perform the operation $P_+$.  Any operators in $S$ that commute with $A$ 
will still fix the state of the system after the measurement and correction.  
Hereafter, in the context of performing operations on encoded states, I will 
usually speak of ``measuring'' $A$ when I mean applying $P_+$ for $A$.

If $A \in S$, there is no need to measure $A$ to perform $P_+$, since the 
state is already an eigenstate of $A$ with eigenvalue $+1$.  If $A$ 
commutes with everything in $S$ but is not in $S$ itself, then measuring 
$A$ will give us information about which state we had that was fixed by $S$.  
However, we do not have an $M$ that anticommutes with $A$, so we 
cannot fix $P_-$ to $P_+$.  If $A$ anticommutes with some element of $S$, say 
$M_1$, then we can choose the remaining $n-k-1$ generators of $S$ to 
commute with $A$ (if $M_i$ anticommutes with $A$, $M_1 M_i$ will 
commute with $A$).  The stabilizer $S'$ after applying $P_+$ will then be 
generated by $A$ and $M_2, \ldots, M_{n-k}$.

We can better understand the operator $P_+$ by looking at the 
transformation it induces from $N(S)/S$ to $N(S')/S'$.  Half of the 
representatives of each coset in $N(S)/S$ will commute with $A$ and half 
will anticommute, since of $N$ and $M_1 N$, one will commute and one 
will anticommute.  If $N \in N(S)$ commutes with $A$, its eigenvectors and 
eigenvalues are left unchanged by measuring $A$.  Therefore the coset 
represented by $N$ in $N(S')/S'$ will act on $P_+ \ket{\psi}$ in the same 
way as the coset in $N(S)/S$ acted on $\ket{\psi}$.  Any representative of 
the same coset in $N(S)/S$ will produce the same coset in $N(S')/S'$ as long 
as it commutes with $A$.  We therefore have a map from $N(S)/S \cong 
\G$ to $N(S')/S' \cong \G$, which is an operation in $N(\G)$.  Using selected 
ancilla preparation and existing tranversal operations, we can use this 
process to create new transversal operations.

A nice example of this formalism, which can be applied independently of
quantum error correction, is a description of quantum 
teleportation~\cite{bennett-teleport}.  We start with three qubits, the first 
in an arbitrary state $\ket{\psi}$, the other two in the Bell state $\ket{00} 
+ \ket{11}$.  This state can be described by the stabilizer $S_1$ generated 
by $I \otimes \X \otimes \X$ and $I \otimes \Z \otimes \Z$.  The cosets of 
$N(S_1)/S_1$ can be represented by $\Xbar = \X \otimes I \otimes I$ and 
$\Zbar = \Z \otimes I \otimes I$.  The third qubit is far away, so we cannot 
perform any quantum gates interacting it with the other two qubits.  
However, we can make measurements on the first two qubits and send the 
information to be used to perform conditional quantum gates just on the 
third qubit.

First, we apply a CNOT from the first qubit to the second qubit.  This 
produces stabilizer $S_2$ generated by $I \otimes \X \otimes \X$ and $\Z 
\otimes \Z \otimes \Z$, with $\Xbar = \X \otimes \X \otimes I$ and $\Zbar 
= \Z \otimes I \otimes I$.  Now measure $\X$ for the first qubit.  This 
produces stabilizer $S_3$ generated by $\X \otimes I \otimes I$ and $I 
\otimes \X \otimes \X$.  The coset representative $\X \otimes \X \otimes 
I$ commutes with the measured operator, so it still represents the new 
coset.  Multiplying by the first generator of $S_3$ still gives a coset 
representative of $\Xbar$ in $N(S_3)/S_3$, so $\Xbar = I \otimes \X 
\otimes I$.  $\Z \otimes I \otimes I$ does not commute with the measured 
operator, but $(\Z \otimes \Z \otimes \Z) (\Z \otimes I \otimes I) = I 
\otimes \Z \otimes \Z$ represents the same coset in $N(S_2)/S_2$ and 
does commute with the measured operator, so it represents the $\Zbar$ 
coset in $N(S_3)/S_3$.  The measurement potentially requires an application 
of $\Z \otimes \Z \otimes \Z$ if it is necessary to correct $P_-$.  This 
provides one of the sets of conditional operations used in quantum 
teleportation.

Now we measure $\Z$ for the second qubit.  This produces the stabilizer 
$S_4$ generated by $\X \otimes I \otimes I$ and $I \otimes \Z \otimes I$.  
This time, the representative of $\Zbar$ commutes with the measured 
operator, so $\Zbar$ for $N(S_4)/S_4$ is $I \otimes \Z \otimes \Z \cong I 
\otimes I \otimes \Z$.  $I \otimes \X \otimes I$ does not commute, but $(I 
\otimes \X \otimes \X) (I \otimes \X \otimes I) = I \otimes I \otimes \X$ 
does, so in $N(S_4)/S_4$, $\Xbar = I \otimes I \otimes \X$.  The operation 
to correct $P_-$ this time is $I \otimes \X \otimes \X$.  This provides the 
second set of conditional operations in teleportation.

Note that $S_4$ completely determines the state of the first two qubits and 
does not restrict the state of the third qubit at all.  In fact, the $\Xbar$ 
operator in $N(S_1)/S_1$, which started as $\X$ for the first qubit, has 
been transformed into $\X$ for the third qubit, and $\Zbar$, which began 
as $\Z$ for the first qubit, has become $\Z$ for the third qubit.  This means 
the final state is $(\ket{0} + \ket{1}) \otimes \ket{0} \otimes \ket{\psi}$, 
and we have teleported the state as desired.

After we measure $\X$, $\Y$, or $\Z$ for a qubit, we have completely 
determined the state of that qubit, so its contribution to the stabilizer will 
just be the operator just measured, and it will not contribute to standard 
representatives of the cosets in $N(S')/S'$ at all.  Therefore, when 
describing how to produce new transversal operations, I will drop qubits 
from the notation after they have been measured.

\section{Producing New Operations in $N(\G)$}
\label{sec-4qubit}

The group $N(\G)$ can be generated by just the operations $R$, $P$, and 
CNOT applied to arbitrary qubits and pairs of qubits.  I will now show that, 
by using measurements, we can, in fact, generate $N(\G)$ using just CNOT.  
Then I will demonstrate that for most known codes, we can apply an 
encoded CNOT transversally.

First, note that by preparing an ancilla in an arbitrary state and measuring 
$\X$, $\Y$, or $\Z$, we can always prepare that ancilla qubit in the $+1$ 
eigenstate of any of these three operators.  Also, there are only six 
interesting operators in $N(\G_1)$: $I$, $R$, $P$ (and $P^\dagger$), $Q$ (and 
$Q^\dagger$), $T$, and $T^2$ (and $T^\dagger$ and $(T^2)^\dagger$), where 
$Q = P^\dagger RP$ switches $\Y$ and $\Z$, and $T = RP^\dagger$ is the cyclic 
permutation of $\X$, $\Y$, and $\Z$.  I have only counted this as six 
operators, since the adjoints produce the same permutations, but with 
different signs distributed among $\X$, $\Y$, and $\Z$.  This effect can also 
be produced by applying $\X$, $\Y$ and $\Z$ themselves.  Any two 
non-identity operators in this set, other than $T$ and $T^2$, will suffice to 
generate all of them.

Suppose we have an arbitrary single-qubit state $\ket{\psi}$.  Let us 
prepare an ancilla qubit in the $+1$ eigenstate of $\Z$, then apply a CNOT 
from the data qubit to the ancilla qubit.  This produces the stabilizer $\Z 
\otimes \Z$, with $\Xbar = \X \otimes \X$ and $\Zbar = \Z \otimes I$.  
Now measure $\Y$ for the ancilla qubit and discard the ancilla.  This leaves 
the first qubit with $\Xbar = -\Y$ and $\Zbar = \Z$, which means we have 
applied $P^\dagger$.

Now prepare the ancilla in the $+1$ eigenstate of $\X$ and apply a CNOT 
from the ancilla qubit to the data qubit.  This produces stabilizer $\X 
\otimes \X$, with $\Xbar = \X \otimes I$ and $\Zbar = \Z \otimes \Z$.  
Measure $\Y$ for the ancilla and discard it, leaving $\Xbar = \X$ and 
$\Zbar = -\Y$.  We have applied $Q^\dagger$.  Along with $P$ from above, this 
suffices to generate $N(\G_1)$ and therefore $N(\G_n)$ for any $n$.

We can also produce $T$ directly by preparing the ancilla in the $+1$ 
eigenstate of $\Y$ and applying a CNOT from the ancilla qubit to the data 
qubit.  This produces a stabilizer of $\X \otimes \Y$, with $\Xbar = \X 
\otimes I$ and $\Zbar = \Z \otimes \Z$.  Measure $\Y$ for the {\em data} 
qubit and discard it, leaving $\Xbar = \Y$ and $\Zbar = \X$, both on the 
former ancilla qubit.   The net result is to apply $T$, but to move the 
data from the data qubit to what began as the ancilla qubit.

Now let us turn our attention to transversal operations on quantum
error-correcting stabilizer codes.  Consider the following four-qubit 
transformation:
\begin{eqnarray}
\X \otimes I \otimes I \otimes I & \rightarrow & \X \otimes \X \otimes \X 
\otimes I \nonumber \\
I \otimes \X \otimes I \otimes I & \rightarrow & I \otimes \X \otimes \X 
\otimes \X \nonumber \\
I \otimes I \otimes \X \otimes I & \rightarrow & \X \otimes I \otimes \X 
\otimes \X \nonumber \\
I \otimes I \otimes I \otimes \X & \rightarrow & \X \otimes \X \otimes I 
\otimes \X \label{eq-4qubit} \\
\Z \otimes I \otimes I \otimes I & \rightarrow & \Z \otimes \Z \otimes \Z 
\otimes I \nonumber \\
I \otimes \Z \otimes I \otimes I & \rightarrow & I \otimes \Z \otimes \Z 
\otimes \Z \nonumber \\
I \otimes I \otimes \Z \otimes I & \rightarrow & \Z \otimes I \otimes \Z 
\otimes \Z \nonumber \\
I \otimes I \otimes I \otimes \Z & \rightarrow & \Z \otimes \Z \otimes I 
\otimes \Z. \nonumber
\end{eqnarray}
Given an element $M$ of an arbitrary stabilizer, this operation applied
bitwise maps
\begin{eqnarray}
M \otimes I \otimes I \otimes I & \rightarrow & M \otimes M \otimes M 
\otimes I \nonumber \\
I \otimes M \otimes I \otimes I & \rightarrow & I \otimes M \otimes M 
\otimes M \label{eq-4qubit-effect} \\
I \otimes I \otimes M \otimes I & \rightarrow & M \otimes I \otimes M 
\otimes M \nonumber \\
I \otimes I \otimes I \otimes M & \rightarrow & M \otimes M \otimes I 
\otimes M. \nonumber
\end{eqnarray}
Each of these images is in the group $S \times S \times S \times S$, so this 
is a valid transversal operation for {\em any} stabilizer code.  Because of 
(\ref{eq-4qubit-effect}), this operation just applies itself to the encoded 
qubits.  When the code has multiple qubits per block, (\ref{eq-4qubit}) 
applies itself to all of the corresponding sets of encoded qubits.

This is very useful, since if we have two logical qubits and prepare two more 
ancilla logical qubits each in the $+1$ eigenstate of $\Z$, and then apply 
(\ref{eq-4qubit}) to these four qubits, we get a stabilizer with generators 
$\Z \otimes I \otimes \Z \otimes \Z$ and $\Z \otimes \Z \otimes I \otimes 
\Z$, and
\begin{eqnarray}
\Xbar_1 & = & \X \otimes \X \otimes \X \otimes I \nonumber \\
\Xbar_2 & = & I \otimes \X \otimes \X \otimes \X \\
\Zbar_1 & = & \Z \otimes \Z \otimes \Z \otimes I \nonumber \\
\Zbar_2 & = & I \otimes \Z \otimes \Z \otimes \Z. \nonumber
\end{eqnarray}
Measure $\X$ for both ancilla qubits and discard them.  This leaves us 
with
\begin{eqnarray}
\Xbar_1 & = & \X \otimes \X \nonumber \\
\Xbar_2 & = & I \otimes \X \\
\Zbar_1 & = & \Z \otimes I \nonumber \\
\Zbar_2 & = & \Z \otimes \Z. \nonumber
\end{eqnarray}
This we can recognize as the CNOT from the first data qubit to the second 
data qubit.  As the CNOT suffices to get every operation in $N(\G)$, we can 
therefore perform any such operation transversally for any stabilizer code 
encoding a single qubit.

There are other operations like (\ref{eq-4qubit}) that work for any 
stabilizer code.  The condition they must satisfy~\cite{rains-orthogonal} is 
for $\X$ tensor any number of copies of the identity to map to the tensor 
product of some number of copies of $\X$ and $I$, and $\Z$ in the same 
position must map to the same tensor product of $\Z$ and $I$.  This means 
any such automorphism can be fully described by an $n \times n$ binary 
matrix (for an $n$-qubit operation).  The image of $\Xs{i}$ must commute 
with the image of $\Zs{j}$ for $i \neq j$.  This means that the binary dot 
product of two different rows of the matrix must be $0$.  Also, the image 
of $\Xs{i}$ must anticommute with the image of $\Zs{i}$.  This means that 
the binary dot product of any row with itself must be $1$.  These two 
conditions combine to say that the matrix must be an element of $O(n, 
{\bf Z}_2)$, the orthogonal group over ${\bf Z}_2$.  The smallest $n$ for 
which this group has an element other than a permutation is $n=4$.  If we
were working with $d$-dimensional states instead of qubits, we would instead
need a matrix in $O(n, {\bf Z}_d)$.  Note that the straightforward
generalization of (\ref{eq-4qubit-effect}) is in $O(n, {\bf Z}_d)$ for
$n=d+2$.

Codes which have single-qubit tranversal operations other than the 
identity will in general have a larger available space of multiple-qubit 
operations.  Any $n$-qubit automorphism that maps $\X$ to the tensor 
product of $I$ with $U_i (\X)$ and $\Z$ to the same tensor product of $I$ 
with $U_i (\Z)$ will be an automorphism of $n$ copies of $S$ if $U_i$ is an 
automorphism of $S$ for all $i$.  Note that $U_i$ may be the identity.  It 
may also be possible for $U_i$ to not be an automorphism of $\G_1$ at all, 
although this will depend on the code.  For instance, for a CSS code, we can 
have $U_i (\X) = \X$, $U_i (\Z) = I$ or $U_i (\X) = I$, $U_i (\Z) = \Z$.

\section{Codes With Multiple Encoded Qubits}
\label{sec-multiple}

For codes encoding more than one qubit per block, we have more work to 
do.  We only know how to perform (\ref{eq-4qubit}) between 
corresponding qubits in different blocks, and furthermore, we must 
perform the operation between {\em all} the encoded qubits in both 
blocks.

The solution to the second problem is straightforward.  If we prepare an 
ancilla qubit in the $+1$ eigenstate of $\X$ and apply a CNOT from the 
ancilla to a single data qubit, we get the stabilizer $\X \otimes \X$, with 
$\Xbar = \X \otimes I$ and $\Zbar = \Z \otimes \Z$.  Then if we measure 
$\Z$ for the data qubit, we are left with $\Xbar = \X$ and $\Zbar = \Z$, 
both for the ancilla qubit.  We have transferred the data qubit to the 
ancilla qubit without changing it.  On the other hand, if we had prepared 
the ancilla qubit in the $+1$ eigenstate of $\Z$ and applied the CNOT, 
nothing in the data qubit would have changed.

We can use this fact to switch individual encoded qubits out of a storage 
block into a temporary holding block.  Prepare the holding block with all 
the encoded qubits in the $+1$ eigenstate of $\Z$, except the $j$th encoded 
qubit, which is in the $+1$ eigenstate of $\X$.  Then use (\ref{eq-4qubit}) 
to apply a CNOT from the holding block to the storage block and measure $\Z$ 
for the $j$th encoded qubit in the storage block.  This switches the $j$th 
encoded qubit from the storage block to the holding block while leaving the 
other qubits in the storage block undisturbed.  The $j$th encoded qubit in the 
storage block is left in the state $\ket{0}$, as are all the encoded 
qubits in the holding block but the $j$th one.

To perform operations between just the $j$th encoded qubits in two (or 
more) different blocks while leaving the other qubits in those blocks alone, 
we can switch both $j$th qubits into new, empty blocks, as above.  Then 
we interact them.  If necessary, we again clear all but the $j$th encoded 
qubit in each temporary block by measuring $\Z$.  Then we can switch the 
qubits back into the initial blocks by applying a CNOT from the holding 
block to the appropriate storage block and measuring $\Xbar_j$ for the 
holding block.

This leaves the questions of interacting the $j$th encoded qubit in one block 
with the $i$th encoded qubit in another block, and of interacting two encoded 
qubits in the same block.  We can partially solve either problem by switching 
the two qubits to be interacted into separate holding blocks.  If we know how 
to swap the $j$th encoded qubit with the first encoded qubit, we can then 
swap both qubits into the first position, interact them as desired, then 
swap them back to their initial positions and switch them back to their 
storage block or blocks.

One way to swap qubits within a block is to perform some nontrivial action 
on a single block.  For a code with trivial automorphism group, this will not 
exist.  However, almost any automorphism will suffice to swap encoded 
qubits as desired.  This is because there are so few two-qubit operations in 
$N(\G)$.  Any automorphism of the code will produce some element of 
$N(\G_k)$ on the $k$ encoded qubits, typically (although certainly not 
always) interacting all of them.  If we perform some measurement on all of 
the encoded qubits in the block except the first and the $j$th, we are left 
with a two-qubit operation between those two encoded qubits.

We can always perform single-qubit operations on any encoded qubit in a 
block by switching the qubit into a fresh block, applying the operation to 
every encoded qubit in the new block, clearing unneccesary qubits and 
switching the qubit back to the first block.  Using this freedom, any 
operation in $N(\G_2)$ can be transformed to map $\X \otimes I$ to one of 
$\X \otimes I$, $\X \otimes \X$, and $I \otimes \X$.  There is still a 
remaining freedom to switch $\Y$ and $\Z$ on either qubit, and we may 
also switch either with $\X$ for any qubit where the image of $\X \otimes 
I$ acts as the identity.  We treat the three possibilities as separate cases:
\begin{itemize}
\item $\X \otimes I \rightarrow \X \otimes I$

The operation preserves the group structure of $\G_2$, so the image of $I 
\otimes \X$ must commute with $\X \otimes I$.  Up to single-qubit 
operations, the possibilities are

\begin{enumerate}
\item $I \otimes \X$: The image of $\Z \otimes I$ must be either $\Z 
\otimes I$ or $\Z \otimes \X$.  In the first case, the image of $I \otimes 
\Z$ is $I \otimes \Z$ and the operation is the identity.  In the second case, 
the image of $I \otimes \Z$ must be $\X \otimes \Z$.  If we apply $R$ to 
the first qubit before the operation and again after it, this produces a CNOT 
from the first qubit to the second qubit.

\item $\X \otimes \X$: The image of $\Z \otimes I$ must be $\Z \otimes 
\Z$ and the image of $I \otimes \Z$ may be either $I \otimes \Z$ or $\X 
\otimes \Y$.  If it is $I \otimes \Z$, the operation is exactly CNOT from the 
second qubit to the first.  If it is $\X \otimes \Y$, we can again get CNOT 
from the second qubit to the first by simply applying $Q$ to the second 
qubit, followed by the operation.
\end{enumerate}

\item $\X \otimes I \rightarrow I \otimes \X$

This case is related to the first one by simply swapping the two qubits.  
Therefore, the possibilities can be reduced to a simple swap, and a CNOT 
either way followed by a swap.

\item $\X \otimes I \rightarrow \X \otimes \X$

Now there are three possibilities for the image of $I \otimes \X$: $I 
\otimes \X$ again, $\X \otimes I$, or $\Z \otimes \Z$.

\begin{enumerate}
\item $I \otimes \X$: The image of $I \otimes \Z$ must be $\Z \otimes \Z$.  
The image of $\Z \otimes I$ may be either $\Z \otimes I$ or $\Y \otimes 
\X$.  As with case two above, if it is $\Z \otimes I$, this is a CNOT from the 
first qubit to the second; if it is $\Y \otimes \X$, we can apply $Q$ to the 
first qubit and then this operation to get a CNOT from the first qubit to the 
second.

\item $\X \otimes I$: This case can be produced from the previous one by 
swapping the two qubits.  Thus, the operation can be converted into a 
CNOT from the first qubit to the second followed by a swap.

\item $\Z \otimes \Z$: In this case, the image of $\Z \otimes I$ can be $\Z 
\otimes I$, $I \otimes \Z$, $\Y \otimes \X$, or $\X \otimes \Y$.  If the 
image of $\Z \otimes I$ is $\Z \otimes I$, the image of $I \otimes \Z$ must 
be $I \otimes \X$ or $\Z \otimes \Y$.  If it is $I \otimes \X$ and we apply 
$R$ to the second qubit and then this operation, it performs a CNOT from the 
first qubit to the second.  If it is $\Z \otimes \Y$, we can apply $T \Z$
to the second qubit, followed by the operation in order to get a CNOT
from the first qubit to the second.  If the image of $\Z \otimes I$ is $I 
\otimes \Z$, we can get it from last case by swapping the qubits, so it can be 
reduced to a CNOT from the first qubit to the second followed by a swap.

If the image of $\Z \otimes I$ is $\Y \otimes \X$, then the image of $I 
\otimes \Z$ may again be either $I \otimes \X$ or $\Z \otimes \Y$.  If it is $I 
\otimes \X$, we can perform $Q$ on the first qubit and $R$ on the second 
qubit, followed by the two-qubit operation.  This produces a CNOT from the 
first qubit to the second one.  If it is $\Z \otimes \Y$, we can perform $Q$ 
on the first qubit and $T \Z$ on the second qubit, followed by the 
two-qubit operation.  This again produces a CNOT from the first qubit to 
the second qubit.

Finally, if the image of $\Z \otimes I$ is $\X \otimes \Y$, we can produce 
the previous case by applying a swap, so the two-qubit operation can be 
converted to a CNOT from the first qubit to the second qubit followed by a 
swap.
\end{enumerate}
\end{itemize}

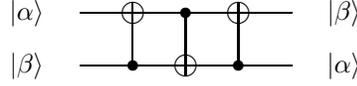
\begin{figure}
\centering
\begin{picture}(160,60)

\put(0,14){\makebox(20,12){$\ket{\beta}$}}
\put(0,34){\makebox(20,12){$\ket{\alpha}$}}
\put(120,14){\makebox(20,12){$\ket{\alpha}$}}
\put(120,34){\makebox(20,12){$\ket{\beta}$}}

\put(30,20){\line(1,0){80}}
\put(30,40){\line(1,0){80}}

\put(50,20){\circle*{4}}
\put(50,20){\line(0,1){24}}
\put(50,40){\circle{8}}

\put(70,40){\circle*{4}}
\put(70,40){\line(0,-1){24}}
\put(70,20){\circle{8}}

\put(90,20){\circle*{4}}
\put(90,20){\line(0,1){24}}
\put(90,40){\circle{8}}

\end{picture}
\caption{Network to swap two qubits using CNOT.}
\label{fig-CNOTtoswap}
\end{figure}

Also, note that $R$ applied to both qubits, followed by a CNOT in one 
direction, followed by $R$ on both qubits, produces a CNOT in the other 
direction.  Therefore, up to application of single-qubit operations, the only 
possible two-qubit operations in $N(\G)$ are the identity, a CNOT, a swap, 
or a CNOT followed by a swap.  We can make a swap out of three CNOTs 
using the simple network from figure~\ref{fig-CNOTtoswap}.

We cannot make a general swap out of CNOT followed by swap.  However, 
if the control qubit of the CNOT begins in the state $\ket{0}$, the operation 
does swap the two qubits.  This is all that is necessary to get all of $N(\G)$, 
since we only need to move a single data qubit around within an otherwise 
empty block.

Even if we have no automorphism to switch the $j$th qubit and the first 
qubit, we can still do it using quantum teleportation~\cite{bennett-teleport}.  
To do this, we will need an EPR pair entangled between the first and $j$th 
encoded qubits.  We can make an unencoded EPR pair and then encode it 
normally.  However, a single error during the encoding can destroy the 
pair.  Therefore, we will need to make a number of EPR pairs and purify 
good ones using an entanglement purification protocol (EPP)
\cite{bennett-tome,bennett-EPP}.  We can interact corresponding qubits in 
the EPR pair using operations in $N(\G)$, which is all that is necessary.  For 
instance, we could make five EPR pairs and use the one-way EPP derived 
from the five-qubit code to purify a single good EPR pair.  It would take 
two independent errors to get an error in this pair.  An easier way to make
the EPR pair is to start with the $+1$ eigenstate of both $\Zbar_1$ and
$\Zbar_j$, then to measure $\Xbar_1 \Xbar_j$, which is an operator in $N(S)$
just like any other.  This leaves the ancilla block in the $+1$ eigenstate
of $\Zbar_1 \Zbar_j$ and $\Xbar_1 \Xbar_j$, which is just an EPR pair.

Once we have a reliable EPR pair, the teleportation process requires only 
operations in $N(\G)$ between corresponding encoded qubits.  This allows us to 
move the $j$th encoded qubit in one otherwise empty block to the first encoded 
qubit in the block that previously held the EPR pair.  This allows us to do 
{\em any} operation in $N(\G)$ for {\em any} stabilizer code.  Essentially
the same procedures will work when the basic unit is the qudit instead of
the qubit \cite{gottesman-qudit}.

\section{The Toffoli Gate}
\label{sec-toffoli}

The group $N(\G)$ is insufficient to allow universal quantum computation.  
In fact, Knill~\cite{knill-normalizer} has shown that a quantum computer 
using only elements from $N(\G)$ and measurements can be simulated 
efficiently on a classical computer.  The argument follows easily from the 
results of the preceding sections.  If we begin with a state initialized to 
$\ket{0\cdots0}$, the stabilizer is $\Zs{1}, \Zs{2}, \ldots$.  Each operation 
in $N(\G)$ produces a well-defined transformation of the stabilizer, which 
can be classically tracked efficiently.  Any measurement will also 
transform the stabilizer in a well-defined way, which is again easy to keep 
track of on a classical computer.  Therefore, we can store and evolve 
complete information on the state of the quantum computer with only 
polynomial classical overhead.

In order to perform truly universal quantum computation, even a single 
gate outside of $N(\G)$ can be sufficient.  For instance, the Toffoli gate (a 
three-qubit gate which flips the third qubit iff both of the first two qubits 
are $\ket{1}$) along with $N(\G)$ suffices for universal computation.  Shor 
gave an implementation of the Toffoli gate~\cite{shor-fault-tol} which can 
be easily adapted to any code allowing $N(\G)$.  Since this is any stabilizer 
code, we can do universal computation for any stabilizer code.  Note that
there are a number of other gates outside $N(\G)$ that we could add to get
a universal set of gates (such as the single-qubit $\pi/8$ rotation), and
for some codes, it may be easier to perform these gates than the Toffoli
gate \cite{knill-concatenate2}.  However, I will just discuss the
implementation of the Toffoli gate.

The Toffoli gate can be expanded using $\G$ as a basis as follows:
\begin{equation}
\frac{1}{4} \left( 3 I + \Zs{1} + \Zs{2} - \Zs{1} \Zs{2} + (I - \Zs{1}) (I - 
\Zs{2}) \Xs{3} 
\right).
\end{equation}
Applying the Toffoli gate to a state therefore produces the following 
transformation on the elements of $\G_3$:
\begin{eqnarray}
\Xs{1} & \rightarrow & \frac{1}{16} \left( 3 I + \Zs{1} + \Zs{2} - \Zs{1} \Zs{2}
+ (I - \Zs{1}) (I - \Zs{2}) \Xs{3} \right) \nonumber \\
& & \mbox{} \times \left( 3 I - \Zs{1} + \Zs{2} + \Zs{1}
\Zs{2} + (I + \Zs{1}) (I - \Zs{2}) \Xs{3} \right) \Xs{1} \nonumber \\
& = & \frac{1}{2} \left(I + \Zs{2} + (I - \Zs{2}) \Xs{3} \right) \Xs{1}
\nonumber \\
\Xs{2} & \rightarrow & \frac{1}{2} \left(I + \Zs{1} + (I - \Zs{1}) \Xs{3}
\right) \Xs{2} \nonumber \\
\Xs{3} & \rightarrow & \Xs{3} \label{eq-toffoli} \\
\Zs{1} & \rightarrow & \Zs{1} \nonumber \\
\Zs{2} & \rightarrow & \Zs{2} \nonumber \\
\Zs{3} & \rightarrow & \frac{1}{16} \left( 3 I + \Zs{1} + \Zs{2} - \Zs{1} \Zs{2}
+ (I - \Zs{1}) (I - \Zs{2}) \Xs{3} \right) \nonumber \\
& & \mbox{} \times \left( 3 I + \Zs{1} + \Zs{2} - \Zs{1} 
\Zs{2} - (I - \Zs{1}) (I - \Zs{2}) \Xs{3} \right) \Zs{3} \nonumber \\
& = & \frac{1}{2} \left(I + \Zs{1} + (I - \Zs{1}) \Zs{2} \right) \Zs{3}.
\nonumber
\end{eqnarray}
This means $\Zs{1}$, $\Zs{2}$, and $\Xs{3}$ stay the same, $\Xs{1}$ becomes 
$\Xs{1}$ tensor a CNOT from qubit two to qubit three, $\Xs{2}$ becomes $\Xs{2}$ 
tensor a CNOT from qubit one to qubit three, and $\Zs{3}$ becomes $\Zs{3}$ 
tensor a conditional sign for qubits one and two.

Suppose we can make the ancilla
\begin{equation}
\ket{A} = \frac{1}{2} (\ket{000} + \ket{010} + \ket{100} + \ket{111}).
\end{equation}
This state is fixed by the three operators
\begin{eqnarray}
M_1 & = & \frac{1}{2} \left(I + \Zs{2} + (I - \Zs{2}) \Xs{3} \right) \Xs{1}
\nonumber \\
M_2 & = & \frac{1}{2} \left(I + \Zs{1} + (I - \Zs{1}) \Xs{3} \right) \Xs{2}
\label{eq-Toffoli-ev} \\
M_3 & = & \frac{1}{2} \left(I + \Zs{1} + (I - \Zs{1}) \Zs{2} \right) \Zs{3}. 
\nonumber
\end{eqnarray}
Now suppose we have three data qubits (numbers four, five, and six) that we 
wish to perform a Toffoli gate on.  We simply apply CNOTs from qubit one to 
qubit four, qubit two to qubit five, and from qubit six to qubit three.  This 
produces the following ``stabilizer'':
\begin{eqnarray}
M_1' & = & \frac{1}{2} \left(I + \Zs{2} + (I - \Zs{2}) \Xs{3} \right) \Xs{1} 
\Xs{4} \nonumber \\
M_2' & = & \frac{1}{2} \left(I + \Zs{1} + (I - \Zs{1}) \Xs{3} \right) \Xs{2} 
\Xs{5} \\
M_3' & = & \frac{1}{2} \left(I + \Zs{1} + (I - \Zs{1}) \Zs{2} \right) \Zs{3}
\Zs{6}. \nonumber
\end{eqnarray}
Then measure $\Zs{4}$, $\Zs{5}$, and $\Xs{6}$ and discard qubits 4--6.  As we 
can see, this produces the transformation (\ref{eq-toffoli}) on the three 
data qubits while moving them to what were formerly the ancilla qubits.  
Note that correcting for measured eigenvalues of $-1$ will require 
applying $M_1$, $M_2$, or $M_3$, which are not elements of $\G$.  They 
are, however, elements of $N(\G)$.

Therefore, in order to perform the Toffoli gate on encoded states, we must 
produce an encoded version of the ancilla $\ket{A}$.  Then we need only 
perform measurements and encoded operations in $N(\G)$ to produce the 
effect of a Toffoli gate.  Below, I will assume $\G$ only encoded one qubit 
per block.  If it encodes more, we can still do the same thing by moving the 
qubits to be interacted into the first encoded qubit in otherwise empty 
blocks.  The $\Xbar$ and $\Zbar$ operators used to create the ancilla are 
just $\Xbar_1$ and $\Zbar_1$.

To produce the encoded ancilla $\ket{A}$, we start with the encoded 
version of the state $\ket{A} + \ket{B}$, where
\begin{equation}
\ket{B} = \frac{1}{2} (\ket{001} + \ket{011} + \ket{101} + \ket{110}).
\end{equation}
Note that $\ket{B}$ is related to $\ket{A}$ by applying $\X$ to the third 
qubit.  Since 
\begin{equation}
\ket{A} + \ket{B} = \Sum_{a=000}^{111} \ket{a} = (\ket{0} + \ket{1})^3,
\end{equation}
we can easily prepare it by measuring $\Xbar$ for each block.  
Henceforth, $\ket{A}$ and $\ket{B}$ will denote the encoded versions of 
themselves.  Now we take an ancilla in a ``cat'' state $\ket{0 \ldots 0} + 
\ket{1 \ldots 1}$, where the number of qubits in the cat state is equal to 
the number of qubits in a single block of the code.  Then we will perform 
an operation that takes
\begin{eqnarray}
\ket{0 \ldots 0} \ket{A} & \rightarrow & \ket{0 \ldots 0} \ket{A} 
\nonumber \\
\ket{1 \ldots 1} \ket{A} & \rightarrow & \ket{1 \ldots 1} \ket{A} 
\label{eq-Toffoli-anc} \\
\ket{0 \ldots 0} \ket{B} & \rightarrow & \ket{0 \ldots 0} \ket{B} 
\nonumber \\
\ket{1 \ldots 1} \ket{B} & \rightarrow & \!\!\! - \ket{1 \ldots 1} \ket{B}. 
\nonumber
\end{eqnarray}
Then under (\ref{eq-Toffoli-anc}),
\begin{equation}
(\ket{0 \ldots 0} + \ket{1 \ldots 1}) (\ket{A} + \ket{B}) \rightarrow (\ket{0 
\ldots 0} + \ket{1 \ldots 1}) \ket{A} + (\ket{0 \ldots 0} - \ket{1 \ldots 1}) 
\ket{B}.
\end{equation}
If we measure $\X \otimes \cdots \otimes \X$ for the cat state, if we get 
$+1$, the rest of the ancilla is in the state $\ket{A}$.  If we get $-1$, the 
rest of the ancilla is in the state $\ket{B}$.  One complication is that a 
single qubit error in the cat state can cause this measurement result to be 
wrong.  Luckily,
\begin{eqnarray}
(\ket{0 \ldots 0} + \ket{1 \ldots 1}) \ket{A} & \rightarrow & (\ket{0 \ldots 
0} + \ket{1 \ldots 1}) \ket{A} \\
(\ket{0 \ldots 0} + \ket{1 \ldots 1}) \ket{B} & \rightarrow & (\ket{0 \ldots 
0} - \ket{1 \ldots 1}) \ket{B}.
\end{eqnarray}
Therefore, if we prepare another cat state and apply (\ref{eq-Toffoli-anc}) 
again, we should again get $+1$ if the ancilla was actually in the state 
$\ket{A}$ after the first measurement and $-1$ if it was actually in the 
state $\ket{B}$.  We can therefore get any desired level of reliability for 
the ancilla state by repeating (\ref{eq-Toffoli-anc}) a number of times.  
Finally, once we are confident we have either $\ket{A}$ or $\ket{B}$, we 
apply $\Xbar$ to the third ancilla qubit if it is $\ket{B}$.  This 
means we will always have prepared the state $\ket{A}$.

To perform (\ref{eq-Toffoli-anc}), we will have to perform the operation 
$\ket{A} \rightarrow \ket{A}$ and $\ket{B} \rightarrow - \ket{B}$ if and
only if the
qubits of the cat state are $\ket{1 \ldots 1}$.  If the qubits of the cat state 
are $\ket{0 \ldots 0}$, then we do nothing to the rest of the ancilla.  I will 
show that we can apply $\ket{A} \rightarrow \ket{A}$ and $\ket{B} 
\rightarrow - \ket{B}$ using a series of transversal operations and 
measurements.  If we apply these operations and measurements 
conditioned on the corresponding qubit from the cat state being $\ket{1}$, 
then we have actually performed (\ref{eq-Toffoli-anc}).  Conditioning the 
operations on the cat state bit will generally involve using Toffoli gates and 
possibly other gates outside $N(\G)$, but they are all gates on {\em single} 
qubits rather than blocks.  We assume we know how to perform universal 
computation on individual qubits, so these gates are available to us.

The state $\ket{A}$ is a $+1$-eigenvector of $M_3$, from equation 
(\ref{eq-Toffoli-ev}).  $\ket{B}$ is a $-1$-eigenvector of the same $M_3$, 
so applying $M_3$ does, in fact, transform $\ket{A} \rightarrow \ket{A}$ 
and $\ket{B} \rightarrow - \ket{B}$.  $M_3$ is just a conditional sign on 
the first two qubits (i.e.\ an overall sign of $-1$ iff both qubits are 
$\ket{1}$) times $\Z$ on the third qubit.  Therefore it is in $N(\G)$ and 
can be performed transversally for any stabilizer code.  Therefore, we can 
perform universal computation using any stabilizer code.

\section{Construction of Gates in $N(\G)$}

In order to use the general fault-tolerant protocols, we need to apply 
three- or four-qubit gates.  Suppose our basic gates are limited to one- and 
two-qubit gates.  These gates are sufficient to give us any gates in $N(\G)$.  
I will now give a construction for any gate in $N(\G)$ using one- and 
two-qubit gates.

The construction will be inductive.  In section~\ref{sec-multiple}, I showed 
that any one- or two-qubit gate could be made using $R$, $P$, and CNOT.  
Suppose we can construct any $n$-qubit gate using one- and two-qubit 
gates, and let $U$ be an $(n+1)$-qubit gate.  Using swaps and one-qubit 
gates, we can guarantee that
\begin{equation}
M = U \Zs{1} U^\dagger = \Xs{1} \otimes M'
\end{equation}
and
\begin{equation}
N = U \Xs{1} U^\dagger = I \otimes N'\ {\rm or}\ \Zs{1} \otimes N'.
\end{equation}
Note that $\{M, N\} = 0$.  Suppose
\begin{equation}
U (\ket{0} \otimes \ket{\psi}) = \ket{0} \otimes \ket{\psi_1} + \ket{1} 
\otimes \ket{\psi_2},
\end{equation}
where $\ket{\psi}$, $\ket{\psi_1}$, and $\ket{\psi_2}$ are states of the 
last $n$ qubits.  The results of section~\ref{sec-measurements} tell us that 
if we measure $\Z$ for the first qubit after applying $U$ and apply 
$M^\dagger$ (which anticommutes with $\Zs{1}$) if the result is $-1$, we 
will get $\ket{0} \otimes \ket{\psi_1}$.  This means that $\ket{\psi_2} = 
M' \ket{\psi_1}$.  Define $U'$ by $U' \ket{\psi} = \ket{\psi_1}$.  Then
\begin{equation}
U \ket{0} \otimes \ket{\psi} = (I + M) (\ket{0} \otimes U' \ket{\psi}).
\end{equation}

Now,
\begin{eqnarray}
U (\ket{1} \otimes \ket{\psi}) & = & U \left[ (\X \ket{0}) \otimes 
\ket{\psi} \right] \\
& = & N U (\ket{0} \otimes \ket{\psi}) \\
& = & N (I + M) (\ket{0} \otimes U' \ket{\psi}) \\
& = & (I - M) N (\ket{0} \otimes U' \ket{\psi}) \\
& = & (I- M) (\ket{0} \otimes N' U' \ket{\psi}).
\end{eqnarray}

\begin{figure}
\centering
\begin{picture}(160,100)

\put(0,20){\line(1,0){10}}
\put(0,30){\makebox(10,20){$\vdots$}}
\put(0,50){\line(1,0){10}}
\put(0,60){\line(1,0){10}}

\put(10,10){\framebox(40,60){{\LARGE $U'$}}}

\put(50,20){\line(1,0){10}}
\put(50,30){\makebox(10,20){$\vdots$}}
\put(50,50){\line(1,0){10}}
\put(50,60){\line(1,0){10}}

\put(0,80){\line(1,0){99}}
\put(99,74){\framebox(12,12){$R$}}
\put(111,80){\line(1,0){49}}

\put(75,80){\circle*{4}}
\put(75,80){\line(0,-1){10}}
\put(60,10){\framebox(30,60){{\LARGE $N'$}}}

\put(90,20){\line(1,0){30}}
\put(90,30){\makebox(30,20){$\vdots$}}
\put(90,50){\line(1,0){30}}
\put(90,60){\line(1,0){30}}

\put(135,80){\circle*{4}}
\put(135,80){\line(0,-1){10}}
\put(120,10){\framebox(30,60){{\LARGE $M'$}}}

\put(150,20){\line(1,0){10}}
\put(150,30){\makebox(10,20){$\vdots$}}
\put(150,50){\line(1,0){10}}
\put(150,60){\line(1,0){10}}

\end{picture}
\caption{Recursive construction of gates in $N(\G)$.}
\label{fig-normalizer}
\end{figure}
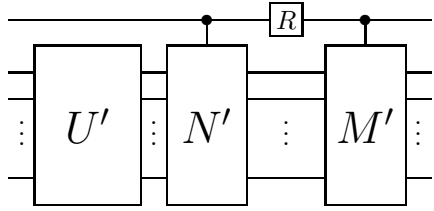

Therefore, if we first apply $U'$ to the last $n$ qubits, followed by applying
$N'$ to the last $n$ qubits conditioned on the first qubit, followed by a 
Hadamard transform $R$ on the first qubit, followed by $M'$ on the last $n$ 
qubits conditioned on the first qubit, we have applied $U$:
\begin{eqnarray}
\ket{0} \otimes \ket{\psi} + \ket{1} \otimes \ket{\phi}
& \rightarrow & \ket{0} \otimes U' \ket{\psi} + \ket{1} \otimes U' 
\ket{\phi} \\
& \rightarrow & \ket{0} \otimes U' \ket{\psi} + \ket{1} \otimes N' U' 
\ket{\phi} \\
& \rightarrow & (\ket{0} + \ket{1}) \otimes U' \ket{\psi} + (\ket{0} - 
\ket{1}) \otimes N' U' \ket{\phi} \nonumber \\ \\
& \rightarrow & (\ket{0} \otimes U' \ket{\psi} + \ket{1} \otimes M' U' 
\ket{\psi}) \nonumber \\
& & \mbox{} + (\ket{0} \otimes N' U' \ket{\phi} - \ket{1} \otimes M' N' U' 
\ket{\phi}) \\
& = & \left[ \ket{0} \otimes U' \ket{\psi} + M (\ket{0} \otimes U' 
\ket{\psi}) \right] \nonumber \\
& & \mbox{} + \left[ \ket{0} \otimes N' U' \ket{\phi} - M (\ket{0} 
\otimes N' U' \ket{\phi}) \right] \\
& = & (I + M) (\ket{0} \otimes U' \ket{\psi}) \nonumber \\
& & \mbox{} + (I - M) (\ket{0} \otimes N' U' \ket{\phi}) \\
& = & U (\ket{0} \otimes \ket{\psi}) + U (\ket{1} \otimes \ket{\phi}) \\
& = & U (\ket{0} \otimes \ket{\psi} + \ket{1} \otimes \ket{\phi}).
\end{eqnarray}
$U'$ is an $n$-qubit gate in $N(\G)$, which, by the inductive hypothesis, we 
can perform using one- and two-qubit gates.  Both $M'$ and $N'$ are in $\G$, so 
applying them conditioned on the first qubit requires only two-qubit gates 
in $N(\G)$.  Therefore, this construction allows us to perform any $U$ in 
$N(\G)$ using only one- and two-qubit gates.  The construction is summarized
in figure~\ref{fig-normalizer}.

To get $M$ and $N$ in the correct form requires only identifying a single 
qubit on which $M$ does not act as the identity and $N$ acts differently 
from $M$.  From there, a single one-qubit gate and a swap between that 
qubit and the first puts $M$ and $N$ in the desired form.  It is not really 
necessary for the construction that the selected qubit be in the first 
position, so we can actually put $M$ and $N$ in the right form using just 
one one-qubit gate.  We also need to perform $R$ on that qubit in the 
middle of the operation.  Applying $M'$ and $N'$ conditioned on the 
selected qubit uses up to $2n$ two-qubit gates.  Therefore, this 
construction of $U$ uses the gates in $U'$ plus up to two one-qubit gates 
and $2n$ two-qubit gates.  Thus, by induction, an $(n+1)$-qubit gate ($n 
\geq 2$) can use up to $2(n-2)$ one-qubit gates and
\begin{equation}
1 + \Sum_{j=3}^{n+1} 2(j-1) = 1 + (n+2) (n-1) = n^2 + n - 1
\end{equation}
two-qubit gates.

Note that this construction can also be used for encoding data into a 
stabilizer code.  The map $U$ will map $\Xs{i} \rightarrow \Xbar_i$ and 
$\Zs{i} \rightarrow \Zbar_i$ ($i = 1, \ldots, k$) for the $k$ data qubits.  The 
remaining $n-k$ qubits start out as $\ket{0}$, so for $i=k+1, \ldots, n$, we 
map $\Zs{i} \rightarrow M_{i-k}$, where $M_j$ ($j = 1, \ldots, n-k$) are 
generators of $S$.  Any remaining freedom for the choice of the image of 
$\Xs{i}$ for $i = k+1, \ldots, n$ is unimportant.  This produces an encoding 
for any stabilizer code using any $\Xbar$ and $\Zbar$ operators in 
$N(\G)$.  In some cases, it may be more efficient than the construction 
given in chapter \ref{chap-encoding}, but the upper bound for efficiency is 
higher.

\section{Refining the Error Correction Algorithm}

Since errors occur while we are measuring the error syndrome, we are 
inevitably led to a race between the errors that are constantly occuring 
and our ability to correct them.  Therefore it is desireable to be able to 
perform error correction as efficiently as possible.  In this section, I will 
discuss a few ways of speeding up error correction.

One significant improvement is to do classical error correction on the 
syndrome bits~\cite{evslin}.  The most basic form of error correction 
described in section~\ref{sec-error-cor} measures the eigenvalues of the 
$n-k$ generators of $S$.  If we treat these as classical bits, we can encode 
them using a classical $[m, n-k, d']$ linear code.  The bits of the classical 
codeword will be linear combinations of the original syndrome bits, which 
means they will correspond to eigenvalues of products of the generators of 
the stabilizer.  This means we need only measure these $m$ new elements 
of the stabilizer.  Then we can do classical error correction on the result to 
extract the actual $(n-k)$-bit syndrome.  If there were less than $d'$ 
errors on the measured syndrome bits, we can still determine the real 
syndrome.  This protects very well against ancilla errors that produce the 
wrong measurement result for a single syndrome bit.  It protects less well 
against data errors that cause the syndrome to change in the middle of 
measurement, but there is a good chance it will warn us when such an 
error has occurred.  If no errors are detected using the classical code, it is 
quite likely we have measured the correct syndrome.  There is still a 
chance that we have not, so we may want to repeat the measurement, but 
we will not have to do it as many times to produce the same level of 
confidence in the result.

Another possible improvement is to reduce the number of qubits needed 
to perform error correction.  Below, I present a method due to 
Steane~\cite{steane-correction}.  This method puts more effort into 
preparing the ancilla, allowing a reduction in the number of operations 
performed on the data.  In some situations, this results in an improvement 
in error tolerance; in other situations, the effort spent in preparing the 
ancilla is too large, and this results in worse tolerance for errors.

Steane's ancilla state uses $2n$ qubits, which are prepared in the sum of 
the states of a classical code.  The specific classical code is formed by 
taking the two matrices in the binary vector space representation of $S$ 
(section~\ref{sec-alternate}) and tacking them together into a single $(n-k) 
\times 2n$ matrix.  The matrix for the $\Z$'s is first.  This is the parity 
check matrix of the classical code.  The ancilla state can be described by a 
stabilizer $S_A$ on $2n$ qubits.  The first $n-k$ generators of the stabilizer 
are the rows of the parity check matrix with $\Z$'s for the $1$s.  The remaining
$n+k$ generators of the stabilizer are the $n+k$ independent tensor 
products of $\X$'s that commute with the first $n-k$ generators.  Note that 
the fact that $S$ is Abelian means that $n-k$ of the new generators will 
also be formed directly from the generators of the stabilizer, this time by 
combining the $\X$ and $\Z$ matrices with the $\X$ one first and 
replacing $1$s with $\X$'s.  There is only a single state in the Hilbert space 
fixed by all $2n$ of these generators, and that is the desired ancilla
state.

For instance, if the original code is a CSS code such as the seven-qubit code, 
the resulting ancilla state is the tensor product of two ancilla states, each 
in the superposition of all the states in one of the two classical codes that 
make up the CSS code.  For the seven-qubit code, that means two copies of 
$\ket{\overline{0}} + \ket{\overline{1}}$, where $\ket{\overline{0}}$ and 
$\ket{\overline{1}}$ are the encoded $0$ and $1$ states for the seven-qubit 
code.  In general, the classical code will be able to identify as many errors 
as the quantum code can, counting errors in both bits $j$ and $j+n$ (for $j 
\leq n$) as a single error.

Once we have this ancilla, we should again verify it, as we did for the ``cat'' 
states in sections~\ref{sec-error-cor} and \ref{sec-toffoli}.  Then we apply 
a CNOT from data qubit~$i$ to ancilla qubit~$i$, followed by a Hadamard 
transform $R$ on the data qubit and a CNOT from the $i$th data qubit to 
the $(n+i)$th ancilla qubit, followed by a final Hadamard transform on the 
data qubit.  Assuming no phase errors in the ancilla, the data qubit ends 
up in its original state.  We can see this by looking at the stabilizer of the 
ancilla.  The last $n+k$ generators $M$ of $S_A$ are all tensor products of 
$\X$'s, so the CNOTs simply map $I \otimes M \rightarrow I \otimes M$, 
which is obviously still in $S \times S_A$.  The first $n-k$ generators are 
tensor products of $\Z$'s, say $M_1 \otimes M_2$ (with $M_1$ and $M_2$ 
$n$-qubit operators).  The CNOTs then map
\begin{equation}
I \otimes (M_1 \otimes M_2) \rightarrow M_1 (R M_2 R^\dagger) \otimes 
(M_1 \otimes M_2).
\end{equation}
But $M_1$ has a $\Z$ anywhere some element $M \in S$ does and $R M_2 
R^\dagger$ has a $\X$ anywhere the same $M$ does, so $M_1 (R M_2 
R^\dagger) = M$, and $M_1 (R M_2 R^\dagger) \otimes (M_1 \otimes 
M_2)$ is in $S \times S_A$.

The effect of the CNOTs on the generators $M$ of $S$ is to copy the $\X$'s 
forward into the first $n$ qubits of the ancilla and the $\Z$'s forward into 
$\X$'s in the last $n$ qubits of the ancilla.  That is, $M \otimes I 
\rightarrow M \otimes (M_1 \otimes M_2)$, where $M_1$ and $M_2$ are 
the product of $\X$'s, and $M_1 \otimes M_2$ is one of the second set of 
$n-k$ generators of $S_A$.  Therefore a correct codeword will have no 
effect on the ancilla.

Measuring $\Z$ on each of the $2n$ ancilla qubits will therefore give us a 
random codeword from the classical code without disturbing the data or 
the quantum code.  A bit flip error in the $j$th qubit of the quantum code 
will carry forward to a bit flip error in the $j$th qubit of the ancilla, and a 
phase error in the $j$th qubit of the quantum code will produce a bit flip 
error in the $(n+j)$th qubit of the ancilla.  Therefore, errors in the quantum 
code will produce bit flip errors in the measured classical codeword.  The 
actual codeword tells us nothing, but the error syndrome will identify the 
error in the quantum code.  As with the cat state method, an incorrect 
ancilla qubit can result in the wrong error syndrome, but repeating the 
error syndrome measurement can give an arbitrarily high confidence level 
to the result.  Single-qubit phase errors in the ancilla will just feed back to 
single-qubit phase or bit flip errors in the data.

\chapter{Concatenated Coding}
\label{chap-concatenation}

\section{The Structure of Concatenated Codes}

Encoding data using a quantum error-correcting code and applying 
fault-tol\-er\-ant operations to it may or may not actually improve the basic 
error rate for the computation.  Since the gates involved in error correction 
are themselves noisy, the process of error correction introduces errors at 
the same time it is fixing them.  If the basic gate error rate is low enough, 
the error correction will fix more errors than it introduces on the average, 
and making a fault-tolerant computation will help rather than harm.  If 
the error rate is too high, attempting to correct errors will introduce more 
errors than are fixed, and error correction is actively doing harm.  Even if 
error correction helps rather than harms, statistical fluctuations will 
eventually produce more errors than the code can correct, resulting in a 
real error in the data.  Furthermore, the extra computational overhead 
required to do fault-tolerant operations may counteract the additional 
resistance to errors provided by the code, so the encoded computer may 
not be able to do longer computations than the original computer.

Nevertheless, if the basic error rate in the quantum computer is low 
enough, we {\em will} be able to do longer computations using quantum 
codes and fault-tolerance than we could without them.  Suppose we can get 
a certain amount of improvement by using a specific code, say the seven-qubit 
code.  We might imagine that by using a code that corrects more errors, we 
could do a longer computation yet, and by increasing the number of errors 
the code corrects indefinitely, we could do arbitrarily long computation.  
However, for arbitrary families of codes, the number of steps required to 
do error correction may increase rapidly with the number of errors 
corrected.  Therefore, the time required to do error correction may 
eventually overwhelm the capability of the code to deal with errors, and 
the performance of the computer will start to decrease again.  To solve this 
problem, we need to find a class of codes where the time to measure the 
error syndrome increases only slowly with the error-correcting capabilities 
of the code.

The desired class of codes is concatenated codes
\cite{aharonov,knill-concatenate2,knill-concatenate1,zalka}.  For a 
concatenated code, the data is encoded using some $[n, k, d]$ code, then 
each qubit in a block is again encoded using an $[n_1, 1, d_1]$ code.  The 
qubits making up blocks in the new code may be further encoded using an 
$[n_2, 1, d_2]$ code, and so on indefinitely.  The result is an $[n n_1 n_2 
\cdots n_{l-1}, k, d d_1 d_2 \cdots d_{l-1}]$ code.  We can find the error 
syndrome of such a code rather rapidly.  We measure the error syndrome 
for the $[n_{l-1}, 1, d_{l-1}]$ code (the {\em first level} of the code) for 
all of the blocks of $n_{l-1}$ qubits at once.  To do this, we must make the 
assumption that we can do parallel computation on different qubits.  Note 
that we need this assumption anyway, or storage errors will always build 
up on some block while we are correcting errors on the other blocks.  
Similarly, we measure the error syndrome for the $[n_{l-2}, 1, d_{l-2}]$ 
code at the second level of the code in parallel for different blocks, and so 
on, for all $l$ levels of the code.  Therefore, we can measure the error 
syndrome for the whole code in only the sum of the number of steps 
required to measure each constituent code, instead of something like the 
product, which would be a more typical complexity for a code of the same 
parameters.

In order to analyze concatenated codes, it is useful to make a few 
simplifying assumptions.  One assumption is that we are using the same 
code at every level.  One particularly good code for this purpose is the 
$[7,1,3]$ code, because any operation in $N(\G)$ can be immediately 
performed transversally, keeping the overhead for fault-tolerant 
computation small.  In addition, it is a small code, so the complexity of 
error correction is not too large.  Allowing varying codes at different levels 
may improve the space efficiency of the code, but it will not change the 
basic results.  The other simplifying assumption is that the operations at 
level $j$ are basically similar to operations at level $j+1$.  Each level feeds 
information about error rates for different gates and storage errors and 
relative times for the different operations to the next lower level, but 
nothing else.  Error correction at each level is an independent process.  
Note that this will impair the error-correction properties of the code, since 
the full minimum distance of the code assumes that we combine 
information about the error syndrome from all the different levels.  
However, even with this assumption, we will find that for low enough basic 
error rates, we can do arbitrarily long computations with arbitrarily low 
real error rates by using sufficiently many levels of concatenation (the 
basic error rate is the rate of errors in actual physical qubits due to gates 
or storage errors; the real error rate is the rate of errors in the encoded 
data).  When the basic error rate is low enough, adding an extra level of
concatenation further reduces the real error rate; if the basic error rate
is too high, adding an extra layer increases the real error rate because of
the extra time spent on error correction and calculation.

In this chapter, I will present a rough calculation of the error threshhold 
below which arbitrarily long computation is possible.  In my discussion, 
the zeroth level of the code consists of the individual physical qubits making 
it up.  These qubits form the blocks of a $[7,1,3]$ code.  Each block of seven 
physical qubits forms a qubit at the first level of the code.  In general, 
qubits at the $j$th level of the code consist of $7^j$ physical qubits.  There 
are a total of $l$ levels in the code.  The qubits at the $l$th level are the 
real data qubits.  We wish to keep the effective error rate on these qubits as 
low as possible.  For this calculation, I will assume that storage errors occur 
independently on different physical qubits with rate $p_{stor}$.  The error 
rate for any one- or two-qubit gate in $N(\G)$ will be $p_g$, and the error 
rate for the Toffoli gate will be $p_{Tof}$.  I assume any gate may produce 
correlated errors on the qubits affected by the gate, but will produce no 
errors on any other qubits.  There will be an additional storage error on 
qubits unaffected by the gate, but the storage error is included in the gate 
error for qubits that are affected by the gate.  All the errors are assumed 
to be stochastically distributed, so the error probabilities for different 
qubits will add instead of the error amplitudes in the quantum states.  In 
addition, the error rates for state preparation and state measurement will 
be important.  I will denote them by $p_{prep}$ and $p_{meas}$, 
respectively.

The computation will call for various operations performed on the qubits 
encoded at various different levels.  After any operation at level $j$, I will 
perform error correction at level $j$.  This means we can give an effective 
error rate to each operation at level $j$.  The fact that a given error rate 
refers to a gate at level $j$ will be noted by a superscript $(j)$.  Thus, 
$p_{stor}^{(0)}$ is the storage error rate on the physical qubits, while 
$p_{g}^{(l)}$ is the effective error rate on the data qubits from performing 
an operation in $N(\G)$.  Only allowing one gate per error correction will 
typically reduce the performance of the code.  Errors created during error 
correction will dominate; an optimized code would perform error 
correction when the expected accumulated chance of errors was roughly 
equal to the chance of errors during error correction.  However, the 
assumption of one gate per error correction is another very useful simplifying 
assumption because it preserves the self-similar character of the 
concatenated code, allowing a relatively straightforward recursive 
calculation of the real error rates.

Some logical operations, such as the Toffoli gate, will require more and 
more physical operations as the level increases.  The basic time required to 
perform a physical operation will be $1$, and the storage error rate (at any 
level) is the error rate per unit time.  The time to perform a Toffoli gate at 
level $j$ will be denoted $t_{Tof}^{(j)}$.  Because operations in $N(\G)$ can 
be performed at any level just by performing a single operation from 
$N(\G)$ in parallel at the next lower level, the time to perform an 
operation in $N(\G)$ at any level is just $1$.  The time to prepare a state 
encoded at the $j$th level is $t_{prep}^{(j)}$ and the time to measure a 
qubit at the $j$th level is $t_{meas}^{(j)}$.  $\tpj{0} = 0$ and $\tmj{0} = 1$.

\section{Threshhold for Storage Errors and Gates From $N(\G)$}

To determine $\pgj{j}$ in terms of quantities at level $j-1$, we note 
that a gate in $N(\G)$ at level $j$ consists of a single gate in $N(\G)$ on 
each of the constituent qubits at level $j-1$ followed by a full error 
correction cycle at level $j-1$.   In order for the level $j$ gate to have an 
error, there must be two errors at level $j-1$, either in the $N(\G)$ gate or 
in the error correction.  I will assume that there is no residual error
that was missed in an earlier error correction step.  A more careful
calculation should consider such leftover errors, which can be significant.  
Suppose the chance of an error occuring in a single 
data qubit during a single measurement of the error syndrome is $p_{EC}$.  
There are a few possible situations that result in an error at level $j$.  Two 
errors at level $j-1$ could occur in any of 
\raisebox{0ex}[0ex][0ex]{$\mbox{\tiny $\pmatrix{7 \cr 2}$} = 21$}
choices of two qubits.  This could occur from two $N(\G)$ gates going 
wrong, with probability $(\pgj{j-1})^2$.  We repeat the error syndrome 
measurement until we get the same result twice.  If there is one error from an 
$N(\G)$ gate and one from either of these measurements of the error syndrome, 
there will be an error at level $j$.  The probability of this is $4 \pgj{j-1} 
p_{EC}$.  Finally, both errors could come from the error correction.  This 
could be two errors in the first or second syndrome measurement, with 
probability $2 p_{EC}^2$.  Given one error in a syndrome measurement, we 
will need to do three syndrome measurements total.  If two of those go 
wrong, it will also produce an error at level $j$.  This has probability $6 
p_{EC}^2$.  There are also a number of possibilities involving an error in the 
ancilla state producing an incorrect syndrome and requiring more 
measurements.  However, I assume the error rates involved are all fairly 
low, so the probability of this situation producing an error at level $j$ is 
smaller by $O(p)$, which I will assume is negligable.  Thus, the total gate 
error rate at level $j$ is
\begin{equation}
\pgj{j} = 21 \left((\pgj{j-1})^2 + 4 \pgj{j-1} p_{EC} + 8 p_{EC}^2 \right).
\label{eq-pg-recursion}
\end{equation}
Similarly, a single time step at level $j$ without a gate involves a single 
time step without a gate at level $j-1$ followed by error correction.  
Therefore,
\begin{equation}
\psj{j} = 21 \left((\psj{j-1})^2 + 4 \psj{j-1} p_{EC} + 8 p_{EC}^2 \right).
\label{eq-ps-recursion}
\end{equation}

The salient aspect of these equations is that the probability of error at 
level $j$ is of the order of the square of the error rate at level $j-1$.  This 
means that $\pgj{l}$ will scale roughly as 
\begin{equation}
\pgj{0} (\pgj{0}/p_{thresh})^{2^l}
\end{equation}
for some threshhold error rate $p_{thresh}$ and 
similarly for $\psj{l}$.  This is a very rapid decrease in $\pgj{l}$ as a 
function of $l$ when $\pgj{0} < p_{thresh}$.  We will thus only need a few 
levels, of order $\log (\log p)$ to bring the real error rate down to $O(p)$ 
per step.  Thus, the number of extra qubits necessary for a fault-tolerant 
computation is only $\mbox{polylog } p$ times the original number, which 
is a very good scaling.  However, while the asymptotic scaling is quite good, 
for vaguely reasonable $p$, the actual number of extra qubits needed is 
quite large.

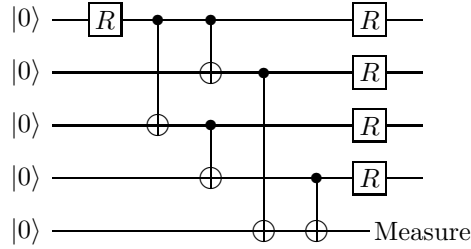
\begin{figure}
\centering
\begin{picture}(210,120)

\put(0,14){\makebox(20,12){$\ket{0}$}}
\put(0,34){\makebox(20,12){$\ket{0}$}}
\put(0,54){\makebox(20,12){$\ket{0}$}}
\put(0,74){\makebox(20,12){$\ket{0}$}}
\put(0,94){\makebox(20,12){$\ket{0}$}}

\put(20,20){\line(1,0){120}}
\put(20,40){\line(1,0){114}}
\put(20,60){\line(1,0){114}}
\put(20,80){\line(1,0){114}}
\put(20,100){\line(1,0){14}}
\put(46,100){\line(1,0){88}}

\put(34,94){\framebox(12,12){$R$}}

\put(60,100){\circle*{4}}
\put(60,100){\line(0,-1){44}}
\put(60,60){\circle{8}}

\put(80,100){\circle*{4}}
\put(80,100){\line(0,-1){24}}
\put(80,80){\circle{8}}
\put(80,60){\circle*{4}}
\put(80,60){\line(0,-1){24}}
\put(80,40){\circle{8}}

\put(100,80){\circle*{4}}
\put(100,80){\line(0,-1){64}}
\put(100,20){\circle{8}}

\put(120,40){\circle*{4}}
\put(120,40){\line(0,-1){24}}
\put(120,20){\circle{8}}

\put(134,34){\framebox(12,12){$R$}}
\put(134,54){\framebox(12,12){$R$}}
\put(134,74){\framebox(12,12){$R$}}
\put(134,94){\framebox(12,12){$R$}}
\put(140,14){\makebox(40,12){Measure}}

\put(146,40){\line(1,0){14}}
\put(146,60){\line(1,0){14}}
\put(146,80){\line(1,0){14}}
\put(146,100){\line(1,0){14}}

\end{picture}
\caption{Cat state construction and verification.}
\label{fig-4qubitcat}
\end{figure}

In order to determine the threshhold $p_{thresh}$, let us calculate 
$p_{EC}$.  I will assume we are using Shor's cat state method to correct 
errors, although another method (such as Steane's) might ultimately lead 
to better performance.  We have to measure six syndrome bits, so we will need 
to prepare six cat states, each using four qubits.  I will assume a limited 
ability to plan ahead in the calculation, so the data qubits will have to wait 
for the first cat state in a single measurement of the error syndrome, but 
the other cat states are being prepared at the same time, so they will be 
ready just when they are needed.  To prepare a cat state, we start with all 
four qubits in the state $\ket{0}$ (encoded using the code at level $j-1$), 
perform a Hadamard rotation $R$ on the first qubit, then a CNOT from the 
first qubit to the third qubit, and then two more CNOTs, from the first 
qubit to the second and from the third to the fourth, as shown in 
figure~\ref{fig-4qubitcat}.
Bit flip errors at this point will become phase errors after the final 
Hadamard transform, so we need to ensure that there is at most one.  
Every way a single gate error earlier in the construction can produce two 
bit flip errors here makes the second and fourth qubits different.  
Therefore, we perform CNOTs from the second and fourth qubits to an 
additional ancilla test qubit and measure the test qubit.  If it is $\ket{0}$, 
we can use the ancilla; if it is $\ket{1}$, there is at least one error in the 
cat state, possibly two.  We throw the cat state out and construct another 
one.  Finally, we must perform a Hadamard transform on each of the four 
qubits in the cat state to get the actual ancilla used in error correction.

An examination of the circuit shows that any bit flip errors before the 
cycle in which there are two CNOTs will cause the test qubit to flip.  
Therefore, only errors at this stage or later will have a chance of affecting 
the actual ancilla used.  For the second and fourth qubits, the error must 
actually occur after (or during) the CNOT to the test qubit.  Therefore, the 
chance of an important error in any single ancilla qubit is $2 p_g + 
p_{stor}$ (for qubits two and four) or $p_g + 2p_{stor}$ (for qubits one and 
three).  Although only phase errors can feed back, the fault-tolerant network 
does not treat $\X$ and $\Z$ errors symmetrically, so in order to be safe, I 
will consider the worst case where every error is of the most  dangerous type.  
However, in no case can an error in the test qubit feed back into the data 
qubits, so I have not included errors from this source.

Now, we can construct a network for error syndrome measurement such 
that each data qubit contributes to at most four syndrome bits.  In 
addition, two Hadamard rotations are necessary.  Therefore, the process of 
syndrome measurement introduces at most an additional probability $6 
p_g + 2 p_{stor}$ of error.  To this, we must add the probability of an error 
feeding back, plus the accumulation of storage errors while we prepare the 
cat state and measure the ancilla.  There is only waiting time for the 
preparation of the first cat state and measurement of the last one, since 
preparation and measurement of the other cat states is taking place in 
parallel.  Feedback is a more serious problem, but we can arrange it so 
that no data qubit interacts with more than two ancilla qubits with error 
rate $2p_g + p_{stor}$, so the total feedback is at most $6 p_g + 6 p_{stor}$. 
Therefore,
\begin{eqnarray}
p_{EC} &  = & (6 p_g + 6 p_{stor}) + (6 p_g + 2 p_{stor}) + 
(6 + t_{prep} + t_{meas}) p_{stor} \\
&  = &  12 p_g + (14 + t_{prep} + t_{meas}) p_{stor}.
\end{eqnarray}

Now, in order to measure a qubit encoded at some level, it is sufficient to 
measure all of the constituent qubits.  At level one, this gives us some 
seven-bit string which is a codeword of the classical Hammng code (possibly 
with some errors).  Whether it is a codeword of even or odd parity will tell us 
whether the corresponding level one qubit is $\ket{0}$ or $\ket{1}$.  We can 
continue to do this at all levels, using classical error correction at each 
level to correct any errors in individual bits.  This will, in general, require 
a fair amount of classical computation.  However, I will assume that classical 
computation is much faster than quantum computation when it can 
perform the same task, and that in the regime of interest, $t_{meas} = 1$.  
No matter what the speed of the classical computer, eventually $\tmj{j}$ 
will become greater than one, but due to the rapid convergence of the double 
exponential, this will have a very small effect on the threshhold.

Preparing encoded $\ket{0}$ states at level $j-1$ does take a fair amount 
of time, however.  Furthermore, the amount of time will increase with 
level.  One way to prepare encoded $0$ states reliably is by performing a 
full error correction cycle for the code with the addition of the $\Zbar$ 
operator $\Zs{5} \Zs{6} \Zs{7}$.  The input state can be anything.  The time to 
do this is at most $4(t_{EC} + 1)$.  Recall that we must get the same error 
syndrome twice before we trust it.  If there is an error in the second 
syndrome measurement, we may have to measure the syndrome twice 
more, for a total of four times.  The chance of two errors is lower order, 
and therefore we ignore it.

The time for one error correction cycle is $t_{EC} = 14 + t_{prep} + 
t_{meas}$, so $\tpj{j} = 64 + 4 \tpj{j-1}$.  In order to cut down the growth 
rate with level, I will assume we can plan ahead enough to prepare the 
ancillas for later syndrome measurements while measuring the earlier 
syndromes.  Then $\tpj{j} = 43 + \tpj{j-1}$.  Recalling that $\tpj{0} = 0$, we 
then get $\tpj{j} = 43 j$.  One benefit of preparing states using error 
correction is that the chance of residual error is minimal.  I will take 
$\ppj{j} = 0$ where it matters.

Finally, we get the result for $p_{EC}$.  The $t_{prep}$ that contributes is 
actually $\tpj{j-1}$, so
\begin{equation}
p_{EC}^{(j)} = 12 \pgj{j-1} + [15 + 43 (j-1)]\,\psj{j-1}.
\end{equation}
Therefore,
\begin{eqnarray}
\pgj{j} & = & 21 \left[(\pgj{j-1})^2 + 4 \pgj{j-1} p_{EC} + 8 p_{EC}^2
\right] \\
& = & 25221\,(\pgj{j-1})^2 + \left[61740 + 176988 (j-1) \right] \pgj{j-1} 
\psj{j-1} \nonumber \\
& & \mbox{} + \left[37800 + 216720 (j-1) + 310632 (j-1)^2 \right] (\psj{j-1})^2
\end{eqnarray}
and
\begin{eqnarray}
\psj{j} & = & 21 \left[(\psj{j-1})^2 + 4 \psj{j-1} p_{EC} + 8 p_{EC}^2
\right] \\
& = & 24192\,(\pgj{j-1})^2 + \left[61488 + 173376 (j-1) \right] \pgj{j-1} 
\psj{j-1} \nonumber \\
& & \mbox{} + \left[39081 + 220332 (j-1) + 310632 (j-1)^2 \right] (\psj{j-1})^2.
\end{eqnarray}
Note a number of things here.  If we perform error correction after every 
time step, whether it has a gate or not, the storage error rate and gate 
error rate at the next level will actually be dominated by the error rate of 
error correction, so they will be very close.  Also, at levels beyond the 
first, the error rate is dominated by storage errors occuring while we wait 
around encoding the ancilla qubits for error correction.  Therefore, the 
algorithm will benefit greatly from a more rapid preparation algorithm, a 
better ability to plan ahead, or both.

First, consider the limit in which storage errors are negligable.  In this 
case, we do not perform error correction after a step without a gate.  
Therefore, $\psj{j} = 0$ at all levels.  Then, $\pgj{j} = 25221\,(\pgj{j-1})^2$,
and the threshhold for a computation involving only operations from $N(\G)$ is 
$p_{thresh} = 1/25200 = 4.0 \times 10^{-5}$.  A second limit would be when 
$\pgj{0} = \psj{0}$, so there are no gate errors beyond the simple storage 
error in the same time step.  Then they should be equal at all other levels, 
as well.  Then
\begin{equation}
\psj{j} = \left[124761 + 393708 (j-1) + 310632 (j-1)^2 \right] (\psj{j-1})^2.
\end{equation}
Then $\psj{1} = 124800\,(\psj{0})^2$, $\psj{2} = 8.3 \times 10^5\,(\psj{1})^2$, 
and $\psj{3} = 2.2 \times 10^6\,(\psj{2})^2$.  For higher $j$, we approximate 
\begin{equation}
\psj{j} = 3.1 \times 10^5\,(j-1)^2 (\psj{j-1})^2 = \left[ (j-1)^2 
\psj{j-1}/ (3.2 \times 10^{-6}) \right] \psj{j-1}.
\end{equation}
To get continual improvement, it is sufficient for $\psj{j}/\psj{j-1} < (j-
1)^2/j^2$.  This will mean $\psj{j} \leq \frac{9}{j^2} \psj{3}$.  It suffices 
for $\psj{4} = \frac{9}{16} \psj{3}$, so $\psj{3} = \frac{1}{16} (3.2 \times 
10^{-6})$.  Following this back, we find that for only storage errors, the 
threshhold is roughly $p_{thresh} = 2.2 \times 10^{-6}$, or slightly more 
than an order of magnitude worse than for just gate errors.

Let us consider another case.  Suppose we can plan ahead well, and 
prepare ancillas for error correction just in time for when they are needed.  
Then $p_{EC} = 12 p_g + 9 p_{stor}$, and
\begin{eqnarray}
\pgj{j} & = & 25221\,(\pgj{j-1})^2 + 37044\,\pgj{j-1} \psj{j-1} + 13608\,
(\psj{j-1})^2 \\
\psj{j} & = & 24192\,(\pgj{j-1})^2 + 37296\,\pgj{j-1} \psj{j-1} + 14385\,
(\psj{j-1})^2.
\end{eqnarray}
For all practical purposes, for $j > 1$, $\pgj{j} = \psj{j} = p^{(j)} = 75873\, 
(p^{(j-1)})^2$.  This means that the threshhold occurs at $p^{(1)} = 1/75873 = 
1.3 \times 10^{-5}$.  At the limit $\psj{0} = 0$, we get a threshhold for 
$p_g$ of $p_{thresh} = 2.3 \times 10^{-5}$.  At the limit $\pgj{0} = \psj{0}$, 
we get a threshhold $p_{thresh} = 1.3 \times 10^{-5}$.

Finally, suppose we do not do error correction after every step, but instead 
attempt to optimize the number of steps $N$ between error corrections.  
Then the chance of error in $N$ steps is $N \pgj{j-1}$ or $N \psj{j-1}$, and 
equations (\ref{eq-pg-recursion}) and (\ref{eq-ps-recursion}) become
\begin{eqnarray}
N \pgj{j} & = & 21 \left[N^2 (\pgj{j-1})^2 + 4 N \pgj{j-1} p_{EC} + 8 p_{EC}^2
\right] \\
N \psj{j} & = & 21 \left[N^2 (\psj{j-1})^2 + 4 N \psj{j-1} p_{EC} + 8 p_{EC}^2
\right].
\end{eqnarray}
The values $\pgj{j}$ and $\psj{j}$ now represent average error rates, 
rather than strict error rates per step.  As long as we do gates from 
$N(\G)$ only or storage only, these values will be accurate representations, 
but if we mix and match, the story will be a bit different.  Optimizing with 
respect to $N$ gives us
\begin{eqnarray}
-\frac{21}{N^2} \left[N^2 (\pgj{j-1})^2 + 4 N \pgj{j-1} p_{EC} + 8 p_{EC}^2
\right] & & \\
\mbox{} + \frac{21}{N} \left[2 N (\pgj{j-1})^2 + 4 \pgj{j-1} p_{EC} \right]
& = & 0
\end{eqnarray}
\vspace{-\belowdisplayskip}
\begin{eqnarray}
N^2 (\pgj{j-1})^2 + 4 N \pgj{j-1} p_{EC} + 8 p_{EC}^2 & = & 2 N^2 
(\pgj{j-1})^2 + 4 N \pgj{j-1} p_{EC} \nonumber \\ \\
N^2 (\pgj{j-1})^2 - 8 p_{EC}^2 & = & 0 \\
N & = & \sqrt{8}\,(p_{EC}/\pgj{j-1}).
\end{eqnarray}
The same is true for storage steps.  The optimum number of steps makes 
the accumulated chance of error during gates $\sqrt{8}$ times the chance 
of error during error correction.  Plugging in this value for $N$ gives us
\begin{equation}
\pgj{j} = \frac{21}{N} (16 + 8 \sqrt{2}) p_{EC}^2.
\end{equation}
Assuming no storage errors, $p_{EC} = 12 \pgj{j-1}$, so $N = 
34$ and $\pgj{j} = 2.4 \times 10^3\,(\pgj{j-1})^2$, so the threshhold is 
$p_{thresh} = 4.1 \times 10^{-4}$.  In practice, we will not be able to perform 
error correction after exactly 34 gates, since there will be Toffoli gates 
occuring at possibly inconvenient times, but if we get close to the right 
frequency of error correction, the actual threshhold will not be too much 
worse than this.

\section{Toffoli Gate Threshhold}

\begin{figure}
\centering
\begin{picture}(335,200)

\put(0,174){\makebox(20,12){$\ket{0}$}}
\put(0,154){\makebox(20,12){$\ket{0}$}}
\put(0,134){\makebox(20,12){$\ket{0}$}}
\put(0,114){\makebox(20,12){$\ket{cat}$}}
\put(0,94){\makebox(20,12){$\ket{cat}$}}
\put(0,74){\makebox(20,12){$\ket{cat}$}}
\put(0,54){\makebox(20,12){$\ket{d_1}$}}
\put(0,34){\makebox(20,12){$\ket{d_2}$}}
\put(0,14){\makebox(20,12){$\ket{d_3}$}}

\put(20,180){\line(1,0){4}}
\put(24,174){\framebox(12,12){$R$}}
\put(20,160){\line(1,0){4}}
\put(24,154){\framebox(12,12){$R$}}
\put(20,140){\line(1,0){4}}
\put(24,134){\framebox(12,12){$R$}}

\put(36,180){\line(1,0){279}}
\put(36,160){\line(1,0){218}}
\put(266,160){\line(1,0){49}}
\put(36,140){\line(1,0){218}}
\put(266,140){\line(1,0){49}}

\put(20,80){\line(1,0){19}}
\put(51,80){\line(1,0){6}}
\put(69,80){\line(1,0){21}}
\put(57,74){\framebox(12,12){$R$}}

\put(39,74){\framebox(12,12){$\Z$}}
\put(45,140){\circle*{4}}
\put(45,140){\line(0,-1){54}}

\put(80,180){\circle*{4}}
\put(80,180){\line(0,-1){104}}
\put(80,160){\circle*{4}}
\put(80,80){\circle{8}}

\put(90,74){\makebox(30,12){Meas.}}

\put(20,100){\line(1,0){69}}
\put(101,100){\line(1,0){6}}
\put(119,100){\line(1,0){21}}
\put(107,94){\framebox(12,12){$R$}}

\put(89,94){\framebox(12,12){$\Z$}}
\put(95,140){\circle*{4}}
\put(95,140){\line(0,-1){34}}

\put(130,180){\circle*{4}}
\put(130,180){\line(0,-1){84}}
\put(130,160){\circle*{4}}
\put(130,100){\circle{8}}

\put(138,94){\makebox(30,12){Meas.}}

\put(20,120){\line(1,0){119}}
\put(151,120){\line(1,0){6}}
\put(169,120){\line(1,0){19}}
\put(157,114){\framebox(12,12){$R$}}

\put(139,114){\framebox(12,12){$\Z$}}
\put(145,140){\circle*{4}}
\put(145,140){\line(0,-1){14}}

\put(180,180){\circle*{4}}
\put(180,180){\line(0,-1){64}}
\put(180,160){\circle*{4}}
\put(180,120){\circle{8}}

\put(190,114){\makebox(30,12){Meas.}}

\put(200,140){\circle{8}}
\put(200,136){\line(0,1){8}}
\put(190,130){\framebox(20,60){}}
\put(150,80){\vector(1,1){50}}
\put(120,80){\line(1,0){30}}
\put(165,100){\line(1,0){5}}
\put(202,122){\vector(0,1){8}}

\put(20,60){\line(1,0){270}}
\put(20,40){\line(1,0){250}}
\put(20,20){\line(1,0){214}}

\put(238,180){\circle*{4}}
\put(238,180){\line(0,-1){124}}
\put(238,60){\circle{8}}

\put(230,160){\circle*{4}}
\put(230,160){\line(0,-1){124}}
\put(230,40){\circle{8}}

\put(222,20){\circle*{4}}
\put(222,20){\line(0,1){124}}
\put(222,140){\circle{8}}

\put(234,14){\framebox(12,12){$R$}}
\put(246,20){\line(1,0){4}}

\put(250,14){\makebox(30,12){Meas.}}
\put(270,34){\makebox(30,12){Meas.}}
\put(290,54){\makebox(30,12){Meas.}}

\put(260,180){\circle*{4}}
\put(260,180){\line(0,-1){14}}
\put(254,154){\framebox(12,12){$\Z$}}
\put(254,134){\framebox(12,12){$\Z$}}
\put(250,130){\framebox(20,60){}}
\put(260,26){\vector(0,1){104}}

\put(276,180){\circle*{4}}
\put(276,180){\line(0,-1){44}}
\put(276,140){\circle{8}}
\put(284,160){\circle{8}}
\put(284,156){\line(0,1){8}}
\put(270,130){\framebox(20,60){}}
\put(280,46){\vector(0,1){84}}

\put(300,180){\circle{8}}
\put(300,176){\line(0,1){8}}
\put(300,160){\circle*{4}}
\put(300,160){\line(0,-1){24}}
\put(300,140){\circle{8}}
\put(290,130){\framebox(20,60){}}
\put(300,66){\vector(0,1){64}}

\put(315,174){\makebox(20,12){$\ket{d_1'}$}}
\put(315,154){\makebox(20,12){$\ket{d_2'}$}}
\put(315,134){\makebox(20,12){$\ket{d_3'}$}}

\end{picture}
\caption[The Toffoli gate construction.]{The Toffoli gate construction.  Each 
line represents seven qubits at the next lower level.}
\label{fig-toffoli}
\end{figure}
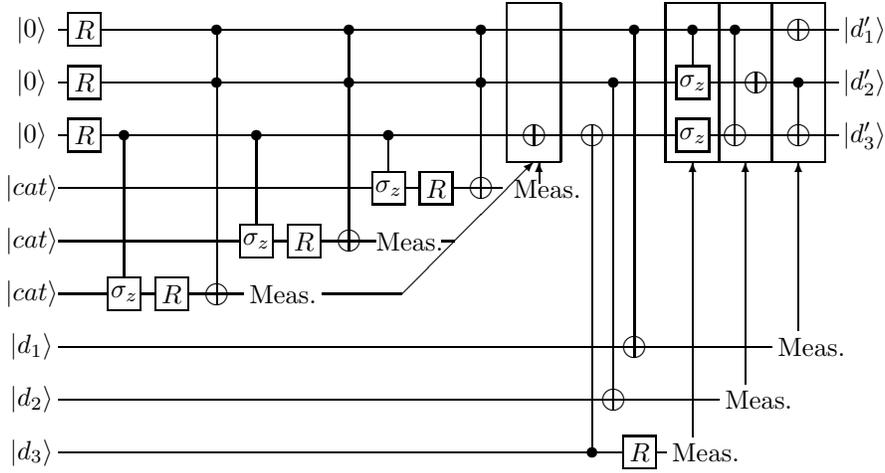

To figure out the recursion relation for the Toffoli gate, look at figure 
\ref{fig-toffoli}, which summarizes the construction in section
\ref{sec-toffoli}.
I will follow each qubit individually in order to figure out the final chance 
of error for that qubit.  This is a construction for the Toffoli gate at level 
$j+1$.  I will assume we do error correction on all three ancilla qubits only 
after the Toffoli gate is completed.  All three ancilla qubits start out with 
$\ppj{j+1}$ chance of error from preparing encoded $\ket{0}$'s.  There are 
actually two types of relevant encoding errors.  There can be errors 
remaining at lower levels.  Since we have just done an error correction 
cycle, I assume that the number of residual errors is negligable.  There is 
also a chance that the qubit will not be an encoded $\ket{0}$, but some 
other encoded state.  This would count as a complete failure of the Toffoli 
gate, since it would produce a real error at level $j+1$.  However, I will 
assume that the chance of this happening is also zero.

Assume the chance of a remaining bit flip error in a cat state is $p_{cat}$ 
and the time to make a cat state is $t_{cat}$.  Only bit flip errors feed back 
from the cat states to the ancilla qubits in this network.  Let $A_1$, $A_2$, 
and $A_3$ be the accumulated chances of error in the three ancilla qubits.
First we have a Hadamard transform on all three of these qubits.  After the 
first gate in the ancilla construction, $A_3 = t_{cat}\,\psj{j} + 2 \pgj{j} + p_{cat}$.  It will have to sit around an additional $1 + \ttj{j}$ 
time steps before the interaction with the next cat state begins.  Thus, after 
the first cat state is finished, $A_3 = (t_{cat} + \ttj{j} + 1) \psj{j} + 
2 \pgj{j} + p_{cat}$.  By the time of the Toffoli gate with the first two 
ancilla qubits, the chance of errors in the cat state which can feed back into 
the main ancilla is at most $p_{cat} + 2 \pgj{j}$.  The first two ancilla 
qubits have already waited a time $t_{cat}+2$, so the overall chance of errors 
in the first two ancilla qubits is 
\begin{equation}
A_1 = A_2 = (t_{cat}+2) \psj{j} + p_{cat} + 3 \pgj{j} + \ptj{j}.
\end{equation}

We repeat the cat state interaction two more times with new cat states, 
which we have been preparing in parallel with the first cat state.  
Therefore, we only need $2+\ttj{j}$ more time steps for each interaction, 
introducing the same amount of error as the equivalent steps in the first 
interaction.  We must also measure the cat states.  We can do it in the basis 
they end up in; we check for odd or even parity.  If two of the three cat 
states have odd parity, we decide the ancilla is in the state $\ket{B}$, and 
we perform $\X$ on the third ancilla qubit.  This process will take an 
additional $\tmj{j} + 1$ time units.  After the ancilla creation is completed, 
the chances of error on the three qubits are
\begin{eqnarray}
A_1 & = & \left(t_{cat} + \tmj{j} + 7 \right) \psj{j} + 3p_{cat} + 7 \pgj{j} + 
3 \ptj{j} \\
A_2 & = & \left(t_{cat} + \tmj{j} + 7 \right) \psj{j} + 3p_{cat} + 7 \pgj{j} + 
3 \ptj{j} \\
A_3 & = & \left(t_{cat} + \tmj{j} + 3 \ttj{j} + 3 \right) \psj{j} + 3p_{cat} + 
5 \pgj{j}.
\end{eqnarray}
The whole ancilla construction has taken a time $t_{cat} + \tmj{j} + 3 \ttj{j} 
+ 7$, during which time the data qubits have been accumulating storage 
errors.  I assume here that $t_{cat} \geq \tpj{j} + 1$.

Now we perform the CNOTs between the data qubits and the ancilla qubits.  
Again we make the conservative assumption that all of the accumulated chance of 
error on the data qubits feeds into the ancilla qubits.  Thus,
\begin{eqnarray}
A_1 & \! = & \! \left(2 t_{cat} + 2 \tmj{j} + 3 \ttj{j} + 14 \right) \psj{j} + 
3 p_{cat} + 8 \pgj{j} + 3 \ptj{j} \\
A_2 &\!  = & \! \left(2 t_{cat} + 2 \tmj{j} + 3 \ttj{j} + 14 \right) \psj{j} + 
3 p_{cat} + 8 \pgj{j} + 3 \ptj{j} \\
A_3 & \! = & \! \left(2 t_{cat} + 2 \tmj{j} + 6 \ttj{j} + 10 \right) \psj{j} + 
3 p_{cat} + 6 \pgj{j}.
\end{eqnarray}
Now we measure $\Z$ for the first two data qubits and $\X$ for the third 
data qubit.  We will add one time step for the Hadamard rotation on the third 
data qubit, plus $\tmj{j}$ to measure.  We should include a chance of the 
Toffoli gate failing because of the wrong result on one of these 
measurements, but I will assume that chance is small compared to the 
accumulated errors on the ancilla qubits.  Before we start doing the 
conditional operations to convert the ancilla states to complete the transfer 
of the data, the chances of error are
\begin{eqnarray}
A_1 & \! = & \! \left(2 t_{cat} + 3 \tmj{j} + 3 \ttj{j} + 15 \right) \psj{j} + 
3 p_{cat} + 8 \pgj{j} + 3 \ptj{j} \\
A_2 & \! = & \! \left(2 t_{cat} + 3 \tmj{j} + 3 \ttj{j} + 15 \right) \psj{j} + 
3 p_{cat} + 8 \pgj{j} + 3 \ptj{j} \\
A_3 & \! = & \! \left(2 t_{cat} + 3 \tmj{j} + 6 \ttj{j} + 11 \right) \psj{j} + 
3 p_{cat} + 6 \pgj{j}.
\end{eqnarray}

I will now assume that all three operations are necessary; this is the 
worst case, and usually there will be fewer gate errors.  The first 
conditional operation interacts ancilla qubits one and two, giving
\begin{eqnarray}
A_1 & \!\! = & \!\! \left(4 t_{cat} + 6 \tmj{j} + 6 \ttj{j} + 30 \right) \psj{j}
+ 6 p_{cat} + 17 \pgj{j} + 6 \ptj{j} \\
A_2 & \!\! = & \!\! \left(4 t_{cat} + 6 \tmj{j} + 6 \ttj{j} + 30 \right) \psj{j}
+ 6 p_{cat} + 17 \pgj{j} + 6 \ptj{j} \\
A_3 & \!\! = & \!\! \left(2 t_{cat} + 3 \tmj{j} + 6 \ttj{j} + 11 \right) \psj{j}
+ 3 p_{cat} + 7 \pgj{j}.
\end{eqnarray}
The second conditional operation interacts ancilla qubits one and three, so
\begin{eqnarray}
A_1 & \!\!\!\! = & \!\!\!\! \left(6 t_{cat} + 9 \tmj{j} + 12 \ttj{j} + 41 
\right) \psj{j} + 9 p_{cat} + 25 \pgj{j} + 6 \ptj{j} \\
A_2 & \!\!\!\! = & \!\!\!\! \left(4 t_{cat} + 6 \tmj{j} + 6 \ttj{j} + 30 
\right) \psj{j} + 6 p_{cat} + 18 \pgj{j} + 6 \ptj{j} \\
A_3 & \!\!\!\! = & \!\!\!\! \left(6 t_{cat} + 9 \tmj{j} + 12 \ttj{j} + 34 
\right) \psj{j} + 9 p_{cat} + 25 \pgj{j} + 6 \ptj{j}.
\end{eqnarray}
The third operation interacts the second and third ancilla qubits.  Much of 
the error from the first and second ancilla qubits has already been 
introduced into the third qubit, so there is no need to add it again.  In fact, 
much of it may cancel out instead.  However, I assume it remains.  The 
only new error for the third ancilla qubit is the gate error on the second 
qubit from the previous operation plus the gate error for this operation.  
Thus,
\begin{eqnarray}
A_1 & \!\!\!\!\! = & \!\!\!\!\! \left(6 t_{cat} + 9 \tmj{j} + 12 \ttj{j} + 41 
\right) \psj{j} + 9 p_{cat} + 26 \pgj{j} + 6 \ptj{j} \\
A_2 & \!\!\!\!\! = & \!\!\!\!\! \left(6 t_{cat} + 9 \tmj{j} + 12 \ttj{j} + 41 
\right) \psj{j} + 9 p_{cat} + 27 \pgj{j} + 6 \ptj{j} \\
A_3 & \!\!\!\!\! = & \!\!\!\!\! \left(6 t_{cat} + 9 \tmj{j} + 12 \ttj{j} + 41 
\right) \psj{j} + 9 p_{cat} + 27 \pgj{j} + 6 \ptj{j}.
\end{eqnarray}
The overall chance of error on a single one of the new data qubits after the  
full Toffoli gate construction is thus 
\begin{equation}
\left(6 t_{cat} + 9 \tmj{j} + 12 \ttj{j} + 41 \right)
\psj{j} + 9 p_{cat} + 27 \pgj{j} + 6 \ptj{j}.
\end{equation}
The time taken to perform this Toffoli gate is
\begin{equation}
\ttj{j+1} = t_{cat} + 2 \tmj{j} + 3 \ttj{j} + 12.
\end{equation}

After error correction, the chance of a real error at level $j+1$ is
\begin{eqnarray}
\ptj{j+1} & \!\!\!\!\! = & \!\!\!\!\! 21\,\Bigg\{ \!\left[(6 t_{cat} + 9 
\tmj{j} + 12 \ttj{j} + 41)\,\psj{j} + 9 p_{cat} + 27 \pgj{j} + 6 
\ptj{j}\right]^2 \nonumber \nopagebreak \\
& & \!\!\!\!\!\! \mbox{} + 4 \! \left[ (6 t_{cat} + 9 \tmj{j} + 12 \ttj{j} + 
41)\,\psj{j} + 9 p_{cat} + 27 \pgj{j} + 6 \ptj{j}\right] \! p_{EC} \nonumber 
\nopagebreak \\
& & \!\!\!\!\!\! \mbox{} + 8 p_{EC}^2 \Bigg\}.
\end{eqnarray}
In order to simplify the recursion relation so that it is easily solvable,
I will only investigate the limit where there are no storage errors.  In
this case, it makes sense to verify the cat state used in the construction
until the chance of errors in it is negligable.  Therefore, I will also
assume that $p_{cat} = 0$.  Then the recursion relation for the Toffoli
gate becomes
\begin{eqnarray}
\ptj{j+1} & = & 21 \left[ (27 \pgj{j} + 6 \ptj{j})^2 + 4\,(27 \pgj{j} +
6 \ptj{j})\,p_{EC} + 8 p_{EC}^2 \right] \\
& = & 66717\,(\pgj{j})^2 + 12852\,\pgj{j} \ptj{j} + 756\,(\ptj{j})^2.
\end{eqnarray}
Recall that in this limit, $\pgj{j} = 25221\,(\pgj{j-1})^2$, so
\begin{equation}
\pgj{j} = 25200^{a(j)} (\pgj{0})^{2^j},
\end{equation}
where $a(j+1) = 1 + 2a(j)$, with $a(1) = 1$.  Therefore, $a(j) = 2^j - 1$,
and
\begin{eqnarray}
\pgj{j} & = & 4.0 \times 10^{-5} 
\left[ \pgj{0}/ (4.0 \times 10^{-5}) \right]^{2^j} \\
& = & p_{thresh} \left( \pgj{0}/p_{thresh} \right)^{2^j}.
\end{eqnarray}
Writing $\epsilon = \pgj{0}/p_{thresh}$, we have
\begin{equation}
\ptj{j+1} = 1.1 \times 10^{-4}\,\epsilon^{2^{j+1}} + 0.51\,\epsilon^{2^j} 
\ptj{j} + 756\,(\ptj{j})^2.
\end{equation}
The first term is often negligable compared to the second term, in which case
\begin{equation}
\ptj{j+1} = \left( 0.51\,\epsilon^{2^j} + 756\,\ptj{j} \right) \ptj{j}.
\end{equation}
In the limit where $\epsilon$ is small, we find a threshhold value of
$\ptj{0} = 1/756 = 1.3 \times 10^{-3}$.

Even when $\epsilon$ is fairly large, the presence of Toffoli gates does
not present much of a problem for the threshhold.  For instance, if we demand 
that $\ptj{0} = \pgj{0} = \epsilon p_{thresh}$, then
\begin{eqnarray}
\ptj{1} & = & 1.1 \times 10^{-4}\,\epsilon^2 + \left[12852\,p_{thresh} \epsilon 
+ 756\,p_{thresh} \epsilon \right] \ptj{0} \\
& \approx & 1.3 \times 10^{-4}\,\epsilon^2, \\
\ptj{2} & = & 1.1 \times 10^{-4}\,\epsilon^4 + \left[0.51\,\epsilon^2 + 
756\,(1.3 \times 10^{-4})\,\epsilon^2 \right] \ptj{1} \\
& = & 1.9 \times 10^{-4}\,\epsilon^4, \\
\ptj{3} & = & 1.1 \times 10^{-4}\,\epsilon^8 + \left[0.51\,\epsilon^4 + 
756\,(1.9 \times 10^{-4})\,\epsilon^4 \right] \ptj{2}, \\
& = & 2.3 \times 10^{-4}\,\epsilon^8.
\end{eqnarray}
If we let $\epsilon^4 = 1.9 / 2.3$, so $\epsilon \approx 0.95$, then $\ptj{3} = 
\ptj{2}$, and as we add levels of concatenation, the Toffoli gate error gate 
will begin to improve.  Therefore, the presence of Toffoli gates with the same 
physical error rate as other gates causes less than a $5\%$ reduction in the 
threshhold.
\pagestyle{myheadings}

\chapter{Bounds on Quantum Error-Correcting Codes}
\label{chap-bounds}
\markright{CHAPTER~\ref{chap-bounds}. \ BOUNDS ON QUANTUM CODES}

\section{General Bounds}
\label{sec-gen-bounds}
\pagestyle{headings}

The question of how efficient an error-correcting code of a given block size 
can be made in terms of both encoded qubits and distance is an interesting 
and important question in the theories of both classical and quantum error 
correction.  In the classical theory, only upper and lower bounds exist on 
the efficiency of codes that must have a given minimum distance between 
all codewords.  The true, achievable bounds on such codes are unknown.  
Better understood in the classical case is the asymptotic efficiency of coding 
(where we only require that the code correct all likely errors).  In the limit 
of infinite bits sent, we usually require the code to correct measure one of 
the errors occuring using some probability measure associated with the channel. 
Classically, Shannon's theorem tells us what the achievable capacity of a 
channel is.  No real quantum analogue of Shannon's theorem is known, despite
extensive work on the subject~\cite{lloyd, schumacher, barnum}.

One simple upper bound on the efficiency of quantum codes is the 
quantum Hamming bound~\cite{ekert}.  For a nondegenerate code with 
basis codewords $\ket{\psi_i}$ and possible errors $E_a$, all of the states 
$E_a \ket{\psi_i}$ are linearly independent for all $a$ and $i$.  If the code 
uses $n$ qubits, there can only be $2^n$ linearly indepedent vectors in the 
Hilbert space, so the number of errors times the number of codewords 
must be less than or equal to $2^n$.  If the code corrects all errors of 
weight $t$ or less and encodes $k$ qubits, this means
\begin{equation}
\Sum_{j=0}^{t} 3^j \pmatrix{n \cr j} 2^k \leq 2^n.
\label{eq-QHB-finite}
\end{equation}
There are \mbox{\tiny $\pmatrix{n \cr j}$} ways to choose $j$ qubits to be 
affected by $j$ errors and $3^j$ ways these errors can be tensor products of 
$\X$, $\Y$, and $\Z$.  This bound is completely analogous to the classical 
Hamming bound, with two differences: the quantum bound has a factor of $3^j$ 
reflecting the additional quantum-mechanical degrees of freedom; and the 
quantum bound only applies to nondegenerate codes.  The distinction 
between degenerate and nondegenerate codes is a purely
quantum-mechanical distinction; there are no classical degenerate codes.  
It is unknown whether there are any degenerate codes that exceed the 
quantum Hamming bound (\ref{eq-QHB-finite}).

If we let the block size $n$ grow arbitrarily large, we should also increase 
the expected number of errors.  Consider the depolarizing channel, which is 
equally likely to have $\X$, $\Y$, and $\Z$ errors.  Suppose there is a 
probability $p$ of having one of these errors on a given qubit and $1-p$ of 
having no error.  The expected number of errors on a block of size $n$ is $t 
= np$.  The number of likely errors will be about the number of errors of 
length $t$, so the quantum Hamming bound becomes
\begin{equation}
3^{np} \pmatrix{n \cr np} 2^k \leq 2^n.
\end{equation}
Taking the logarithm and rearranging gives us
\begin{equation}
\frac{k}{n} \leq 1 - p \log_2 3 - H(p).
\label{eq-QHB}
\end{equation}
Again, $H(x) = - x \log_2 x - (1 - x) \log_2 (1-x)$, as with the asymptotic 
form of the classical Hamming bound (\ref{eq-Hamming}).  As with the 
classical case, we can achieve the quantum Hamming bound by using 
random codes.  Unlike the classical case, this is not always the most 
efficient use of the channel, so (\ref{eq-QHB}) does not give the actual 
channel capacity of the quantum channel.  I will discuss this question in 
greater detail in section~\ref{depolarizing}.

For minimum distance codes, it is not in general possible to achieve the 
quantum Hamming bound.  We can set a lower bound, the quantum 
Gilbert-Varshamov bound.  Recall that
\begin{equation}
\bra{\psi_i} E_a^\dagger E_b \ket{\psi_j} = C_{ab} \delta_{ij}
\end{equation}
for a quantum code correcting errors $\{E_a\}$ with basis states 
$\ket{\psi_i}$.  The matrix $C_{ab}$ is Hermitian, but is further 
constrained by the algebraic relationships of the operators $E_a^\dagger 
E_b$.  It is better to consider $C_{ab}$ as a function of operators $O = 
E_a^\dagger E_b$.  When the possible errors are all operators of up to 
weight $t$, $O$ can be any operator of weight $\leq 2t$.  Slightly more 
generally, for a code of distance $d$, $O$ is any operator of weight less 
than $d$.  Therefore, the statement
\begin{equation}
\bra{\psi} E_a^\dagger E_b \ket{\psi} = C_{ab}
\label{eq-Cab}
\end{equation}
is actually 
\begin{equation}
N = \Sum_{j=0}^{d-1} 3^j \pmatrix{n \cr j}
\end{equation}
constraints on the state $\ket{\psi}$.  For generic $C_{ab}$ (satisfying the 
appropriate algebraic constraints) and generic linear subspace $V$ with 
dimension larger than $N$, there will be states $\ket{\psi}$ satisfying 
equation (\ref{eq-Cab}).

Suppose we choose generic $C_{ab}$ and a generic state $\ket{\psi_1}$ 
satisfying (\ref{eq-Cab}).  Now restrict attention to the subspace 
orthogonal to $\ket{\psi_1}$ and to all $O \ket{\psi_1}$ for operators $O$ of 
weight less than $d$.  For an $n$-qubit Hilbert space, this subspace has 
dimension $2^n - N$.  Choose a generic state $\ket{\psi_2}$ in this subspace 
satisfying (\ref{eq-Cab}).  Now restrict attention to the subspace 
orthogonal to both $O \ket{\psi_1}$ and $O \ket{\psi_2}$.  We can again 
pick $\ket{\psi_3}$ in this subspace satisfying (\ref{eq-Cab}), and so on.  
Choose $\ket{\psi_i}$ orthogonal to all $O \ket{\psi_j}$ ($j \leq i-1$) and 
satisfying (\ref{eq-Cab}).  We can continue doing this as long as
\begin{equation}
\Sum_{j=0}^{d-1} 3^j \pmatrix{n \cr j} i < 2^n.
\end{equation}
Therefore, we can always find a distance $d$ quantum code encoding $k$ 
qubits in $n$ qubits satisfying
\begin{equation}
\Sum_{j=0}^{d-1} 3^j \pmatrix{n \cr j} 2^k \geq 2^n.
\label{eq-QGV}
\end{equation}
This is the quantum Gilbert-Varshamov bound.  In the limit where $t = pn 
= d/2$, with $n$ large, this becomes
\begin{equation}
\frac{k}{n} \geq 1 - 2p \log_2 3 - H(2p).
\end{equation}

The quantum Hamming bound only limits the efficiency of nondegenerate 
codes.  For degenerate codes, we can still set a bound, but it will not be as 
restrictive.  For an $[n, k, d]$ code, we can choose any $d-1$ qubits and 
remove them.  The remaining $n-d+1$ qubits must contain enough 
information to reconstruct not only the $2^k$ possible codewords, but the state 
of the missing qubits as well.  Because the missing qubits can be any 
qubits, we can choose them to have maximum entropy.  Then
\begin{eqnarray}
n-d+1 & \geq & d-1+k \\
n & \geq & 2(d-1) + k.
\end{eqnarray}
This is the Knill-Laflamme bound~\cite{knill-laflamme-theory,cerf-cleve}. 
It is a quantum analog of the classical Singleton bound. 
A code to correct $t$ errors must have distance $d=2t+1$, so for such a 
code, $n \geq 4t + k$.  This bound holds for any code with a given 
minimum distance, whether it is degenerate or nondegenerate.  For 
instance, this bound demonstrates that the smallest one-error-correcting 
quantum code uses five qubits.

\section{Weight Enumerators and Linear Programming Bounds}
\label{sec-enumerators}

In the classical theory of error-correcting codes, the distribution of 
codeword weights contains a great deal of information about the code.  
This distribution is often encoded in the coefficients of a polynomial, and 
algebraic relationships between these polynomials, known as {\em weight 
enumerators}, can be very useful for setting bounds on classical codes.  
Many of the same ideas can be adapted for use with quantum 
error-correcting codes~\cite{rains-shadow, shor-laflamme-QMW, 
rains-enumerators, rains-poly-invariants}.

Let $A_d$ be the number of elements of the stabilizer $S$ with weight $d$, 
and let $B_d$ be the number of elements of $N(S)$ with weight $d$ (ignoring
overall phases).  Note that $B_d \geq A_d \geq 0$.  Define polynomials
\begin{eqnarray}
A (z) & = & \Sum_{d=0}^{n} A_d z^d \\
B (z) & = & \Sum_{d=0}^{n} B_d z^d.
\end{eqnarray}
$A_0 = B_0 = 1$ always.  For a code of distance $d$, $B_{d'} = A_{d'}$ for 
all $d' < d$.  For a nondegenerate code, $B_{d'} = A_{d'} = 0$ for $d' < d$.  A 
degenerate code has $B_{d'} = A_{d'} > 0$ for at least one $d' < d$.  $A(z)$ 
and $B(z)$ are the weight enumerators of $S$ and $N(S)$.

The polynomials $A(z)$ and $B(z)$ satisfy the quantum MacWilliams 
identity \cite{shor-laflamme-QMW}:
\begin{equation}
B(z) = \frac{1}{2^{n-k}} (1+3z)^n A \left( \frac{1-z}{1+3z} \right).
\label{eq-QMW}
\end{equation}
In other words,
\begin{equation}
\Sum_{d=0}^{n} B_d z^d = \frac{1}{2^{n-k}} \Sum_{d=0}^{n} A_d (1-z)^d 
(1+3z)^{n-d}.
\end{equation}
Matching coefficients of $z^d$, we find
\begin{equation}
B_d = \frac{1}{2^{n-k}} \Sum_{d'=0}^{n} \left[ \Sum_{s=0}^{d} (-1)^s 3^{d-s} 
\pmatrix{d' \cr s} \pmatrix{n-d' \cr d-s} \right] A_{d'}.
\end{equation}

To prove this, note that an operator $E \in \G$ of weight $d$ will either 
commute with every operator $M \in S$ or it will commute with exactly 
half of the operators in $S$.  Therefore, if we sum
\begin{equation}
\Sum_{M \in S} (-1)^{f_M (E)},
\end{equation}
we will get zero if $E \notin N(S)$ and $2^{n-k}$ if $E \in N(S)$ (recall that 
$f_M (E)$ is $0$ if $M$ and $E$ commute and $1$ if they do not).  
Therefore, we can write $B_d$ as follows:
\begin{equation}
B_d = \frac{1}{2^{n-k}} \Sum_{E} \Sum_{M \in S} (-1)^{f_M (E)},
\end{equation}
where the sum over $E$ is taken over all $E \in \G$ of weight $d$.  We 
reverse the order of summation and break up the sum over $M$ to the 
sum over $d'$ and the sum over $M \in S$ of weight $d'$ to get
\begin{equation}
B_d = \frac{1}{2^{n-k}} \Sum_{d'=0}^{n} \Sum_M \Sum_E (-1)^{f_M (E)}.
\end{equation}
Now, any given $M$ and $E$ will both act nontrivially on some set of $s$ 
qubits.  Of those $s$, they will act as different Pauli matrices on $t$ qubits 
and as the same Pauli matrix on $s-t$ qubits.  Now,
\begin{equation}
(-1)^{f_M (E)} = (-1)^t.
\end{equation}
The number of operators $E$ that agree with $M$ on $s-t$ qubits and 
disagree on $t$ qubits is
\begin{equation}
1^{s-t} 2^t 3^{d-s} \pmatrix{s \cr t} \pmatrix{d' \cr s} \pmatrix{n-d' \cr d-
s}.
\end{equation}
Note that this does not depend on $M$.  Thus,
\begin{eqnarray}
B_d & \!\! = & \!\! \frac{1}{2^{n-k}} \Sum_{d'=0}^{n} \Sum_M \Sum_{s=0}^{d} 
\Sum_{t=0}^{s} \left[1^{s-t} (-2)^t \pmatrix{s \cr t}\right] 3^{d-s} 
\pmatrix{d' \cr s} \pmatrix{n-d' \cr d-s} \\
& \!\! = & \!\! \frac{1}{2^{n-k}} \Sum_{d'=0}^{n} \Sum_M \Sum_{s=0}^{d} (1 - 
2)^s 3^{d-s} \pmatrix{d' \cr s} \pmatrix{n-d' \cr d-s} \\
& \!\! = & \!\! \frac{1}{2^{n-k}} \Sum_{d'=0}^{n} \Sum_M \Sum_{s=0}^{d} (-1)^s 
3^{d-s} \pmatrix{d' \cr s} \pmatrix{n-d' \cr d-s} \\
& \!\! = & \!\! \frac{1}{2^{n-k}} \Sum_{d'=0}^{n} \left[ \Sum_{s=0}^{d} (-1)^s 
3^{d-s} \pmatrix{d' \cr s} \pmatrix{n-d' \cr d-s} \right] A_{d'}.
\end{eqnarray}

This proves the quantum MacWilliams identity (\ref{eq-QMW}) for 
stabilizer codes.  The coefficients $A_d$ and $B_d$ can also be defined 
for non-stabilizer codes, and equation (\ref{eq-QMW}) will still hold, so 
any bounds derived strictly from the quantum MacWilliams identity will 
hold for any quantum code, not just stabilizer codes.  For any code of 
distance $d$, the coefficients $A_d$ and $B_d$ satisfy the additional 
constraints
\begin{eqnarray}
B_0 & = & A_0 = 1 \\
B_{d'} & = & A_{d'}\ (d' < d) \\
B_{d'} & \geq & A_{d'} \geq 0\ (\forall\,d').
\end{eqnarray}
For a nondegenerate code, $A_{d'} = B_{d'} = 0$ for $d' < d$.  These 
constraints along with equation (\ref{eq-QMW}) restrict the allowed values 
of $A_d$ and $B_d$.  The constraints are all linear, so standard linear 
programming techniques will find solutions.  If there are no possible 
integer values of $A_d$ and $B_d$ satisfying all of the constraints, there is 
no $[n, k, d]$ code.  Otherwise, the possible solutions will give us 
parameters of possible codes.  For instance, applying the constraints for a 
$[5, 1, 3]$ code produces the unique solution $A_i = (1, 0, 0, 0, 15, 0)$ and 
$B_i = (1, 0, 0, 30, 15, 18)$~\cite{shor-laflamme-QMW}.  Therefore, the 
usual five-qubit code is essentially the only $[5,1,3]$ code.  There are thus 
no degenerate five-qubit codes.

Even tighter linear programming bounds than those produced by the 
quantum MacWilliams identity are possible.  This can be done using the 
quantum shadow enumerator~\cite{rains-shadow}.  The {\em shadow} $Sh(S)$ of 
a code $S$ is defined as the set of $E \in \G$ satisfying
\begin{equation}
f_M (E) \equiv {\rm wt} (M) \pmod{2}
\end{equation}
for all $M \in S$ (where ${\rm wt} (M)$ is the weight of $M$).  Define 
$S_d$ to be the number of elements of $Sh(S)$ of weight $d$ (again, ignoring
overall phases), and
\begin{equation}
S (z) = \Sum_{d=0}^{n} S_d z^d.
\end{equation}
$S(z)$ is the {\em shadow enumerator} of $S$.  Then
\begin{equation}
S(z) = \frac{1}{2^{n-k}} (1+3z)^n A \left( \frac{z-1}{1+3z} \right).
\label{eq-shadow}
\end{equation}

If $S$ contains only operators of even weight, then $E \in Sh(S)$ iff $f_M (E) 
= 0$ for all $M \in S$, so $Sh(S) = N(S)$, and $S_d = B_d$.  Furthermore, in 
this case, $A(z)$ is an even function, so
\begin{eqnarray}
S(z) & = & B(z) = \frac{1}{2^{n-k}} (1+3z)^n A \left( \frac{1-z}{1+3z} \right) 
\\
& = & \frac{1}{2^{n-k}} (1+3z)^n A \left( \frac{z-1}{1+3z} \right).
\end{eqnarray}

If $S$ contains an element of odd weight, consider the subset $S' \subset 
S$ of even weight operators.  Then $S'$ has exactly $2^{n-k-1}$ elements.  
This is true because in order for $M, M' \in S$ to commute, they must overlap
and disagree only on an even number of qubits.  Thus, ${\rm wt}(MM') \equiv
{\rm wt}(M) + {\rm wt}(M') \pmod{2}$.
The shadow of $S$ is just $Sh(S) = N(S') - N(S)$.  Let $B'(z)$ and $A'(z)$ be 
the weight enumerators of $S'$ and $N(S')$.  Then
\begin{eqnarray}
S(z) & = & B' (z) - B(z) \\
& = & \frac{1}{2^{n-k-1}} (1+3z)^n A' \left( \frac{1-z}{1+3z} \right) - 
\frac{1}{2^{n-k}} (1+3z)^n A \left( \frac{1-z}{1+3z} \right) \nonumber \\ \\
& = & \frac{1}{2^{n-k}} (1+3z)^n \left[ 2 A' \left( \frac{1-z}{1+3z} \right) - 
A \left( \frac{1-z}{1+3z} \right) \right].
\end{eqnarray}
Now, $A'_d = A_d$ for even $d$ and $A'_d = 0$ for odd $d$, so $A(z) + A(-
z) = 2 A'(z)$, and
\begin{equation}
S(z) = \frac{1}{2^{n-k}} (1+3z)^n A \left( \frac{z-1}{1+3z} \right).
\end{equation}

Again, the shadow enumerator can be defined for non-stabilizer codes and 
satisfies the same relationship with $A(z)$ as for stabilizer codes.  In both 
the stabilizer and non-stabilizer case, $S_d \geq 0$.  Along with 
(\ref{eq-shadow}), this provides additional constraints for the linear 
programming bound restricting the parameters of any code.  These bounds 
have been applied to all possible codes with $n \leq 30$
\cite{rains-shadow,calderbank-GF4}.  Among other things, they show that 
the smallest possible distance five code is an $[11,1,5]$ code and that 
degenerate codes in this region all fall below the quantum Hamming 
bound.  The shadow enumerator can also be used to show that any nondegenerate
code on $n$ qubits can correct at most $\lfloor \frac{n+1}{6} \rfloor$
errors~\cite{rains-shadow}.

\section{Bounds on Degenerate Stabilizer Codes}

It is still unknown whether there are any degenerate codes that exceed the 
limits set by the quantum Hamming bound, but for certain restricted cases, 
we can show that there are not.  For codes using fewer than 30 qubits, the 
linear programming bounds of the previous section show this.  In this 
section, I will show that the statement also is true for all stabilizer codes 
that correct one or two errors.  The results can be extended slightly 
beyond stabilizer codes, but do not apply to the most general possible code.

For a one-error-correcting degenerate code, the stabilizer $S$ will contain 
one or more operators of weight one or two.  Weight one operators totally 
constrain a qubit and both the operator and the qubit can be eliminated, 
converting an $[n, k, d]$ code into an $[n-1, k, d]$.  If the latter satisfies 
the quantum Hamming bound, the former will as well.  Suppose there are $l$ 
independent weight two operators $M_1, \ldots, M_l$ in $S$.  Let $D$ be the 
group generated by $M_1, \ldots, M_l$.  Note that $S - D$ will contain no 
operators of weight less than three.  The weight two operators in $D$ tell us 
which errors produce the same states.  For instance, if $M_1 = \Zs{1} \Zs{2}$, 
$\Zs{1} \ket{\psi} = \Zs{2} \ket{\psi}$ for any codeword $\ket{\psi}$.

Any operator in $N(D)$ will take states fixed by $D$ to states fixed by $D$.  
The total dimensionality of the subspace fixed by $D$ is $2^{n-l}$.  Suppose 
that none of the operators in $D$ acts on some qubit $j$.  Then all of the 
three operators $\Xs{j}$, $\Ys{j}$, and $\Zs{j}$ are in $N(D)$, and they are
not degenerate.  Therefore, they must produce orthogonal states in the 
subspace fixed by $D$ for each basis codeword.  There are always at least 
$n-2l$ qubits not affected by $D$, since each generator of $D$ can add at 
most two qubits.  Therefore,
\begin{eqnarray}
\left[1 + 3(n-2l) \right] 2^k & \leq & 2^{n-l} \\
k & \leq & n - l - \log_2 [1+3(n-2l)]. \label{eq-QHB-deg1}
\end{eqnarray}
Recall that the quantum Hamming bound says that
\begin{equation}
k \leq n - \log_2 (1+3n),
\end{equation}
so (\ref{eq-QHB-deg1}) is more restrictive when
\begin{eqnarray}
l + \log_2 [1+3(n-2l)] & \geq & \log_2 (1+3n) \\
l & \geq & \log_2 \left[ \frac{1+3n}{1+3(n-2l)} \right] \\
& = & \log_2 \left[ 1 + \frac{6l}{1 + 3(n-2l)} \right]. \label{eq-QHB-deg1'}
\end{eqnarray}
Assuming $n \geq 2l$, we see that the quantum Hamming bound will still 
hold if $l \geq \log_2 (1+6l)$.  This is true for $l \geq 5$.  For $l=4$, 
(\ref{eq-QHB-deg1'}) holds for $n \geq 9$; for $l=3$, it holds for $n \geq 
7$.  For $l=2$, (\ref{eq-QHB-deg1'}) holds for $n \geq 5$, and for $l=1$, it 
holds for $n \geq 4$.  The remaining possibilities with $n \geq 2l$ are 
ruled out by the linear programming bounds of section
\ref{sec-enumerators}.  On the other hand, if $l > n/2$, then $k \leq n-l 
\leq n/2$.  For $n \geq 13$, the quantum Hamming bound is less 
restrictive than this, so in conjunction with the linear programming 
bounds, we can conclude that there are no distance three degenerate stabilizer 
codes that exceed the quantum Hamming bound.

We can make a similar argument for codes to correct two errors.  Now let $D$ 
be generated by the operators of weight four or less in $S$.  There must be 
at least $n-4l$ qubits that are unaffected by operators in $D$.  All the 
possible weight one and two errors on those qubits give orthogonal states, so
\begin{eqnarray}
\left[1 + 3(n-4l) + \frac{9}{2} (n-4l) (n-4l-1)\right] 2^k & \leq & 2^{n-l} \\
\left[1 - \frac{3}{2} n + \frac{9}{2} n^2 + 6 l (1 + 12 l - 6 n)\right] 2^l & \leq & 2^{n-k}.
\end{eqnarray}
The quantum Hamming bound will still hold if
\begin{eqnarray}
\left[1 - \frac{3}{2}n + \frac{9}{2} n^2 + 6 l (1 + 12 l - 6 n)\right] 2^l & 
\geq & 1 - \frac{3}{2}n + \frac{9}{2} n^2 \\
\left[ 1 - \frac{6l (6n - 12 l - 1)}{1 - 3n/2 + 9n^2/2} \right] 2^l & \geq & 1.
\label{eq-QHB-deg2}
\end{eqnarray}
Now, $l (6n - 12 l - 1) = -12 [l^2 - (6n-1) l /12]$ is maximized for $l = (6n-
1)/24$.  That means (\ref{eq-QHB-deg2}) will be satisfied when
\begin{eqnarray}
\left[ 1 - \frac{(6n - 1)^2}{8 - 12n + 36n^2} \right] 2^l & \geq & 1 \\
\frac{7}{8 - 12n + 36n^2}\,2^l & \geq & 1 \\
7 \cdot 2^{l-2} & \geq & 9n^2 - 3n + 2.
\end{eqnarray}
If this is true, the code will satisfy the quantum Hamming bound.  If it is 
{\em not} true, then
\begin{eqnarray}
l & \leq & 2 - \log_2 7 + \log_2 (9n^2 - 3n + 2) \\
& \leq & 3 + 2 \log_2 n.
\end{eqnarray}
Then $l (6n - 12l - 1) \leq 6 n l \leq 6 n (3 + 2 \log_2 n)$, so equation 
(\ref{eq-QHB-deg2}) will again be satisfied when
\begin{equation}
\left[ 1 - \frac{6 n (3 + 2 \log_2 n)}{1 - 3n/2 + 9n^2/2} \right] 2^l \geq 1.
\end{equation}
However, for $n \geq 30$,
\begin{equation}
\frac{6 n (3 + 2 \log_2 n)}{1 - 3n/2 + 9n^2/2} \leq 0.58,
\end{equation}
so (\ref{eq-QHB-deg2}) will be satisfied for any $l$ with $1 < l \leq n/4$ 
in the regime of interest.  When $l=1$, (\ref{eq-QHB-deg2}) becomes
\begin{equation}
1 - \frac{6 (6n - 13)}{1 - 3n/2 + 9n^2/2} \geq 1/2.
\end{equation}
However, for $n \geq 30$,
\begin{equation}
\frac{6 (6n - 13)}{1 - 3n/2 + 9n^2/2} \leq 0.26,
\end{equation}
so (\ref{eq-QHB-deg2}) is satisfied for $l=1$ as well.

Therefore, we are left with $l > n/4$.  Again, this implies that $k \leq n-l < 
3n/4$.  This is at least as restrictive than the quantum Hamming bound for $n 
\geq 52$.  For $n=31$, the quantum Hamming bound says $k \leq n-13$.  
Therefore, for $31 \leq n \leq 51$, the only remaining region of interest, 
the code must have $l \leq n/4 + 5$ to violate the quantum Hamming bound.  
The only possibility for $l > n/4 + 4$ is $l=12$, $n=31$.  Assume for the 
moment that $l \leq n/4 + 4$.  Then there are at least $n - 16$ qubits in 
the code that are affected by at most one of the generators of $D$.  This is 
more than $l+3$, so either at least two of the generators of $D$ must each
affect two qubits that are fixed by all of the other generators, or one generator fixes four qubits that are unaffected by all of the other generators. 
The second case will be more restrictive to the code than the first one, so I 
will assume the first case holds.  Assume without loss of generality that the 
two generators are $M_{l-1}$ and $M_l$.  Then errors on the four qubits 
affected only by these generators leave the codewords within the subspace fixed 
by $D'$, the group generated by $M_1, \ldots, M_{l-2}$.  There are 67 errors of 
weight zero, one and two on the four qubits, so
\begin{eqnarray}
67 \cdot 2^k & \leq & 2^{n-(l-2)} \\
k & \leq & n - l - 5.
\end{eqnarray}
This is at least as restrictive as the quantum Hamming bound for any $n$ 
between 31 and 51.

That leaves the case $l=12$, $n=31$.  Even in this case, there must be at 
least fourteen qubits that are affected by at most one of the generators of 
$D$.  As before, this is enough to ensure that we can pick two generators of 
$D$ that will together act on four qubits unaffected by any of the other 
generators.  Again, $k \leq n - l - 5$, which is more restrictive than the quantum Hamming bound.  Therefore, there are no two-error-correcting degenerate 
stabilizer codes exceeding the quantum Hamming bound.

The methods of this section could be adapted and perhaps applied to codes 
correcting three or more errors, but it gets more difficult for each 
additional error, since the cases with $l > n/(2t)$ must be treated on a 
special basis, and the range of $n$ for which this could violate the 
quantum Hamming bound grows rapidly with $t$.  Eventually, it might 
well be true that some code with enough degeneracies does violate the 
quantum Hamming bound.

Even though we cannot rule out the possibility of a sufficiently large 
degenerate code violating the quantum Hamming bound, we can still set
a less restrictive bound on degenerate stabilizer codes by constructing
a classical code from the quantum code~\cite{cleve-classical}.  Since bounds 
on the efficiencies of classical codes are known, we can therefore get
bounds on the possible parameters of quantum codes.

To produce a classical code from a quantum code, first put the code in
standard form, as per (\ref{eq-standard-form}).  In particular, note the
$r \times k$ matrix $A_2$.  $r \leq n-k$, but by performing single qubit
rotations from $N(\G)$, we can always convert one generator to the product of 
$\Z$'s, so we can ensure that $r \leq n-k-1$.  If we look at the classical code 
$C$ with $k \times (r+k)$ generator matrix $(A_2^T | I)$, then $C$ encodes
$k$ bits in at most $n-1$ bits.  If the original quantum code could correct
$t$ quantum errors, it turns out that the classical code $C$ can correct $t$
classical bit flip errors, whether the quantum code was degenerate or 
nondegenerate.  Therefore, the existence of an $[n, k, d]$ quantum code
implies that an $[n-1, k, d]$ classical code exists.

\section{Error-Correcting Codes and Entanglement Purification Protocols}

Before discussing bounds on the channel capacity, I will discuss another 
way of looking at quantum codes that is sometimes helpful for thinking 
about the channel capacity.  Consider the situation where Alice prepares a 
number of EPR pairs and sends one member of the pair to Bob.  In general, 
both the qubits that Alice keeps and the qubits she sends to Bob may be 
subject to errors and decoherence.  This means that Alice and Bob will 
share a number of imperfect pairs.  If Alice attempts to teleport a state 
using these imperfect EPR pairs, for instance, the state that Bob receives 
will be incorrect.  Alice and Bob wish to perform some local operations on 
their halves of the imperfect pairs so that they are left with a smaller 
number of perfect pairs (or at least better ones).  A protocol to do this is 
called an {\em entanglement purification protocol} (or EPP) 
\cite{bennett-tome,bennett-EPP}.

Depending on the situation, Bob and Alice may or may not be allowed to 
communicate with each other and perform operations conditioned on the 
results of measurements by the other one.  If both Bob and Alice can 
communicate with each other via classical communication channels, the 
possible protocols they can implement are called two-way error purification 
protocols (or 2-EPPs).  If Bob can only receive classical information (as well 
as qubits) from Alice, but not transmit, then Bob and Alice are restricted to 
using one-way error purification protocols (or 1-EPPs).  In principle, there is 
another possibility.  Bob and Alice might not be able to communicate 
classically at all.  However, it turns out that the protocols available for 
them in this case are equivalent to the 1-EPPs.  On the other hand, it is 
known that in some circumstances, 2-EPPs allow more good pairs to be 
purified than 1-EPPs do~\cite{bennett-tome}.

One remarkable fact about 1-EPPs is that they are equivalent to 
quantum error-correcting codes.  Suppose we have a quantum code.  We 
can make a 1-EPP out of it as follows: Alice encodes the qubits she is going 
to send to Bob using the code, then Bob corrects and decodes.  The encoded 
qubits that are thus preserved in the channel retain their entanglement 
with the qubits Alice kept, and thus form part of a good EPR pair.  The 
number of good pairs is just equal to the number of encoded qubits.

Conversely, suppose we have a 1-EPP that distills $k$ good pairs from $n$ 
noisy pairs and we wish to make a quantum code.  In this case Alice is the 
encoder and Bob is the decoder for the code.  Alice creates $n$ EPR pairs 
and sends them to Bob, then performs her half of the 1-EPP.  Since she 
cannot receive transmissions from Bob, she does not need to wait until Bob 
receives the qubits to do this.  This is why a quantum code is equivalent to 
a 1-EPP and not a 2-EPP.  After she has performed her half of the 
purification protocol, sending any necessary classical information, she 
takes the $k$ qubits she wishes to protect and performs her half of the 
teleportation protocol using her half of what will be the $k$ good pairs.  
Again, she sends the classical information about the measurement results 
to Bob.  Bob now receives the qubits, plus all the classical information.  He 
completes the purification protocol, purifying $k$ good pairs.  Since they 
are good EPR pairs, when he then completes the teleportation protocol, the 
resulting state is the correct one, and the whole process acts like a code 
encoding $k$ qubits in $n$ qubits.

\section{Capacity of the Erasure Channel}

Most quantum channels are very difficult to analyze.  However, the 
channel capacity is known for at least one simple channel of interest.  The 
{\em erasure channel} is the channel for which every qubit sent through the 
channel has some chance $p$ of being totally randomized.  However, when 
this happens, we always know on which qubit it occurred.  The capacity of 
the erasure channel for both quantum codes and 2-EPPs is straightforward 
to calculate~\cite{bennett-erasure}.

The capacity for 2-EPPs is particularly straightforward.  If Alice sends $n$ 
EPR pairs through the channel, $pn$ of them will be destroyed, but $(1-
p)n$ will remain intact.  Furthermore, Bob will know which pairs remain 
intact, so he tells Alice and they discard the useless pairs.  This achieves a 
rate of $1-p$.  Clearly, it is impossible to do better than this.  This means 
that the capacity for a 2-EPP is just $1-p$.

With a 1-EPP or quantum code, we cannot do as well, because Bob cannot 
tell Alice which pairs she should keep and which she should throw away.  In 
fact, we can set an upper bound on the capacity of $1-2p$.  Suppose the erasure 
rate of $p$ in the channel is actually caused by Charlie, who steals any 
given qubit with probability $p$, replaces any stolen qubits with random 
ones, and then tells Bob which qubits he stole.  When $p = 1/2$, Bob has 
exactly the same number of valid pairs as Charlie.  If there were any 
operations Alice could make without consulting Bob that enabled him to 
purify even a single valid pair, Charlie could do the same thing as Bob, also 
giving a valid pair.  Now when Alice attempts to teleport something to Bob, 
she is also teleporting it to Charlie.  This would allow the cloning of a 
quantum state.  Therefore, the rate for $p>1/2$ is zero.  For $p<1/2$, we 
can imagine Alice somehow knows $n(1-2p)$ of the pairs that will not be 
stolen by Charlie.  The remaining $2pn$ pairs she is uncertain about.  Of 
them, $pn$ will be stolen by Charlie, again leaving him with the same 
number of good pairs from this set as Bob has.  If Alice attempts to purify 
more than $n(1-2p)$ pairs with Bob, she will therefore also be purifying 
pairs with Charlie, again leading to state cloning.  Therefore, the capacity is 
bounded above by $1-2p$.

This is, in fact, the actual achievable capacity for this channel.  Suppose we 
take a random Abelian subgroup of $\G_n$ with $n-k$ generators.  This 
subgroup will act as the stabilizer $S$ of a code.  If we encode $k$ qubits 
using this code, and then send them through the erasure channel, for large 
$n$, with high probability, $pn$ known qubits will have been randomized.  
We need to distinguish between the $4^{pn}$ possible errors on these 
qubits.  Since the error operators are all on the same $pn$ qubits, there are 
again $4^{pn}$ products of these operators.  If measure one of these products
anticommute with some element of $S$, then we will be able to correct the 
errors and decode the $k$ qubits, with fidelity approaching one for large $n$.  
Since the generators are chosen randomly, each one will commute with half of 
the possible operators of weight $pn$ and anticommute with half of the possible 
operators.  The different generators commute and anticommute with operators 
independently, so the number of operators that commute with all $n-k$ 
generators is
\begin{equation}
4^{pn} / 2^{n-k} = 2^{k - (1-2p)n} = 2^{(r - 1 + 2p) n},
\end{equation}
where $r$ is the rate: $k = rn$.  As long as $r < 1 - 2p$, the chance of not 
being able to distinguish all the likely errors goes to zero as $n \rightarrow 
\infty$.  Therefore, a random stabilizer code can give us rate $1-2p$.  Since 
this coincides with the upper bound on the capacity, it is the actual 
capacity of the erasure channel.

\section{Capacity of the Depolarizing Channel}
\label{depolarizing}

The {\em depolarizing channel} is a very natural channel to consider.  In this 
channel, with probability $1-p$, each qubit is left alone.  In addition, there 
are equal probabilities $p/3$ that $\X$, $\Y$, or $\Z$ affects the qubit.  We 
can apply similar methods to the depolarizing channel as with the erasure 
channel to place upper and lower bounds on its capacity.  However, 
currently these bounds do not meet, so the actual capacity of the 
depolarizing channel is unknown.

The depolarizing channel can also simulated by imagining Charlie is 
randomly stealing some qubits from the channel.  If Charlie steals a qubit 
with probability $q$ and replaces it with a random qubit (not telling Bob 
which one was stolen), there is still a $1/4$ chance that Charlie happens to 
replace the stolen qubit with one in the same state.  There is only a chance 
$q/4$ of Charlie applying each of $\X$, $\Y$, and $\Z$.  Therefore, this 
situation corresponds to the depolarizing channel with $p = 3q/4$.  We can 
make a cloning argument just as with the erasure channel to set an upper 
bound on the capacity.  Again we find that the capacity is limited by $1-2q 
= 1- 8p/3$.  When $p > 3/8$, the rate of transmission is necessarily zero.

Actually, we can set a tighter upper bound than this.  Randomly stealing qubits 
is not the best eavesdropping method available to Charlie that will look like 
the depolarizing channel.  The best eavesdropping method actually allows 
him to produce the same state as Bob whenever $p > 1/4$~\cite{fuchs-KL}.  
This means that the rate is limited to $1-4p$.  This is the asymptotic form 
of the Knill-Laflamme bound, which was derived for codes with a fixed 
minimum distance in section~\ref{sec-gen-bounds}.

We can set a lower bound for the achievable rate by again considering the 
rate for a random stabilizer code.  If we encode $k$ qubits in $n$ qubits 
using a random stabilizer $S$, the expected number of errors is $pn$.  We 
need measure one of the errors to be distinguishable from each other.  The 
errors $E$ and $F$ are distinguishable if $E^\dagger F$ anticommutes with 
some elements of $S$, and are not if they do not.  The typical product 
$E^\dagger F$ actually does not have weight $2pn$.  There is a chance 
$p^2$ that $E$ and $F$ will both have nontrivial action on a given 
qubit.  If they act as different Pauli matrices, the product will still act on 
that qubit.  If they act as the same Pauli matrix, the product will not act on 
that qubit at all.  The probability of having both act as the same Pauli 
matrix is $p^2/3$.  Therefore, the expected length of the product 
$E^\dagger F$ is $(2p - 4p^2/3)n$.  Let $x = 2p - 4p^2/3$.

Let the number of errors of weight $w$ be $N(w)$.  Then the number of 
different products of weight $xn$ is $N(xn)$, and therefore the number of 
typical products that commute with everything in $S$ is $N(xn) / 2^{n-k}$.  
Now, there are $N(pn)$ likely errors, so the number of ways we can pair 
them into products is $N(pn) [N(pn) -1]/2$.  This means that the number of 
ways of getting any given operator $O$ of weight $xn$ is
\begin{equation}
\left. \pmatrix{ N(pn) \cr 2} \right/ N(xn).
\end{equation}
For each of the pairs that gives one of the $N(xn)/2^{n-k}$ products that 
commute with $S$, we must remove one of the errors in the pair from the 
group of likely errors.  Therefore, we must remove
\begin{equation}
\left. \pmatrix{ N(pn) \cr 2} \right/ 2^{n-k}
\end{equation}
errors.  We want to remove only measure zero of the errors, so we wish this 
number to be small compared to $N(pn)$ for large $n$.  Thus,
\begin{eqnarray}
N (pn) / 2^{n-k+1} & \ll & 1 \\
N (pn) & \ll & 2^{n-k+1} \\
k/n & < & 1 - \frac{1}{n} \log_2 N (pn) = 1 - p \log_2 3 - H(p).
\end{eqnarray}
This is just the quantum Hamming bound (\ref{eq-QHB}).  In other words, 
a random code saturates the quantum Hamming bound.

However, the quantum Hamming bound only limits the efficiency of 
nondegenerate codes.  The typical element of a random stabilizer will have 
weight $3n/4$, which is much larger than $pn$ for any $p$ where the rate 
could possibly be nonzero.  Therefore, a random code will have a 
negligable number of degenerate errors, and the quantum Hamming bound 
will still apply.  However, if we choose the stabilizer to be of a restricted 
form rather than totally random, we can choose it to have very many 
degeneracies, and the quantum Hamming bound may be exceeded
\cite{shor-smolin}, although existing codes only allow us to exceed the 
rate of a random code by a very small amount.  Shor and Smolin showed that
by concatenating a random code with a simple repetition code ($\ket{0}$
becomes the tensor product of $\ket{0}$'s and $\ket{1}$ becomes the tensor
product of $\ket{1}$'s), the rate of the code is improved slightly near the
zero-rate limit.  The optimum block size for repetition turns out to be five.

We can still set an upper bound on the efficiency of a degenerate stabilizer 
code using similar arguments to those that gave us the capacity of a 
random stabilizer code.  Note that this upper bound does not necessarily 
apply to all codes, so it may not be a strict upper bound on the capacity.  
However, non-stabilizer codes are very difficult to work with, so it does 
provide a practical upper bound on the capacity.

\begin{figure}
\epsfig{file=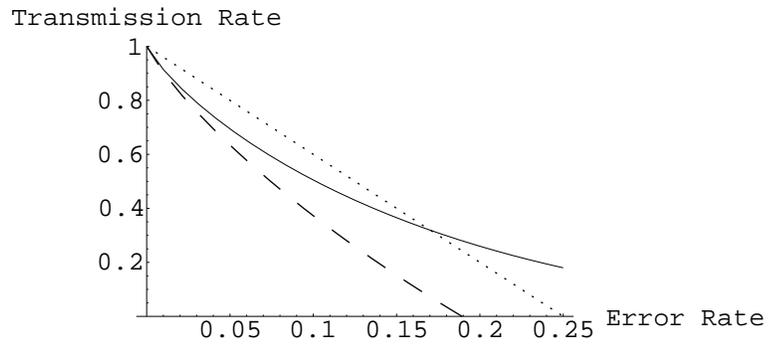}
\caption[The quantum Hamming bound, the Knill-Laflamme bound, and the bound
from equation~(\ref{eq-deg-bound})]{The quantum Hamming bound (dashed), the 
Knill-Laflamme bound (dotted), and the bound from equation~(\ref{eq-deg-bound}) 
(solid).}
\label{CCBounds}
\end{figure}

To give this bound, assume that every element of $S$ actually has weight 
$xn$.  This bound is unlikely to be achievable, since the product of two 
operators of weight $xn$ will only rarely have weight $xn$ again.  There 
are at least $N(xn)/2^{n-k}$ operators of weight $n$ that commute with $S$, 
but $2^{n-k}$ of them are in $S$.  Therefore, in the best case, there are only 
$N(xn)/2^{n-k} - 2^{n-k}$ operators that can potentially cause a problem.  In 
the limit where $n$ and $k = rn$ are both large, either $N(xn)/2^{n-k}$ will 
dominate the number of troublesome operators, or $N(xn)/2^{n-k} \ll 2^{n-
k}$.  In the first case, the calculation goes through as for a completely 
random stabilizer, giving us a capacity only at the quantum Hamming 
bound.  In the second case,
\begin{eqnarray}
N(xn) & \ll & 2^{2(n-k)} \\
r = k/n & < & 1 - \frac{1}{2n} \log_2 N(xn) = 1 - \frac{x}{2} \log_2 3 - 
\frac{1}{2} H(x).
\label{eq-deg-bound}
\end{eqnarray}
Since $x = 2p - 4p^2/3$, this is higher than the quantum Hamming bound.  
Equation (\ref{eq-deg-bound}) gives an upper bound on the capacity of the 
depolarizing channel achievable using stabilizer codes.  It is shown in
figure~\ref{CCBounds} along with the Knill-Laflamme bound and the quantum
Hamming bound.  Cleve has also proved a bound on the capacity achievable
using degenerate stabilizer codes \cite{cleve-classical}, but it is slightly
worse than (\ref{eq-deg-bound}) everywhere in the region of interest, so
it is not shown in the figure.

\chapter{Examples of Stabilizer Codes}
\label{chap-examples}

There are many known stabilizer codes \cite{shor-9qubit, steane-7qubit, 
bennett-tome, vaidman, grassl, leung, gottesman-stab, calderbank-stab, 
laflamme-5qubit, gottesman-pasting, calderbank-GF4, steane-8qubit, steane-RM,
calderbank-CSS, steane-CSS, knill-qudit, chau-d^2, chau-5qudit, 
rains-orthogonal, bennett-EPP}.  I will not attempt to list them all 
here, but will instead concentrate on a few interesting individual codes and 
classes of codes.  In a number of cases, I will not just describe the 
stabilizers of the codes, but will also discuss the normalizers and 
automorphism groups of the stabilizers, since these are important to realizing 
fault-tolerant computation in the most efficient possible way.

\section{Distance Two Codes}
\label{sec-dist2}

For even $n$, there is always an $[n, n-2, 2]$ code.  The stabilizer $S$ has 
two generators, one the product of all $n$ $\X$'s and one the product of all 
the $\Z$'s.  For even $n$, these commute.  $N(S)$ consists of tensor 
products in $\G$ that contain an even number of $\X$'s, an even number of 
$\Y$'s, and an even number of $\Z$'s.  We can write
\begin{eqnarray}
\Xbar_i & = & \Xs{1} \Xs{(i+1)} \\
\Zbar_i & = & \Zs{(i+1)} \Zs{n},
\end{eqnarray}
for $i = 1, \ldots, n-2$.

The automorphism group $\A(S)$ contains all possible 
permutations of the qubits and the Hadamard rotation $R$ applied to all 
$n$ qubits at once.  If $n$ is a multiple of four, any single-qubit operation 
in $N(\G)$ applied to all the qubits gives an element of $\A(S)$.  The order of 
$\A(S)$ is thus either $2 n!$ or $6 n!$.  Swapping qubit $i$ with qubit $j$ 
switches the $(i-1)$th encoded qubit with the $(j-1)$th encoded qubit (for 
$1 < i, j < n$).  Swapping qubit $1$ with qubit $i+1$ ($i =1, \ldots, n-2$) 
transforms
\begin{eqnarray}
\Xbar_i & \rightarrow & \Xbar_i \nonumber \\
\Xbar_j & \rightarrow & \Xbar_i \Xbar_j\ (i \neq j) \nonumber \\
\Zbar_i & \rightarrow & \Zbar_1 \Zbar_2 \cdots \Zbar_{n-2} \\
\Zbar_j & \rightarrow & \Zbar_j\ (i \neq j). \nonumber
\end{eqnarray}
Similarly, swapping qubit $n$ with qubit $i+1$ ($i = 1, \ldots, n-2$) 
transforms
\begin{eqnarray}
\Xbar_i & \rightarrow & \Xbar_1 \Xbar_2 \cdots \Xbar_{n-2} \nonumber \\
\Xbar_j & \rightarrow & \Xbar_j\ (i \neq j) \nonumber \\
\Zbar_i & \rightarrow & \Zbar_i \\
\Zbar_j & \rightarrow & \Zbar_i \Zbar_j\ (i \neq j). \nonumber
\end{eqnarray}
Swapping the first qubit with the $n$th qubit performs the transformation
\begin{eqnarray}
\Xbar_i & \rightarrow & \Xbar_1 \cdots \Xbar_{i-1} \Xbar_{i+1} \cdots 
\Xbar_{n-2} \nonumber \\
\Zbar_i & \rightarrow & \Zbar_1 \cdots \Zbar_{i-1} \Zbar_{i+1} \cdots 
\Zbar_{n-2}.
\end{eqnarray}
Performing $R$ on every qubit performs the same transformation as 
swapping the first and $n$th qubits, but also performs $R$ on every 
encoded qubit.  For $n$ a multiple of four, performing $P$ on every qubit 
performs the following operation:
\begin{eqnarray}
\Xbar_i & \rightarrow & - \Xbar_i \Zbar_1 \cdots \Zbar_{i-1} \Zbar_{i+1} 
\cdots \Zbar_{n-2} \nonumber \\
\Zbar_i & \rightarrow & \Zbar_i.
\end{eqnarray}
Because these codes are of the CSS form, a CNOT applied to every qubit 
transversally between two blocks is also a valid fault-tolerant operation, 
and performs CNOTs between the corresponding encoded qubits.  

The case of $n=4$, the smallest distance two code, is of particular interest.  
The code from figure~\ref{fig-droplast} can be converted into the form of
the codes currently under consideration using single-qubit rotations,
although the $\Xbar$ and $\Zbar$ operators will need to redefined.  It 
can be used to detect a single error~\cite{vaidman} or to correct a single 
erasure~\cite{grassl}.  In this case,
\begin{eqnarray}
\Xbar_1 & = & \Xs{1} \Xs{2} \nonumber \\
\Xbar_2 & = & \Xs{1} \Xs{3} \nonumber \\
\Zbar_1 & = & \Zs{2} \Zs{4} \\
\Zbar_2 & = & \Zs{3} \Zs{4}. \nonumber
\end{eqnarray}
Switching the second and third qubits or switching the first and fourth qubits 
both swap the two encoded qubits.  Swapping the first and second qubits or the 
third and fourth qubits produces the transformation
\begin{eqnarray}
\Xbar_1 & \rightarrow & \Xbar_1 \nonumber \\
\Xbar_2 & \rightarrow & \Xbar_1 \Xbar_2 \nonumber \\
\Zbar_1 & \rightarrow & \Zbar_1 \Zbar_2 \\
\Zbar_2 & \rightarrow & \Zbar_2. \nonumber
\end{eqnarray}
This is just a CNOT from the second encoded qubit to the first encoded 
qubit.  Similarly, swapping the first and third qubits or the second and fourth 
qubits performs a CNOT from the first encoded qubit to the second encoded 
qubit.  The transversal Hadamard rotation in this case performs the 
Hadamard rotations on both qubits and switches them.  Applying $P$ to all 
four qubits performs the gate
\begin{eqnarray}
\Xbar_1 & \rightarrow & - \Xbar_1 \Zbar_2 \nonumber \\
\Xbar_2 & \rightarrow & - \Zbar_1 \Xbar_2 \nonumber \\
\Zbar_1 & \rightarrow & \Zbar_1 \\
\Zbar_2 & \rightarrow & \Zbar_2. \nonumber
\end{eqnarray}
We can recognize this as the encoded conditional sign gate followed by an 
encoded $\Zs{1} \Zs{2}$.

A more extensive discussion of the properties of distance two codes (and a
few codes of greater distances) appears in \cite{rains-dist2}.

\section{The Five-Qubit Code}

The five-qubit code is the shortest possible quantum code to correct one 
error, and is therefore of immense interest~\cite{bennett-tome, 
laflamme-5qubit}.  Its stabilizer is given in table~\ref{table-5qubit}.  Recall 
that the stabilizer is simply generated by cyclic permutations of $\X 
\otimes \Z \otimes \Z \otimes \X \otimes I$.  There are five cyclic 
permutations of this, but only four produce independent generators.  The 
stabilizer has sixteen elements: the identity, and the $3 \times 5$ cyclic 
permutations of $\X \otimes \Z \otimes \Z \otimes \X \otimes I$, $\Y 
\otimes \X \otimes \X \otimes \Y \otimes I$, and $\Z \otimes \Y \otimes 
\Y \otimes \Z \otimes I$.  $\Xbar$ is just the tensor product of five $\X$'s 
and $\Zbar$ is the tensor product of the five $\Z$'s.

As I noted in section~\ref{sec-alternate}, the five-qubit code is a linear 
GF(4) code.  Therefore, the operation
\begin{equation}
T: \X \rightarrow \Y, \ \Z \rightarrow \X
\end{equation}
applied transversally is a valid fault-tolerant operation and performs an 
encoded version of itself.  We can use this operation to derive a valid 
three-qubit operation for the five-qubit code:
\begin{eqnarray}
\X \otimes I \otimes I & \rightarrow & \X \otimes \Y \otimes \Z 
\nonumber \\
I \otimes \X \otimes I & \rightarrow & \Y \otimes \X \otimes \Z 
\nonumber \\
I \otimes I \otimes \X & \rightarrow & \X \otimes \X \otimes \X \\
\Z \otimes I \otimes I & \rightarrow & \Z \otimes \X \otimes \Y \nonumber 
\\
I \otimes \Z \otimes I & \rightarrow & \X \otimes \Z \otimes \Y \nonumber 
\\
I \otimes I \otimes \Z & \rightarrow & \Z \otimes \Z \otimes \Z. \nonumber
\end{eqnarray}
We can, of course, permute the qubits on the right and apply $T$ or $T^2$ 
to any or all of them and still get a valid three-qubit operation.

Using measurements and this three-qubit operation, we can generate 
directly a number of additional one- and two-qubit operations.  We can 
always get such gates using the protocol described in section \ref{sec-4qubit}, 
but it may be more efficient to get some gates using this three-qubit 
operation.  Suppose we place the data qubit in the third place and prepare 
the first two qubits in encoded $\ket{0}$ states.  Then apply the 
three-qubit operation and measure $\Y$ on the first two qubits.  The effect is 
to perform a Hadamard rotation $R$ on the data qubit.  Alternatively, prepare 
the first two qubits in $+1$ eigenstates of $\X$, apply the three-qubit gate, 
and measure $\Z$ on the first two qubits.  This performs $P$ on the data qubit.
By preparing a single ancilla qubit, applying the three-qubit operation, and
making a single measurement, we can also get a variety of two-qubit operations.

\section{A Class of Distance Three Codes}
\label{sec-2toj}

The eight-qubit code of table~\ref{table-8qubit} is just one of a class of 
codes with parameters $[2^j, 2^j - j - 2, 3]$~\cite{gottesman-stab}.  Note 
that according the quantum Hamming bound, this is the maximal number 
of encoded qubits for $n=2^j$, $d=3$.  These codes are related to the 
classical Reed-Muller codes~\cite{steane-RM}, but are more efficient than 
CSS codes formed from the classical Reed-Muller codes.  Like the classical
Reed-Muller codes, the codes described in this section allow us to efficiently
compute the actual error occuring from the measured error syndrome.

The first two generators of these codes are always the same.  One is the 
product of $2^j$ $\X$'s and the second is the product of $2^j$ $\Z$'s.  We 
will call these generators $M_X$ and $M_Z$, and the remaining $j$ 
generators will be $M_1$ through $M_j$.  The stabilizers of these codes 
always include the distance two codes discussed in section~\ref{sec-dist2}.  
This is convenient when correcting errors --- we can measure the first two 
generators and use them to detect whether any error has occurred.  If not, 
we do not need to go any further.

It will be convenient to construct the codes by describing the error 
syndromes of the $3n$ possible one-qubit errors.  I will show that they 
are all distinct and then that the generators that give those error 
syndromes all commute.  For these codes, the error syndrome $f(E)$ for 
error $E$ is a $(j+2)$-bit number.  Recall that each bit corresponds to a 
generator of $S$, and the $i$th bit is $0$ iff $E$ commutes with generator 
$M_i$.  $f(E)$ is a group homomorphism from $\G$ to $\left({\bf Z}_2 
\right)^{j+2}$.

Because of the form of the first two generators, the first two bits of 
$f(\Xs{i})$ are always $01$, the first two bits of $f(\Zs{i})$ are always $10$, 
and the first two bits of $f(\Ys{i})$ are always $11$, as they must be to
preserve the group structure of $\G$.  For the remaining bits of the error 
syndrome, we will number the qubits from $0$ to $n-1$ and write the number in 
base two.  Then
\begin{eqnarray}
f(\Xs{i}) & = & 01 \oplus i \\
f(\Zs{i}) & = & 10 \oplus \sigma (i) \\
f(\Ys{i}) & = & 11 \oplus (i + \sigma (i)).
\end{eqnarray}
The function $\sigma(i)$ is some as yet undefined additive group 
automorphism on $\left( {\bf Z}_2 \right)^j$.  We will be able to completely 
describe it by defining its action on $0 \ldots 01$, $0 \ldots 010$, \ldots, 
$10 \ldots 0$.

For this to give a distance three code, the error syndrome must have the 
property that $f(E) \neq 0$ for any weight two operator $E \in \G$.  By 
including the stabilizer of a distance two code, we have already insured that 
any weight one operator has non-zero error syndrome.  We can immediately 
see that $f(E) \neq 0$ unless $E$ is the product of two Pauli 
matrices of the same type.  Therefore, we need to consider
\begin{eqnarray}
f(\Xs{l} \Xs{m}) & = & 00 \oplus (l+m) \\
f(\Zs{l} \Zs{m}) & = & 00 \oplus \sigma (l+m) \\
f(\Ys{l} \Ys{m}) & = & 00 \oplus (l+m) + \sigma (l+m),
\end{eqnarray}
for $l \neq m$.  The second and third equations follow because 
$\sigma$ is a group homomorphism.  Since $i = l+m$ can be anything but $0$, 
$\sigma (l+m)$ will not be $0$ either, and we need only choose $\sigma$ 
so that $\sigma (i) \neq i$ for any $i \neq 0$.

The actual function $\sigma$ we want to use will depend on whether $j$ is 
even or odd.  For even $j$, consider the following function $\sigma$:
\begin{eqnarray}
\sigma (0 \ldots 0 0 0 1) & = & 1 1 \ldots 1 1  \nonumber \\
\sigma (0 \ldots 0 0 1 0) & = & 0 \ldots 0 0 1 \nonumber \\
\sigma (0 \ldots 0 1 0 0) & = & 0 \ldots 0 1 0 \label{eq-sigma-even}\\
& \vdots & \nonumber \\
\sigma (1 0 0 0 \ldots 0) & = & 0 1 0 \ldots 0. \nonumber
\end{eqnarray}
Then clearly $\sigma (i) = i/2$ for any nonzero $i$ ending in $0$.  If $i$ 
does end in $1$, for $\sigma (i)$ to end in $1$ also, the previous bit must 
have been $0$, which means that the bit before that must have been $1$, 
and so on.  Therefore, the only possible number for which $i = \sigma (i)$ 
is $i = 0 1 0 \ldots 1 0 1$.  Because $j$ is even, the first bit must be $0$.  
But $\sigma(l)$ always begins in $1$ for any $l$ ending in $1$, so even for 
this particular $i$, $\sigma (i) \neq i$.  Therefore, the error syndrome 
produces a distance three code.  The smallest case is a $[16,10,3]$ code, which 
is given in table~\ref{table-16qubit}.
\begin{table}
\centering
{\setlength{\tabcolsep}{0.4em}
\begin{tabular}{c|cccccccccccccccc}
$M_X$ & $\X$ & $\X$ & $\X$ & $\X$ & $\X$ & $\X$ & $\X$ & $\X$ & $\X$ 
& $\X$ & $\X$ & $\X$ & $\X$ & $\X$ & $\X$ & $\X$ \\
$M_Z$ & $\Z$ & $\Z$ & $\Z$ & $\Z$ & $\Z$ & $\Z$ & $\Z$ & $\Z$ & $\Z$ & 
$\Z$ & $\Z$ & $\Z$ & $\Z$ & $\Z$ & $\Z$ & $\Z$ \\
$M_1$ & $I$ & $\X$ & $I$ & $\X$ & $I$ & $\X$ & $I$ & $\X$ & $\Z$ & 
$\Y$ & $\Z$ & $\Y$ & $\Z$ & $\Y$ & $\Z$ & $\Y$ \\
$M_2$ & $I$ & $\X$ & $I$ & $\X$ & $\Z$ & $\Y$ & $\Z$ & $\Y$ & $\X$ & 
$I$ & $\X$ & $I$ & $\Y$ & $\Z$ & $\Y$ & $\Z$ \\
$M_3$ & $I$ & $\X$ & $\Z$ & $\Y$ & $\X$ & $I$ & $\Y$ & $\Z$ & $I$ & 
$\X$ & $\Z$ & $\Y$ & $\X$ & $I$ & $\Y$ & $\Z$ \\
$M_4$ & $I$ & $\Y$ & $\X$ & $\Z$ & $I$ & $\Y$ & $\X$ & $\Z$ & $I$ & 
$\Y$ & $\X$ & $\Z$ & $I$ & $\Y$ & $\X$ & $\Z$
\end{tabular}
\caption{The stabilizer for a $[16,10,3]$ code.}
\label{table-16qubit}}
\end{table}

We do still need to verify that it is an actual code by verifying that 
there are commuting generators that give these error syndromes.  The 
first two generators $M_X$ and $M_Z$ will always commute with the 
other $j$ generators, since $f(\Xs{i})$ and $f(\Zs{i})$ each have a $0$ in the 
$r$th position for $n/2$ $i$'s and a $1$ in the $r$th position for $n/2$ $i$'s.
When the $r$th bit of $f(\Xs{i})$ is $0$  and the $r$th bit of $f(\Zs{i})$ is
$1$, then the $r$th generator is the tensor product of $\Xs{i}$ with something
else (thus, this generator commutes with $\Xs{i}$ and anticommutes with
$\Zs{i}$).  Other combinations will produce $I$, $\Ys{i}$, or $\Zs{i}$, and
we can determine the complete form of $M_r$ in this way.

We need only check that $M_r$ and $M_s$ commute.  Let $f_r (E)$ be the 
$(r+2)$th bit of $f(E)$, that is, the bit corresponding to $M_r$.  I assume 
without loss of generality that $s > r$.  The binary matrix representation of 
$S$ is closely related to the error syndrome, and $M_r$ and $M_s$ 
commute iff
\begin{equation}
\Sum_{i=0}^{n} \left( f_r (\Xs{i}) f_s (\Zs{i}) + f_r (\Zs{i}) f_s (\Xs{i}) 
\right) = 0.
\end{equation}
There are a few possible cases to consider:
\begin{itemize}
\item $j > s > r+1 > 2$: In this case, $f_s (\Zs{i})$ is equal to the sum of 
the $j$th bit of $i$ and the $(s-1)$th bit and $f_r (\Zs{i})$ is the sum of the 
$j$th bit of $i$ and the $(r-1)$th bit.  On the other hand, $f_r(\Xs{i})$ is
just equal to the $r$th bit of $i$ and $f_s(\Zs{i})$ is equal to the $s$th
bit of $i$.  The $j$th, $(r-1)$th, and $(s-1)$th bits are distinct from bits 
$r$ and $s$.  Therefore, the $f_r (\Xs{i}) f_s (\Zs{i})$ term contributes to 
the sum when the $r$th bit of $i$ is 1 and the $j$th and $(s-1)$th bits of
$i$ are different.  This is true for $n/4$ values of $i$.  The $f_r(\Zs{i})
f_s (\Xs{i})$ term similarly contributes to the sum for $n/4$ $i$'s.  Since 
$n/4 + n/4$ is even, $M_r$ and $M_s$ commute.

\item $j > s > r+1 = 2$: In this case, $f_s (\Zs{i})$ is still equal to the sum 
of the $j$th bit of $i$ and the $(s-1)$th bit, but $f_r (\Zs{i})$ is just equal 
to the $j$th bit of $i$.  However, both the $f_r (\Xs{i}) f_s (\Zs{i})$ and the 
$f_r (\Zs{i}) f_s (\Xs{i})$ terms still contribute to the sum for $n/4$ $i$'s, 
so $M_r$ and $M_s$ still commute.

\item $j = s > r+1 > 2$: Both $f_s (\Zs{i})$ and $f_r (\Zs{i})$ are given as in 
the first case.  $f_r (\Xs{i}) f_s (\Zs{i})$ still contributes to $n/4$ terms 
in the sum.  Now, however, $f_r (\Zs{i}) f_s (\Xs{i})$ can only contribute when 
the $j$th bit of $i$ is $1$.  Since we also need $f_r (\Zs{i}) = 1$, this term 
only contributes when the $j$th bit of $i$ is $1$ and the $(r-1)$th bit is $0$. 
This still contributes to $n/4$ terms in the sum, so $M_r$ and $M_s$ again 
commute.

\item $j > s = r+1 > 2$: Now, the $(s-1)$th bit is equal to the $r$th bit. That 
means $f_r (\Xs{i}) f_s (\Zs{i})$ only contributes when the $r$th bit of $i$ is 
$1$ and the $j$th bit of $i$ is $0$.  This contributes to $n/4$ terms in the 
sum, as does $f_r (\Zs{i}) f_s (\Xs{i})$, so $M_r$ and $M_s$ commute in this 
case as well.

\item $j = s = r+1 > 2$: This is a combination of the previous two cases.  
$f_r (\Xs{i}) f_s (\Zs{i})$ only contributes when the $r$th bit of $i$ is $1$ 
and the $j$th bit of $i$ is $0$ and $f_r (\Zs{i}) f_s (\Xs{i})$ contributes 
when the $j$th bit of $i$ is $1$ and the $(r-1)$th bit is $0$.  Again, this is 
an even number of contributing terms, so $M_r$ and $M_s$ commute.

\item $j > s = r+1 = 2$: $f_r (\Zs{i})$ is again equal to the $j$th bit of 
$i$.  However, this does not affect $f_r (\Xs{i}) f_s (\Zs{i})$, which
contributes to $n/4$ terms in the sum, as in the previous two cases.  It does
affect $f_r (\Zs{i}) f_s (\Xs{i})$, but this term still contributes to $n/4$
terms, so $M_r$ and $M_s$ commute.

\item $j = s > r+1 = 2$: As before, $f_r (\Xs{i}) f_s (\Zs{i})$ contributes to 
$n/4$ terms in the sum.  Now, however, $f_r (\Zs{i}) f_s (\Xs{i})$ contributes 
whenever the $j$th bit of $i$ is $1$.  This means it contributes to $n/2$ 
terms instead of $n/4$.  Therefore, there are a total of $3n/4$ contributing 
terms.  However, since $j \geq 3$, $n/4$ is still even, and $M_1$ and 
$M_j$ commute too.

\item $j = s = r+1 = 2$: Since $j \geq 3$, this case is impossible.
\end{itemize}

For the case of odd $j$, we do something very similar.  Now let
\begin{eqnarray}
\sigma (0 \ldots 0 0 0 1) & = & 1 1 \ldots 1 1  \nonumber \\
\sigma (0 \ldots 0 0 1 0) & = & 0 \ldots 0 0 1 \nonumber \\
\sigma (0 \ldots 0 1 0 0) & = & 0 \ldots 0 1 0 \label{eq-sigma-odd}\\
& \vdots & \nonumber \\
\sigma (0 1 0 0 \ldots 0) & = & 0 0 1 \ldots 0 \nonumber \\
\sigma (1 0 0 0 \ldots 0) & = & 1 0 1 \ldots 1. \nonumber
\end{eqnarray}
An example of a code using this $\sigma$ is the $[8,3,3]$ code given in 
table~\ref{table-8qubit}.  In this case, if the first bit is 0, the last bit
must also be 0 for the first bits of $i$ and $\sigma(i)$ to match.  However,
$\sigma (i)$ is certainly not equal to $i$ for any $i$ with both first and last 
bits $0$.  If the first bit is $1$, the last bit must be $0$ in order for the 
first bits of $i$ and $\sigma (i)$ to match. Thus, the second bit must be $0$, 
which means the third bit must be $1$, and so on.  However, since $j$ is odd, 
this progression would mean that the $j$th bit would have be $1$, while we 
already know it must be $0$.  
Therefore, there is no $i$ for which $\sigma (i) = i$.  Again, we have a 
distance three code.

We again need to check that the generators commute.  As for even $j$, 
everything immediately commutes with $M_X$ and $M_Z$.  We consider 
similar cases to see if $M_r$ and $M_s$ commute:
\begin{itemize}

\item $j > s > r+1 > 3$: Here, $f_r (\Zs{i})$ is the sum of the first, $j$th, 
and $(r-1)$th bits of $i$, and $f_s (\Zs{i})$ is the sum of the first, $j$th, 
and $(s-1)$th bits of $i$.  This still leads to both $f_r (\Xs{i}) f_s (\Zs{i})$
and $f_r (\Zs{i}) f_s (\Xs{i})$ contributing to $n/4$ terms each in the sum, so 
$M_r$ and $M_s$ commute.

\item $j > s > r+1 = 3$: Now $f_r (\Zs{i})$ is just equal to the $j$th bit of 
$i$, as in the case $j > s > r+1 = 2$ for even $j$.  As then, $M_r$ and $M_s$ 
commute.

\item $j > s > r+1 = 2$: Now $f_r (\Zs{i})$ is the sum of the first and $j$th 
bits of $i$, and $f_r (\Xs{i}) f_s (\Zs{i})$ contributes only when the first 
bit of $i$ is $1$ and the $(s-1)$th and $j$th bits of $i$ agree, but this still 
contributes to $n/4$ terms in the sum, so $M_r$ and $M_s$ still commute.

\item $j = s > r+1 > 3$: In this case, $f_r (\Zs{i}) f_s (\Xs{i})$ only 
contributes when the $j$th bit of $i$ is $1$ and the first and $(r-1)$th bits 
are the same.  This still occurs for $n/4$ $i$'s, so $M_r$ and $M_s$ commute.

\item $j > s = r+1 > 3$: Now, $f_r (\Xs{i}) f_s (\Zs{i})$ contributes when the 
$r$th bit of $i$ is $1$ and the first and $j$th bits are the same.  This 
occurs for $n/4$ $i$'s, so $M_r$ and $M_s$ commute.

\item $j = s = r+1 > 3$: $f_r (\Xs{i}) f_s (\Zs{i})$ contributes to $n/4$ terms
in the sum, as in the previous case, and $f_r (\Zs{i}) f_s (\Xs{i})$ does too, 
as in the case before that.  Therefore, $M_r$ and $M_s$ still commute.

\item $j > s = r+1 = 3$: As with the previous two cases, $f_r (\Xs{i}) f_s 
(\Zs{i})$ contributes to $n/4$ terms in the sum.  $f_r (\Zs{i})$ is equal to 
the $j$th bit of $i$, so $f_r (\Zs{i}) f_s (\Xs{i})$ contributes only when the 
$s$th and $j$th bits of $i$ are both $1$. This is still $n/4$ values of $i$, so 
$M_r$ and $M_s$ again commute.

\item $j > s = r+1 = 2$: In this case, $f_s (\Zs{i})$ is the $j$th bit of $i$ 
and $f_r (\Zs{i})$ is the sum of the first and $j$th bits.  That means $f_r 
(\Xs{i}) f_s (\Zs{i})$ contributes when the first and $j$th bits of $i$ are $1$,
and $f_r (\Zs{i}) f_s (\Xs{i})$ contributes when the second bit of $i$ is $1$ 
and the first and $j$th bits are different.  Both of these terms therefore 
contribute to $n/4$ terms in the sum, so $M_r$ and $M_s$ commute.

\item $j = s > r+1 = 3$: As usual, $f_r (\Xs{i}) f_s (\Zs{i})$ contributes to 
$n/4$ terms in the sum.  $f_r (\Zs{i}) f_s (\Xs{i})$ contributes whenever the 
$j$th bit of $i$ is $1$.  This means it contributes to $n/2$ terms in the sum, 
for a total of $3n/4$ nonzero terms.  Again, since $j \geq 3$, $3n/4$ is even, 
so $M_r$ and $M_s$ commute.

\item $j = s > r+1 = 2$: Now, $f_r (\Xs{i}) f_s (\Zs{i})$ contributes whenever 
the first bit of $i$ is $1$ and the $j$th and $(j-1)$th bits agree.  This is 
true for $n/4$ $i$'s. $f_r (\Zs{i}) f_s (\Xs{i})$ contributes when the first 
bit of $i$ is 0 and the $j$th bit of $i$ is $1$, which is again true for $n/4$ 
$i$'s.  Therefore, $M_r$ and $M_s$ commute.

\item $j = s = r+1 = 3$: This case only arises for the $[8, 3, 3]$ code, so we
can just check it by looking at table~\ref{table-8qubit}.  Again, the case
$j = s = r+1 = 2$ does not arise at all.

\end{itemize}

Now I will describe the $\Xbar$ and $\Zbar$ operators for these codes.  I 
will choose all of the $\Xbar$ operators to be of the form $\Xs{a} \Xs{i}$ (for 
some $i \neq a$) times the product of $\Z$'s.  In order to do this, we just 
need to find a set $K$ of $j+1$ $\Z$'s (not including $\Zs{a}$) for which 
$f(\Zs{l})$ over the $\Zs{l} \in K$ form a spanning set of binary vectors in 
$\left({\bf Z}_2 \right)^{j+1}$ (skipping $M_Z$, which $\Z$ will never 
anticommute with).  Then we will be able to pick some operator $E$ that is 
a product of these $\Z$'s so that $\Xbar_i = \Xs{a} \Xs{i'} E$ commutes with 
all the generators of $S$, and another operator $E'$ so that $\Zbar_i = 
\Zs{i'} E'$ also is in $N(S)$. If we choose the possible values of $i'$ so that 
they do not overlap with the qubits $l$ for which $\Zs{l} \in K$, then $\{ 
\Xbar_i, \Zbar_i \} = 0$ and $[\Xbar_i, \Zbar_m] = 0$ for $i \neq m$.

For even $j$, $K$ will consist of $\Zs{2^l}$ for $l = 1, \ldots, j-1$, plus 
$\Zs{0}$ and $\Zs{(n-1)}$ (recall the qubits are numbered $0$ to $n-1$).  $f 
(\Zs{0}) = 1 0 \oplus 0 \ldots 0$, $f (\Zs{(n-1)}) = 10 \oplus 1 0 \ldots 0$, 
and $f (\Zs{2^l})$ is $1 0$ followed by the binary representation of $2^{l-1}$. 
This set $K$ has the desired properties.  We pick $a=1$.

For odd $j$, $K$ will again include $\Zs{2^l}$, but only for $l = 1, \ldots j-
2$.  The remaining elements of $K$ will be $\Zs{0}$, $\Zs{(2^{(j-1)} + 1)}$, 
and $\Zs{(n-2)}$.  Now, $f(\Zs{(2^{(j-1)}+1)}) = 1 0 \oplus 0 1 0 \ldots 0$, 
and $f (\Zs{(n-2)}) = 1 0 \oplus 1 0 \ldots 0$, so again $K$ will have the 
desired property.  We again pick $a=1$.  Note that for the eight-qubit code, 
this will actually give us a different definition of $\Xbar_i$ and $\Zbar_i$ 
than in table~\ref{table-8qubit}.

I will conclude this section with a brief discussion of the automorphism 
groups of these codes.  There will not generally be a simple transversal 
operation in $\A(S)$ for one of these codes, but they have a number of 
symmetries when we allow permutations of the qubits.  One simple but 
large class of symmetries switches qubit $i$ with qubit $i + l$, where the 
addition is bitwise binary.  For instance, we might swap the first $n/2$ 
qubits with the last $n/2$ qubits, or the first $n/4$ qubits with the second 
$n/4$ and the third $n/4$ with the last $n/4$.  The effect of this swap is 
to add $1$ to any bit $r$ of $f(\Xs{i})$ (for all $i$) where $l$ is $1$ in the 
$r$th bit.  This much is equivalent to multiplying $M_r$ by $M_Z$.  We 
also add $1$ to any bit $r$ of $f(\Zs{i})$ (for all $i$) where $\sigma (l)$ is 
$1$ in the $r$th bit.  This is equivalent to multiplying $M_r$ by $M_X$.  
Whether we multiply by $M_X$, $M_Z$, or both, the product is still in $S$, 
so the operation preserves $S$ and is a valid fault-tolerant operation.  
There may be other symmetries of these codes, as well.

\section{Perfect One-Error-Correcting Codes}
\label{sec-perfect}

A {\em perfect} quantum code is a nondegenerate code for which the 
inequality of the quantum Hamming bound becomes an equality.  For
one-error-correcting codes, that means $(1+3n)2^k = 2^n$.  The possibility 
of a perfect code therefore exists whenever $1+3n$ is a power of two (up 
to $2^n$).  For instance, the five-qubit code is a perfect code.  $1+3n$ will 
be a power of two iff $n = (2^{2j} - 1)/3$ for some $j$.  Therefore there could 
be perfect codes for $n=5$, $n=21$, $n=85$, and so on, with parameters 
$[(2^{2j}-1)/3, (2^{2j}-1)/3 - 2j, 3]$.  In fact, perfect codes do exist for 
all these parameters.

One construction of these codes uses the Hamming codes over 
GF(4)~\cite{calderbank-GF4}.  Another construction is to paste together one 
of the codes from the previous section with an earlier perfect code.  The 
stabilizer $S_1$ of any code from section~\ref{sec-2toj} contains the 
stabilizer $R_1 = \{I, M_X, M_Z, M_X M_Z\}$ for a distance two code.  To 
make the perfect code for $j \geq 3$, let $S_1$ be the stabilizer for the 
$[2^{2j - 2}, 2^{2j - 2} - 2j, 3]$ code, and $S_2$ be the stabilizer for the
perfect code for $j-1$, with parameters $[(2^{2j-2}-1)/3, (2^{2j-2}-1)/3 - 2j + 
2, 3]$.  For $j=2$, $S_2$ is the stabilizer for the five-qubit code.  Then 
using trivial $R_2$ (which still has distance one), the pasting construction of 
section~\ref{sec-construction} gives us a new code of distance three.  The 
total number of qubits used by the code is 
\begin{equation}
2^{2j - 2} + (2^{2j-2}-1)/3 = (4\ 2^{2j-2} - 1)/3 = (2^{2j} - 1)/3.
\end{equation}
It encodes $(2^{2j}-1)/3 - 2j$ qubits, and therefore is the perfect code for 
$j$.

\section{A Class of Distance Four Codes}

We can extend the stabilizers of the codes from section~\ref{sec-2toj} to 
get distance four codes.  The parameters of these distance four codes will be 
$[2^j, 2^j - 2j - 2, 4]$.  The first two generators of $S$ will again be $M_X$ 
and $M_Z$.  The next $j$ generators of $S$ are the generators $M_1$ through 
$M_j$ from section \ref{sec-2toj}, so $S$ includes the stabilizer for a 
distance three code.  The last $j$ generators of $S$ are $N_i = R M_i R$ for 
$i = 1, \ldots, j$, where $R$ is applied to all $2^j$ qubits.  As with the
codes of section~\ref{sec-2toj}, the error occuring for these codes can
be efficiently determined from the error syndrome.

We can summarize this by writing the error syndromes for $\Xs{i}$ and 
$\Zs{i}$:
\begin{eqnarray}
f (\Xs{i}) & = & 01 \oplus i \oplus \sigma (i) \\
f (\Zs{i}) & = & 10 \oplus \sigma (i) \oplus i.
\end{eqnarray}
Since $S$ includes the stabilizer of a distance three code, it automatically 
has distance at least three.  We need to check that $f(E) \neq 0$ for any 
weight three operator $E$.  The only form of an operator $E$ for which the 
first two bits of $f(E)$ could be $00$ is $E = \Xs{a} \Ys{b} \Zs{c}$.  Then
\begin{eqnarray}
f (E) & = & 00 \oplus (a + \sigma (b) + b + \sigma (c) ) \oplus (\sigma (a) + 
b + \sigma (b) + c) \\
& = & 00 \oplus (a+b + \sigma (b+c) ) \oplus (b+c + \sigma (a+b) ).
\end{eqnarray}
If $r = a+b$ and $s = b+c$, then $f(E)$ is nonzero as long as $r \neq 
\sigma (s)$ or $s \neq \sigma (r)$.  This means that we need
\begin{equation}
s \neq \sigma (\sigma (s)) = \sigma^2 (s)
\end{equation}
for all nonzero $s$ (when $r=s=0$, $E=I$).  To see that this is true, note 
that for even $j$,
\begin{eqnarray}
\sigma^2 (0 \ldots 0 0 0 1) & = & 1 0 \ldots 0 0  \nonumber \\
\sigma^2 (0 \ldots 0 0 1 0) & = & 1 1 \ldots 1 1 \nonumber \\
\sigma^2 (0 \ldots 0 1 0 0) & = & 0 \ldots 0 0 1 \\
& \vdots & \nonumber \\
\sigma^2 (1 0 0 0 \ldots 0) & = & 0 0 1 \ldots 0. \nonumber
\end{eqnarray}
If $s$ has a $0$ in the next-to-last bit, it cannot have $\sigma^2 (s) = s$ 
unless $s=0$.  If $s$ has a $1$ in the next-to-last bit, it must have a $0$ for 
the fourth-from-the-last bit, and so on.  If $j$ is a multiple of four, we find 
that the first bit must be a $0$, which means that the last bit of $s$ must be 
a $1$.  This in turn implies that the third-from-the-last bit is $0$, and so on 
until we reach the second bit of $s$, which must be $0$, so $s = 0 0 1 1 0 0 
\ldots 1 1$.  However, the second bit of $\sigma^2 (s)$ is $1$ because the 
next-to-last bit is.  Therefore, $\sigma(s) \neq s$ in this case.  
If $j$ is even, but not a multiple of four, the first bit 
of $s$ must be $1$, which means that the last bit is $0$.  Again we follow the 
chain of logic back to the second bit of $s$ and again find that it must be 
$0$, again giving a contradiction.  Therefore $\sigma^2 (s) \neq s$ for any 
nonzero $s$ for any even $j$.  An example for even $j$ is the $[16,6,4]$ code 
given in table \ref{table-16qubit-dist4}.
\begin{table}
\centering
{\setlength{\tabcolsep}{0.4em}
\begin{tabular}{c|cccccccccccccccc}
$M_X$ & $\X$ & $\X$ & $\X$ & $\X$ & $\X$ & $\X$ & $\X$ & $\X$ & $\X$ 
& $\X$ & $\X$ & $\X$ & $\X$ & $\X$ & $\X$ & $\X$ \\
$M_Z$ & $\Z$ & $\Z$ & $\Z$ & $\Z$ & $\Z$ & $\Z$ & $\Z$ & $\Z$ & $\Z$ & 
$\Z$ & $\Z$ & $\Z$ & $\Z$ & $\Z$ & $\Z$ & $\Z$ \\
$M_1$ & $I$ & $\X$ & $I$ & $\X$ & $I$ & $\X$ & $I$ & $\X$ & $\Z$ & 
$\Y$ & $\Z$ & $\Y$ & $\Z$ & $\Y$ & $\Z$ & $\Y$ \\
$M_2$ & $I$ & $\X$ & $I$ & $\X$ & $\Z$ & $\Y$ & $\Z$ & $\Y$ & $\X$ & 
$I$ & $\X$ & $I$ & $\Y$ & $\Z$ & $\Y$ & $\Z$ \\
$M_3$ & $I$ & $\X$ & $\Z$ & $\Y$ & $\X$ & $I$ & $\Y$ & $\Z$ & $I$ & 
$\X$ & $\Z$ & $\Y$ & $\X$ & $I$ & $\Y$ & $\Z$ \\
$M_4$ & $I$ & $\Y$ & $\X$ & $\Z$ & $I$ & $\Y$ & $\X$ & $\Z$ & $I$ & 
$\Y$ & $\X$ & $\Z$ & $I$ & $\Y$ & $\X$ & $\Z$ \\
$N_1$ & $I$ & $\Z$ & $I$ & $\Z$ & $I$ & $\Z$ & $I$ & $\Z$ & $\X$ & $\Y$ 
& $\X$ & $\Y$ & $\X$ & $\Y$ & $\X$ & $\Y$ \\
$N_2$ & $I$ & $\Z$ & $I$ & $\Z$ & $\X$ & $\Y$ & $\X$ & $\Y$ & $\Z$ & 
$I$ & $\Z$ & $I$ & $\Y$ & $\X$ & $\Y$ & $\X$ \\
$N_3$ & $I$ & $\Z$ & $\X$ & $\Y$ & $\Z$ & $I$ & $\Y$ & $\X$ & $I$ & 
$\Z$ & $\X$ & $\Y$ & $\Z$ & $I$ & $\Y$ & $\X$ \\
$N_4$ & $I$ & $\Y$ & $\Z$ & $\X$ & $I$ & $\Y$ & $\Z$ & $\X$ & $I$ & 
$\Y$ & $\Z$ & $\X$ & $I$ & $\Y$ & $\Z$ & $\X$
\end{tabular}
\caption{The stabilizer for a $[16,6,4]$ code.}
\label{table-16qubit-dist4}}
\end{table}

If $j$ is odd,
\begin{eqnarray}
\sigma^2 (0 \ldots 0 0 0 1) & = & 0 1 1 1 \ldots 1 1  \nonumber \\
\sigma^2 (0 \ldots 0 0 1 0) & = & 1 1 1 1 \ldots 1 1 \nonumber \\
\sigma^2 (0 \ldots 0 1 0 0) & = & 0 0 0 \ldots 0 0 1 \nonumber \\
\sigma^2 (0 \ldots 1 0 0 0) & = & 0 0 0 \ldots 0 1 0 \\
& \vdots & \nonumber \\
\sigma^2 (0 1 0 \ldots 0 0) & = & 0 0 0 1 \ldots 0 0 \nonumber \\
\sigma^2 (1 0 0 0 \ldots 0) & = & 0 1 0 1 \ldots 1 1. \nonumber
\end{eqnarray}
In order to have $\sigma^2 (s) = s$, we cannot have the first bit and last 
two bits of $s$ all $0$.  If the first bit of $s$ is $1$, then the next-to-last 
bit of $s$ must also be $1$.  Then if the last bit is $0$, the
third-from-the-last bit must be $0$ and the fourth-from-the-last bit must 
be $1$.  Also, the second bit is $0$ and the third bit is $1$.  After the third 
bit, they must continue to alternate $0$ and $1$ until the next-to-last bit.  
This means odd numbered bits are $1$ and even numbered bits are $0$.  
However, the fourth-from-the-last bit is an even numbered bit, giving a 
contradiction.  Therefore, if the first bit of $s$ is $1$, the last two bits 
must both be $1$ also.  That means the third-from-the-last and
fourth-from-the-last bits must both be $0$.  However, it also means that 
the second bit of $s$ is $1$ and the third bit of $s$ is $0$.  The fourth bit 
is $0$ again, but the fifth bit is $1$, and after that they alternate until the 
last two bits.  This contradicts the fact that the third- and
fourth-from-the-last bits must both be $0$.

That leaves the possibility that the first bit of $s$ is $0$.  Then the
next-to-last bit is $0$ too, so the last bit must be $1$.  That means the 
third-from-the-last bit is $0$ and the fourth-from-the-last bit is $1$.  Also, 
the second and third bits of $s$ are both $1$.  The next two bits are both 
$0$, and the two after that are both $1$.  The bits pair up to be the same, 
with the pairs alternating between $0$ and $1$.  However, the fourth- and 
third-from-the-last bits form one of these pairs, and they are different, 
giving another contradiction.  Therefore, $\sigma^2 (s) \neq s$ for any 
nonzero $s$ for odd $j$ as well as for even $j$.  An example for odd $j$ is 
the $[8, 0, 4]$ code shown in table~\ref{table-8qubit-dist4}.
\begin{table}
\centering
\begin{tabular}{c|cccccccc}
$M_X$ & $\X$ & $\X$ & $\X$ & $\X$ & $\X$ & $\X$ & $\X$ & $\X$ \\
$M_Z$ & $\Z$ & $\Z$ & $\Z$ & $\Z$ & $\Z$ & $\Z$ & $\Z$ & $\Z$ \\
$M_1$ & $I$ & $\X$ & $I$ & $\X$ & $\Y$ & $\Z$ & $\Y$ & $\Z$ \\
$M_2$ & $I$ & $\X$ & $\Z$ & $\Y$ & $I$ & $\X$ & $\Z$ & $\Y$ \\
$M_3$ & $I$ & $\Y$ & $\X$ & $\Z$ & $\X$ & $\Z$ & $I$ & $\Y$ \\
$N_1$ & $I$ & $\Z$ & $I$ & $\Z$ & $\Y$ & $\X$ & $\Y$ & $\X$ \\
$N_2$ & $I$ & $\Z$ & $\X$ & $\Y$ & $I$ & $\Z$ & $\X$ & $\Y$ \\
$N_3$ & $I$ & $\Y$ & $\Z$ & $\X$ & $\Z$ & $\X$ & $I$ & $\Y$ \\
\end{tabular}
\caption{The stabilizer for the $[8, 0, 4]$ code.}
\label{table-8qubit-dist4}
\end{table}

To show that this set of generators forms the stabilizer for a code, we still 
have to show that they all commute.  From the fact that $M_r$ and $M_s$ 
commute with each other and $M_X$ and $M_Z$, we can immediately 
conclude that $N_r$ and $N_s$ commute with each other and $M_X$ and 
$M_Z$.  Also, $M_r$ and $N_r$ commute, since they get one sign of $-1$ 
for each $\X$ or $\Z$ in $M_r$, and there are an even number of $\X$'s and 
$\Z$'s.  We must show that $M_r$ commutes with $N_s$ for $r \neq s$.  Now,
\begin{equation}
f_{M_r} (N_s) = \Sum_{i=0}^{n-1} \left[i^{(r)} i^{(s)} + \sigma(i)^{(r)} 
\sigma(i)^{(s)} \right].
\end{equation}
Here, $x^{(r)}$ is the $r$th bit of $x$.  Now, $\sigma$ is a permutation of 
$0$ through $n-1$, so the second term in the sum is equal to the first term 
in the sum.  Therefore, the sum is automatically zero, and these generators 
do form a stabilizer.

\section{CSS Codes}
\label{sec-CSS}

As discussed in section~\ref{sec-stab-examples}, a CSS code 
\cite{calderbank-CSS,steane-CSS} is one where
some of the generators are tensor products of $\X$'s and the rest are tensor
products of $\Z$'s.  The $\X$ generators and the $\Z$ generators correspond 
to the parity check matrices of two classical codes $C_1$ and $C_2$, with
$C_1^{\perp} \subseteq C_2$.  For instance, the classical Reed-Muller codes
can be used to create a number of good quantum codes.  CSS codes cannot be
as efficient as the most general quantum code, but they can still be quite
good.  We can set upper and lower bounds using adaptations of the classical
Hamming bound and Gilbert-Varshamov bound.  This argument shows that the
rate $k/n$ of a CSS code to correct $t$ arbitrary errors is asymptotically 
limited by 
\begin{equation}
1 - 2 H(2t/n) \leq k/n \leq 1 - 2 H(t/n).
\end{equation} 

The CSS codes are a particularly interesting class of codes for two reasons: 
First, they are built using classical codes, which have been more heavily
studied than quantum codes, so it is fairly easy to construct useful quantum
codes simply by looking at lists of classical codes.  Second, because of
the form of the generators, the CSS codes are precisely those for which a
CNOT applied between every pair of corresponding qubits in two blocks performs
a valid fault-tolerant operation (see section~\ref{sec-normalizer}).  This
makes them particularly good candidates for fault-tolerant computation.

In order to get universal fault-tolerant computation for a code, the first
step is to produce the encoded CNOT for the code.  For the most general
stabilizer code, this requires performing a four-qubit operation using
two ancilla qubits and making two measurements.  In a CSS code, this process
is reduced to a single transversal operation.  Next, in order to produce
one-qubit operations, we need to use one ancilla qubit, perform a CNOT, and
make a measurement.  For the most general CSS code, we will still have to
do this.  However, if the code has the property that $C_1 = C_2$ (so
$C_1^{\perp} \subseteq C_1$), then the $\X$ generators have the same form
as the $\Z$ generators, so a transversal Hadamard rotation is also a valid
fault-tolerant operation.  If we further have the property that the parity 
check matrix of $C_1$ has a multiple of four 1s in each row, then the 
transversal phase $P$ is a valid fault-tolerant operation too.  For a general 
CSS code satisfying these conditions, these operations will perform some 
multiple-qubit gate on the qubits encoded in a single block.  However, if each 
block only encodes a single qubit, we can choose the $\Xbar$ and $\Zbar$ 
operators so that transversal Hadamard performs an encoded Hadamard rotation, 
and so that the transversal $P$ performs an encoded $P$ or $P^\dagger$.  In
particular, when $C_1$ is a punctured doubly-even self-dual classical code,
all these conditions are satisfied, and we can perform any operation in
$N(\G)$ by performing a single transversal operation \cite{shor-fault-tol}.  
In order to get universal computation, we will also need the Toffoli gate or 
some other gate outside $N(\G)$, and this will almost always require a more 
complicated construction.

\section{Amplitude Damping Codes}

Suppose we restrict attention to the amplitude damping channel.  In this
channel, each qubit behaves independently according to one of the following
matrices:
\begin{equation}
\pmatrix{1 & \ 0 \cr 0 & \sqrt{1-\epsilon^2}} {\rm\ or\ }
\pmatrix{0 & \epsilon \cr 0 & 0}.
\end{equation}
It is difficult to create efficient codes that will deal with the exact 
evolution produced by this channel.  However, when $\epsilon$ is fairly small, 
it is sufficient to merely satisfy equation (\ref{eq-condition}) approximately
\cite{leung}.  If we wish to correct the equivalent of one error, corrections
of $O(\epsilon^3)$ will not matter, since that would be equivalent to
distinguishing one error from two errors.  Let us expand
\begin{equation}
\pmatrix{1 & \ 0 \cr 0 & \sqrt{1-\epsilon^2}} = 
I - \frac{1}{4} \epsilon^2 (I - \Z) + O(\epsilon^4).
\end{equation}
All of the higher order corrections to this equation will be powers of
$I - \Z$.  Therefore, if we let
\begin{equation}
A = \X (I - \Z) = \frac{2}{\epsilon} \pmatrix{0 & \epsilon \cr 0 & 0},
\end{equation}
and
\begin{equation}
B = I - \Z,
\end{equation}
we need to consider all terms of the form
\begin{equation}
\bra{\psi_i} E^\dagger F \ket{\psi_j},
\end{equation}
where $E$ and $F$ are products of $A$ and $B$.  We get one factor of $\epsilon$
for each $A$ and one factor of $\epsilon^2$ for each $B$.  We only need to
consider those terms that have total order less than $\epsilon^d$ to have an
effectively distance $d$ code.  This corrects $t$ errors where $d= 2t+1$.

One possible way to achieve this is to have a CSS code for which the $\Z$
generators can {\em correct} $t$ $\X$ errors and the $\X$ generators can 
{\em detect} $t$ $\Z$ errors.  For instance, the code given in table 
\ref{table-XZcode} will work if we first map $\Z \rightarrow \X$ and $\Y 
\rightarrow \Z$.  For such a code, we are correcting $I$ and $\Z$ rather than 
$B$.  Since $B$ is in the linear span of $\Z$ and the identity, it is handled 
by these codes as well.

We can expand the range of possible codes by taking the actual linear 
combination of $I$ and $\Z$ that appears in $A$ and $B$ into account.  For
instance, consider the code from table~\ref{table-amplitude} \cite{leung}.
\begin{table}
\centering
\begin{tabular}{c|cccc}
$M_1$ & $\X$ & $\X$ & $\X$ & $\X$ \\
$M_2$ & $\Z$ & $\Z$ & $I$ & $I$ \\
$M_3$ & $I$ & $I$ & $\Z$ & $\Z$ \\
\hline
\low{$\Xbar$} & \low{$\X$} & \low{$\X$} & \low{$I$} & \low{$I$} \\
\low{$\Zbar$} & \low{$\Z$} & \low{$I$} & \low{$\Z$} & \low{$I$}
\end{tabular}
\caption{A four-qubit code for the amplitude damping channel.}
\label{table-amplitude}
\end{table}
This code can correct one amplitude damping error (i.e., it satisfies
(\ref{eq-condition}) to $O(\epsilon^3)$).  We can instantly see that
(\ref{eq-condition}) is satisfied for $E^\dagger F = A_i$ (the subscript
indicates the affected qubit) or $E^\dagger F = A_i^\dagger A_j$, where
$(i, j) \neq (1, 2), (3, 4)$.  When $(i, j) = (1,2)$ (or $(3,4)$), something
interesting and unusual happens:
\begin{eqnarray}
\bra{\psi_i} A_1^\dagger A_2 \ket{\psi_j} & = & 
\bra{\psi_i} (I - \Zs{1}) \Xs{1} \Xs{2} (I - \Zs{2}) \ket{\psi_j} \\
& = & \bra{\psi_i} \Xs{1} \Xs{2} (I + \Zs{1})\,(I - \Zs{2}) \ket{\psi_j}.
\end{eqnarray}
Now, $\Zs{1} \Zs{2} \ket{\psi_j} = \ket{\psi_j}$, so
\begin{eqnarray}
\bra{\psi_i} \Xs{1} \Xs{2} (I + \Zs{1})\,(I - \Zs{2}) \ket{\psi_j}
& = & \bra{\psi_i} \Xs{1} \Xs{2} (I + \Zs{1})\,(I - \Zs{1}) \ket{\psi_j} 
\nonumber \\
\\
& = & 0,
\end{eqnarray}
since $(I + \Zs{1})\,(I - \Zs{1}) = 0$.  We also need to consider the terms
$E^\dagger F = B$ and $E^\dagger F = A_i^\dagger A_i = I - \Zs{i} = B$.  In
this case, we can again separate $B$ into $I$ and $\Z$, and the latter is
handled by the generator $M_1$.

By applying similar principles, we can see that Shor's nine-qubit code (table
\ref{table-9qubit}) can be used to correct two amplitude damping errors.
We need to consider products of one through four $A$'s and products of one
or two $B$'s, as well as the product of a $B$ with one or two $A$'s.  Shor's
code breaks down into three blocks of three.  If for any block of three,
we have one or two $A$'s acting on that block, $E^\dagger F$ will anticommute
with one of the $\Z$ generators for that block, and $\bra{\psi_i} E^\dagger
F \ket{\psi_j} = 0$.  This takes care of all possible operators $E^\dagger F$
involving one, two, or four $A$'s.  We still need to consider $A_1^\dagger A_2
A_3$ (and similar terms) and products of one or two $B$'s.  The products of 
$B$'s we again expand into $I$ and $\Z$, producing products of zero, one, and 
two $\Z$'s.  Operators with one $\Z$ or with two $\Z$'s in different blocks of 
three will anticommute with one of the $\X$ operators.  Operators such as 
$\Zs{1} \Zs{2}$ that act on two qubits in the same block of three are in the 
stabilizer and are thus equivalent to the identity.  Finally, operators such as
$A_1^\dagger A_2 A_3$ are dealt with similarly to $A_1^\dagger A_2$ for the
four qubit code above:
\begin{eqnarray}
\bra{\psi_i} A_1^\dagger A_2 A_3 \ket{\psi_j} & \!\! = &
\!\! \bra{\psi_i} (I - \Zs{1}) \Xs{1} \Xs{2} (I - \Zs{2}) \Xs{3} (I - \Zs{3}) 
\ket{\psi_j} \\
& \!\! = & \!\! \bra{\psi_i} \Xs{1} \Xs{2} \Xs{3} (I + \Zs{1})\,(I - \Zs{2})\,
(I - \Zs{3}) \ket{\psi_j} \\
& \!\! = & \!\! \bra{\psi_i} \Xs{1} \Xs{2} \Xs{3} (I + \Zs{1})\,(I - \Zs{1})\,
(I - \Zs{3}) \ket{\psi_j} \\
& \!\! = & \!\! 0.
\end{eqnarray}
Thus, the nine qubit code can correct two amplitude damping errors.

Fault tolerance for these codes must be handled carefully.  Transversal
operations of any sort will not respect the form of the error operators, so
we need to be sure the code will be able to correct the new error operators.
For instance, the CNOT applied to $I \otimes A$ produces $(I \otimes \X)\,
(I \otimes I - \Z \otimes \Z)$.  This cannot be written as the tensor product
of $A$'s and $B$'s.  However, $I \otimes A_i$ is still distinguishable from the 
images of $I \otimes A_j$ (since $(I \otimes I + \Z \otimes \Z)\,(I \otimes I
- \Z \otimes \Z) = 0$) and $A_j \otimes I$.  Therefore, transversal CNOT is
a valid fault-tolerant operation for the four-qubit code as long as we correct 
errors taking its effects into account.

\section{Some Miscellaneous Codes}

In this section I present a few more codes that do not fit easily
into any of the classes I have already discussed.  Figure~\ref{table-11qubit}
shows an $[11,1,5]$ code, the smallest code to correct two errors
\cite{calderbank-GF4}.  Figure \ref{table-XZcode} gives a code that can 
correct one $\X$ error or one $\Z$ error, but not a $\Y$ error.  This code is 
better than any possible distance three code, and is another example 
illustrating the utility of stabilizer codes for more general channels than the 
depolarizing channel.  It is based on the classical Hamming code with an 
additional generator to distinguish between $\X$ and $\Z$ errors.  In fact, 
this code also detects if a $\Y$ error has occurred, although it cannot tell us 
where the error occurred.

\begin{table}
\centering
\begin{tabular}{c|ccccccccccc}
$M_1$ & $\Z$ & $\Z$ & $\Z$ & $\Z$ & $\Z$ & $\Z$ & $I$ & $I$ & $I$ & $I$ & $I$ 
\\
$M_2$ & $\X$ & $\X$ & $\X$ & $\X$ & $\X$ & $\X$ & $I$ & $I$ & $I$ & $I$ & $I$ 
\\
$M_3$ & $I$ & $I$ & $I$ & $\Z$ & $\X$ & $\Y$ & $\Y$ & $\Y$ & $\Y$ & $\X$ & $\Z$
\\
$M_4$ & $I$ & $I$ & $I$ & $\X$ & $\Y$ & $\Z$ & $\Z$ & $\Z$ & $\Z$ & $\Y$ & $\X$
\\
$M_5$ & $\Z$ & $\Y$ & $\X$ & $I$ & $I$ & $I$ & $\Z$ & $\Y$ & $\X$ & $I$ & $I$ 
\\
$M_6$ & $\X$ & $\Z$ & $\Y$ & $I$ & $I$ & $I$ & $\X$ & $\Z$ & $\Y$ & $I$ & $I$ 
\\
$M_7$ & $I$ & $I$ & $I$ & $\Z$ & $\Y$ & $\X$ & $\X$ & $\Y$ & $\Z$ & $I$ & $I$
\\
$M_8$ & $I$ & $I$ & $I$ & $\X$ & $\Z$ & $\Y$ & $\Z$ & $\X$ & $\Y$ & $I$ & $I$ 
\\
$M_9$ & $\Z$ & $\X$ & $\Y$ & $I$ & $I$ & $I$ & $\Z$ & $\Z$ & $\Z$ & $\X$ & $\Y$
\\
$M_{10}$ & $\Y$ & $\Z$ & $\X$ & $I$ & $I$ & $I$ & $\Y$ & $\Y$ & $\Y$ & $\Z$ & 
$\X$ \\
\hline
\low{$\Xbar$} & \low{$I$} & \low{$I$} & \low{$I$} & \low{$I$} & \low{$I$} & 
\low{$I$} & \low{$\X$} & \low{$\X$} & \low{$\X$} & \low{$\X$} & \low{$\X$} \\
\low{$\Zbar$} & \low{$I$} & \low{$I$} & \low{$I$} & \low{$I$} & \low{$I$} & 
\low{$I$} & \low{$\Z$} & \low{$\Z$} & \low{$\Z$} & \low{$\Z$} & \low{$\Z$} 
\end{tabular}
\caption{The stabilizer for an $[11,1,5]$ code.}
\label{table-11qubit}
\end{table}

\begin{table}
\centering
\begin{tabular}{c|ccccccc}
$M_1$ & $\Z$ & $\Z$ & $\Z$ & $\Z$ & $\Z$ & $\Z$ & $\Z$ \\
$M_2$ & $\Y$ & $\Y$ & $\Y$ & $\Y$ & $I$ & $I$ & $I$ \\
$M_3$ & $\Y$ & $\Y$ & $I$ & $I$ & $\Y$ & $\Y$ & $I$ \\
$M_4$ & $\Y$ & $I$ & $\Y$ & $I$ & $\Y$ & $I$ & $\Y$ \\
\hline
\low{$\Xbar_1$} & \low{$\X$} & \low{$\X$} & \low{$I$} & \low{$I$} & \low{$I$} & 
\low{$I$} & \low{$\Z$} \\
\low{$\Xbar_2$} & \low{$\X$} & \low{$I$} & \low{$\X$} & \low{$I$} & \low{$I$} &
\low{$\Z$} & \low{$I$} \\
\low{$\Xbar_3$} & \low{$\X$} & \low{$I$} & \low{$I$} & \low{$\Z$} & \low{$\X$} &
\low{$I$} & \low{$I$} \\
\low{$\Zbar_1$} & \low{$I$} & \low{$\Z$} & \low{$I$} & \low{$\Z$} & \low{$I$} &
\low{$\Z$} & \low{$I$} \\
\low{$\Zbar_2$} & \low{$I$} & \low{$I$} & \low{$\Z$} & \low{$\Z$} & \low{$I$} &
\low{$I$} & \low{$\Z$} \\
\low{$\Zbar_3$} & \low{$I$} & \low{$I$} & \low{$I$} & \low{$I$} & \low{$\Z$} & 
\low{$\Z$} & \low{$\Z$}
\end{tabular}
\caption{The stabilizer for a code to correct one $\X$ or $\Z$ error.}
\label{table-XZcode}
\end{table}

The set of all possible codes includes many codes that are not equivalent
to stabilizer codes.  Currently, however, only one is known that is better
than any stabilizer code~\cite{rains-nonstab}.  This code has distance two
and encodes six states using five qubits, whereas any distance two stabilizer 
code could only encode two qubits (four states) with five qubits.  It can be 
given in terms of the projection $P$ onto the subspace of valid codewords:
\begin{eqnarray}
\lefteqn{P = 1/16\ [ 3\ I \otimes I \otimes I \otimes I \otimes I
+ (I \otimes \Z \otimes \Y \otimes \Y \otimes \Z)_{\rm cyc}} \nonumber \\
& & \mbox{} + (I \otimes \X \otimes \Z \otimes \Z \otimes \X)_{\rm cyc} 
- (I \otimes \Y \otimes \X \otimes \X \otimes \Y)_{\rm cyc} \\
& & \mbox{} + 2\ (\Z \otimes \X \otimes \Y \otimes \Y \otimes \X)_{\rm cyc}
- 2\ \Z \otimes \Z \otimes \Z \otimes \Z \otimes \Z ]. \nonumber
\end{eqnarray}
The subscript ``cyc'' means that we actually add the five cyclic permutations
of the indicated term.  Note that this means the projection operator, and
therefore the code, is itself cyclic.  The trace of $P$ is six, so $P$ projects
onto a six-dimensional space and the code can therefore be used to encode
six basis states.  Conjugation of $P$ by $\X$, $\Y$, or $\Z$ on any single 
qubit will produce $P'$ with $P P' = 0$, so the code for this projection 
operator satisfies (\ref{eq-condition}) for a distance two code, with $C_{ab} = 
\delta_{ab}$.

\appendix

\chapter{Quantum Gates}
\label{app-gates}

It is usually helpful to think of a quantum computer as performing a series
of {\em gates}, drawn from some fairly small basic set of physically
implementable unitary transformations.  The net transformation applied
to the quantum computer is the product of the unitary transformations
associated with the gates performed.  In order to have a universal quantum
computer, it should be possible to get arbitrarily close to any unitary
transformation.  This property makes no guarantees about how many gates
are required to get within $\epsilon$ of the desired unitary operation,
and figuring out how to get a given operator with the minimum
number of basic gates is the goal of quantum algorithm design.

There are a number of known sets of universal quantum gates 
\cite{lloyd-universal, gates}.  For instance, all single-qubit unitary
operators and the controlled-NOT together comprise a universal set.  The
controlled-NOT gate (or CNOT) is a two-qubit operator that flips the second 
qubit iff the first qubit is $\ket{1}$.  It has the matrix
\begin{equation}
\pmatrix{1 & 0 & 0 & 0 \cr 0 & 1 & 0 & 0 \cr 0 & 0 & 0 & 1 \cr 0 & 0 & 1 & 0}.
\end{equation}
In fact, the controlled-NOT and one single-qubit operator are sufficient,
as long as the the single-qubit rotation acts by an angle incommensurate
with $2 \pi$.  Another finite universal set of quantum gates consists of the
Hadamard rotation $R$,
\begin{equation}
R = \frac{1}{\sqrt{2}} \pmatrix{1 & \ 1 \cr 1 & -1},
\end{equation}
the phase gate $P$,
\begin{equation}
P = \pmatrix{1 & 0 \cr 0 & i},
\end{equation}
the controlled-NOT, and the Toffoli gate, which is a three-qubit gate which
flips the third qubit iff the first two qubits are in the state $\ket{11}$.

In addition to the gates mentioned above, I refer to a number of other simple
gates in this thesis.  For instance, the simple NOT gate, the sign gate, and
the combined bit and sign flip gate (which are equal to $\X$, $\Z$, and $\Y$,
respectively) play a crucial role in the stabilizer formalism.  I also
refer to two other single-qubit gates related to $P$ and $R$.  They are
\begin{equation}
Q = \frac{1}{\sqrt{2}} \pmatrix{\ 1 & \ i \cr -i & -1},
\end{equation}
and
\begin{equation}
T = \frac{1}{\sqrt{2}} \pmatrix{ 1 & -i \cr 1 & \ i }.
\end{equation}
I also occasionally refer to the ``conditional sign'' gate, which is a 
two-qubit gate that gives the basis state $\ket{11}$ a sign of $-1$ and
leaves the other three basis states alone.  The conditional sign gate is
equivalent to the controlled-NOT via conjugation of one qubit by $R$.  The
conditional sign gate is effectively a controlled-$\Z$ gate, where $\Z$
gets applied to one qubit iff the other qubit is $\ket{1}$.  I also use
an analogous controlled-$\Y$ operator.  The CNOT is the controlled-$\X$.

To describe a series of gates, it is usually helpful to draw a diagram
of the gate array.  Horizontal lines represent the qubits of the quantum
computer, which enter at the left and leave from the right.  A summary
of the symbols I use for the various gates is given in figure
\ref{fig-gates}.
\begin{figure}
\centering
\begin{picture}(320, 160)

\put(35,150){\line(1,0){20}}
\put(45,150){\circle{8}}
\put(45,146){\line(0,1){8}}
\put(55,144){\makebox(50,12){$\X$}}

\put(130,150){\line(1,0){4}}
\put(146,150){\line(1,0){4}}
\put(134,144){\framebox(12,12){$\Y$}}
\put(150,144){\makebox(50,12){$\Y$}}

\put(230,150){\line(1,0){4}}
\put(246,150){\line(1,0){4}}
\put(234,144){\framebox(12,12){$\Z$}}
\put(250,144){\makebox(50,12){$\Z$}}

\put(35,110){\line(1,0){20}}
\put(35,90){\line(1,0){20}}
\put(45,110){\circle*{4}}
\put(45,110){\line(0,-1){24}}
\put(45,90){\circle{8}}
\put(0,104){\makebox(35,12){Control}}
\put(0,84){\makebox(35,12){Target}}
\put(55,94){\makebox(70,12){Controlled-NOT}}

\put(130,110){\line(1,0){20}}
\put(130,90){\line(1,0){4}}
\put(146,90){\line(1,0){4}}
\put(140,110){\circle*{4}}
\put(140,110){\line(0,-1){14}}
\put(134,84){\framebox(12,12){$\Y$}}
\put(150,94){\makebox(70,12){Controlled-$\Y$}}

\put(230,110){\line(1,0){20}}
\put(230,90){\line(1,0){4}}
\put(246,90){\line(1,0){4}}
\put(240,110){\circle*{4}}
\put(240,110){\line(0,-1){14}}
\put(234,84){\framebox(12,12){$\Z$}}
\put(250,94){\makebox(70,12){Controlled-$\Z$}}

\put(35,50){\line(1,0){20}}
\put(35,30){\line(1,0){20}}
\put(35,10){\line(1,0){20}}
\put(45,50){\circle*{4}}
\put(45,30){\circle*{4}}
\put(45,50){\line(0,-1){44}}
\put(45,10){\circle{8}}
\put(0,44){\makebox(30,12){Control}}
\put(0,24){\makebox(30,12){Control}}
\put(0,4){\makebox(30,12){Target}}
\put(55,24){\makebox(70,12){Toffoli gate}}

\put(130,50){\line(1,0){4}}
\put(146,50){\line(1,0){4}}
\put(134,44){\framebox(12,12){$P$}}
\put(150,44){\makebox(50,12){$P$}}

\put(230,50){\line(1,0){4}}
\put(246,50){\line(1,0){4}}
\put(234,44){\framebox(12,12){$Q$}}
\put(250,44){\makebox(50,12){$Q$}}

\put(130,10){\line(1,0){4}}
\put(146,10){\line(1,0){4}}
\put(134,4){\framebox(12,12){$R$}}
\put(150,4){\makebox(70,12){Hadamard $R$}}

\put(230,10){\line(1,0){4}}
\put(246,10){\line(1,0){4}}
\put(234,4){\framebox(12,12){$T$}}
\put(250,4){\makebox(50,12){$T$}}

\end{picture}
\caption{Various quantum gates.}
\label{fig-gates}
\end{figure}

\chapter{Glossary}
\label{app-glossary}

\begin{description}

\item[additive code] Another name for a stabilizer code.  Often contrasted
with linear quantum codes, which are a subclass of additive codes.

\item[amplitude damping channel] A channel for which the $\ket{1}$ state
may relax to the $\ket{0}$ state with some probability.  An example is
a two-level atom relaxing via spontaneous emission.

\item[cat state] The $n$-qubit entangled state $\ket{0 \ldots 0} + \ket{1
\ldots 1}$.  Cat states act as ancillas in many fault-tolerant operations.

\item[coding space] The subset of the Hilbert space corresponding to correctly
encoded data.  The coding space forms a Hilbert space in its own right. 

\item[concatenation] The process of encoding the physical qubits making
up one code as the logical qubits of a second code.  Concatenated codes
are particularly simple to correct, and can be used to perform arbitrarily
long fault-tolerant computations as long as the physical error rate is
below some threshhold.

\item[CSS code] Short for Calderbank-Shor-Steane code.  A CSS code is formed
from two classical error-correcting codes.  CSS codes can easily take
advantage of results from the theory of classical error-correcting codes
and are also well-suited for fault-tolerant computation.  See sections
\ref{sec-stab-examples} and \ref{sec-CSS}. 

\item[cyclic code] A code that is invariant under cyclic permutations of
the qubits.

\item[decoherence] The process whereby a quantum system interacts with its
environment, which acts to effectively measure the system.  The world looks
classical at large scales because of decoherence.  Decoherence is likely to
be a major cause of errors in quantum computers.

\item[degenerate code] A code for which linearly independent correctable 
errors acting on the coding space sometimes produce linearly dependent states.  
Degenerate codes bypass many of the known bounds on efficiency of quantum 
codes and have the potential to be much more efficient than any 
nondegenerate code.

\item[depolarizing channel] A channel that produces a random error on each
qubit with some fixed probability.

\item[distance] The minimum weight of any operator $E_a^\dagger E_b$ such that
equation~(\ref{eq-condition}) is {\em not} satisfied for an orthonormal
basis of the coding space.  A quantum code with distance $d$ can detect
up to $d-1$ errors, or it can correct $\lfloor (d-1)/2 \rfloor$ general
errors or $d-1$ located errors. 
 
\item[entanglement] Nonlocal, nonclassical correlations between two quantum
systems.  The presence of entangled states gives quantum computers their
additional computational power relative to classical computers.

\item[entanglement purification protocol] Often abbreviated EPP.  An EPP is
a protocol for producing high-quality EPR pairs from a larger number of
low-quality EPR pairs.  EPPs are classified depending on whether they use
one-way or two-way classical communication.  A 1-way EPP (or 1-EPP) is
equivalent to a quantum error-correcting code.

\item[EPR pair] Short for Einstein-Podalsky-Rosen pair.  An EPR pair is
the entangled state $(1/ \sqrt{2}) \left(\ket{00} + \ket{11} \right)$, and
acts as a basic unit of entanglement.

\item[erasure channel] A channel that produces one or more located errors.

\item[error syndrome] A number classifying the error that has occurred.  For
a stabilizer code, the error syndrome is a binary number with a 1 for each
generator of the stabilizer the error anticommutes with and a 0 for each
generator of the stabilizer the error commutes with.

\item[fault-tolerance] The property (possessed by a network of gates) that an 
error on a single physical qubit or gate can only produce one error in any 
given block of an error-correcting code.  A fault-tolerant network can be
used to perform computations that are more resistant to errors than the
physical qubits and gates composing the computer, provided the error rate
is low enough to begin with.  A valid fault-tolerant operation should also
map the coding space into itself to avoid producing errors when none existed
before.

\item[leakage error] An error in which a qubit leaves the allowed computational
space.  By measuring each qubit to see if it is in the computational space,
a leakage error can be converted into a located error.

\item[linear code] A stabilizer code that, when described in the GF(4)
formalism (section~\ref{sec-alternate}), has a stabilizer that is invariant
under multiplication by $\omega$.  Often contrasted with an additive code.

\item[located error] Sometimes called an erasure.  A located error is an
error which acts on a known qubit in an unknown way.  A located error is
easier to correct than a general error acting on an unknown qubit.

\item[nice error basis] A basis which shares certain essential properties
with the Pauli matrices and can be used to define a generalized stabilizer
code.  See section~\ref{sec-qudits}.

\item[nondegenerate code] A code for which linearly independent correctable
errors acting on the coding space always produce linearly independent states.
Nondegenerate codes are much easier to set bounds on than degenerate codes.

\item[pasting] A construction for combining two quantum codes to make a
single larger code.  See section~\ref{sec-construction}.

\item[perfect code] A code for which every error syndrome corresponds to a
correctable error.  See section~\ref{sec-perfect} for a construction of
the distance three perfect codes.

\item[quantum error-correcting code] Sometimes abbreviated QECC.  A QECC
is a set of states that can be restored to their original state after some
number of errors occur.  A QECC must satisfy equation~(\ref{eq-condition}).

\item[qubit] A single two-state quantum system that serves as the fundamental
unit of a quantum computer.  The word ``qubit'' comes from ``quantum bit.''

\item[qudit] A $d$-dimensional generalization of a qubit.

\item[shadow] The set of operators in $\G$ which commute with the 
even-weight elements of the stabilizer and anticommute with the odd-weight
elements of the stabilizer.

\item[shadow enumerator] The weight enumerator of the shadow.  It is useful
for setting bounds on the existence of quantum codes.

\item[stabilizer] The set of tensor products of Pauli matrices that fix
every state in the coding space.  The stabilizer is an Abelian subgroup
of the group $\G$ defined in section~\ref{sec-general-prop}.  The stabilizer
contains all of the vital information about a code.  In particular, operators
in $\G$ that anticommute with some element of the stabilizer can be
detected by the code.

\item[stabilizer code] A quantum code that can be described by giving its
stabilizer.  Also called an additive code or a GF(4) code.

\item[teleportation] A process whereby a quantum state is destroyed and exactly 
reconstructed elsewhere.  Quantum teleportation of a single qubit requires
one EPR pair shared between the source and destination, and involves two
measurements on the source qubit.  The two bits from the measurements must be 
classically transmitted to the destination in order to reconstruct the original 
quantum state. 

\item[threshhold] The error rate below which a suitably configured quantum
computer can be used to perform arbitrarily long computations.  Current
methods for proving the existence of a threshhold use concatenated codes.
Most estimates of the threshhold lie in the range $10^{-6}$ -- $10^{-4}$.

\item[transversal operation] An operation applied in parallel to the
various qubits in a block of a quantum error-correcting code.  Qubits
from one block can only interact with corresponding qubits from another
block or from an ancilla.  Any transversal operation is automatically
fault-tolerant.

\item[weight] A property of operators only defined on operators which
can be written as the tensor product of single-qubit operators.  For such
an operator, the weight is the number of single-qubit operators in the
product that are not equal to the identity.

\item[weight enumerator] A polynomial whose coefficients $c_n$ are the number
of elements of weight $n$ in some set, such as the stabilizer or
the normalizer of the stabilizer.  Weight enumerators are very helpful
in setting bounds on the possible existence of quantum error-correcting
codes through identities such as the quantum MacWilliams identities (equation
(\ref{eq-QMW})).

\end{description}


\begin{thebibliography}{99}

\bibitem{church-turing} A.~Church, ``An unsolvable problem of elementary
number theory,'' Amer.\ J.\ Math {\bf 58}, 345 (1936); A.~M.~Turing, ``On
computable numbers, with an application to the Entscheidungsproblem,''
Proc.\ Lond.\ Math.\ Soc.\ (2) {\bf 42}, 230 (1936) and Proc.\ Lond.\ Math.\
Soc.\ (2) {\bf 43}, 544 (1937).
\bibitem{feynman} R.~P.~Feynman, ``Simulating physics with computers,''
Int.\ J.\ Theor.\ Phys.\ {\bf 21}, 467 (1982).
\bibitem{shor-factoring} P.~Shor, ``Algorithms for quantum computation:
discrete logarithms and factoring,'' Proceedings, 35th Annual Symposium
on Fundamentals of Computer Science, (1994).
\bibitem{grover} L.~K.~Grover, ``A fast quantum mechanical algorithm for
database search,'' Proceedings, 28th ACM Symposium on Theory of Computation,
212 (1996).
\bibitem{bennett-strengths} C.~B.~Bennett, E.~Bernstein, G.~Brassard, and
U.~Vazirani, ``Strengths and weaknesses of quantum computing,''
quant-ph/9701001 (1997).
\bibitem{cirac-zoller} J.~I.~Cirac and P.~Zoller, ``Quantum computations
with cold trapped ions,'' Phys.\ Rev.\ Lett.\ {\bf 74}, 4091 (1995).
\bibitem{wineland} C.~Monroe, D.~M.~Meekhof, B.~E.~King, W.~M.~Itano, and
D.~J.~Wineland, ``Demonstration of a fundamental quantum logic gate,'' Phys.\ 
Rev.\ Lett.\ {\bf 75}, 4714 (1995).
\bibitem{kimble} Q.~A.~Turchette, C.~J.~Hood, W.~Lange, H.~Mabuchi, and
H.~J.~Kimble, ``Measurement of conditional phase shifts for quantum logic,'' 
Phys.\ Rev.\ Lett.\ {\bf 75}, 4710 (1995).
\bibitem{gershenfeld} N.~Gershenfeld and I.~Chuang, ``Bulk spin resonance
quantum computation,'' Science {\bf 275}, 350 (1997).
\bibitem{shor-9qubit} P.~Shor, ``Scheme for reducing decoherence in quantum
memory,'' Phys.\ Rev.\ A {\bf 52}, 2493 (1995).
\bibitem{steane-7qubit} A.~M.~Steane, ``Error correcting codes in quantum
theory,'' Phys.\ Rev.\ Lett.\ {\bf 77}, 793 (1996).
\bibitem{cohen-tannoudji} C.~Cohen-Tannoudji, {\it Quantum Mechanics},
Wiley, New York (1977).
\bibitem{macwilliams-sloane} F.~J.~MacWilliams and N.~J.~A.~Sloane, {\it
The Theory of Error-Correcting Codes}, North-Holland Publishing Company,
New York (1977).
\bibitem{no-cloning} W.~K.~Wootters and W.~H.~Zurek, ``A single quantum
cannot be cloned,'' Nature {\bf 299}, 802 (1982).
\bibitem{shannon} C.~E.~Shannon, ``A mathematical theory of communication,''
Bell Sys.\ Tech.\ J.\ {\bf 27}, 379, 623 (1948).
\bibitem{knill-laflamme-theory} E.~Knill and R.~Laflamme, ``A theory of
quantum error-correcting codes,'' Phys.\ Rev.\ A {\bf 55}, 900 (1997).
\bibitem{bennett-tome} C.~Bennett, D.~DiVincenzo, J.~Smolin, and W.~Wootters,
``Mixed state entanglement and quantum error correction,'' Phys.\ Rev.\ A
{\bf 54}, 3824 (1996).
\bibitem{vaidman} L.~Vaidman, L.~Goldenberg, and S.~Wiesner, ``Error prevention
scheme with four particles,'' Phys.\ Rev.\ A {\bf 54}, 1745R (1996).
\bibitem{grassl} M.~Grassl, Th.~Beth, and T.~Pellizzari, ``Codes for the
quantum erasure channel,'' quant-ph/9610042 (1996).
\bibitem{leung} D.~W.~Leung, M.~A.~Nielsen, I.~L.~Chuang, Y.~Yamamoto,
``Approximate quantum error correction can lead to better codes,''
quant-ph/9704002 (1997).
\bibitem{gottesman-stab} D.~Gottesman, ``Class of quantum error-correcting
codes saturating the quantum Hamming bound,'' Phys.\ Rev.\ A {\bf 54}, 1862
(1996).
\bibitem{calderbank-stab} A.~R.~Calderbank, E.~M.~Rains, P.~W.~Shor, and
N.~J.~A.~Sloane, ``Quantum error correction and orthogonal geometry,'' Phys.\
Rev.\ Lett.\ {\bf 78}, 405 (1997).
\bibitem{rains-shadow} E.~Rains, ``Quantum shadow enumerators,''
quant-ph/9611001 (1996).
\bibitem{laflamme-5qubit} R.~Laflamme, C.~Miquel, J.~P.~Paz, and W.~Zurek,
``Pefect quantum error correction code,'' Phys.\ Rev.\ Lett.\ {\bf 77}, 198 
(1996).
\bibitem{gottesman-pasting} D.~Gottesman, ``Pasting quantum codes,''
quant-ph/9607027 (1996).
\bibitem{calderbank-GF4} A.~R.~Calderbank, E.~M.~Rains, P.~W.~Shor, and
N.~J.~A.~Sloane, ``Quantum error correction via codes over GF(4),''
quant-ph/9608006 (1996).
\bibitem{steane-8qubit} A.~Steane, ``Simple quantum error correcting codes,''
Phys.\ Rev.\ A {\bf 54}, 4741 (1996).
\bibitem{steane-RM} A.~Steane, ``Quantum Reed-Muller codes, ''
quant-ph/9608026 (1996).
\bibitem{calderbank-CSS} A.~R.~Calderbank and P.~W.~Shor, ``Good quantum
error-correcting codes exist,'' Phys.\ Rev.\ A {\bf 54}, 1098 (1996).
\bibitem{steane-CSS} A.~Steane, ``Multiple particle interference and quantum
error correction,'' Proc.\ Roy.\ Soc.\ Lond.\ A {\bf 452}, 2551 (1996).
\bibitem{knill-qudit} E.~Knill, ``Non-binary error bases and quantum codes,''
quant-ph/9608048 (1996); E.~Knill, ``Group representations, error bases and
quantum codes,'' quant-ph/9608049 (1996).
\bibitem{chau-d^2} H.~F.~Chau, ``Correcting quantum errors in higher spin 
systems,'' quant-ph/9610023 (1996)
\bibitem{chau-5qudit} H.~F.~Chau, ``Five quantum register error correction
code for higher spin systems,'' quant-ph/9702033 (1997).
\bibitem{aharonov} D.~Aharonov and M.~Ben-Or, ``Fault-tolerant quantum
computation with constant error,'' quant-ph/9611025 (1996).
\bibitem{rains-orthogonal} E.~Rains, ``Nonbinary quantum codes,'' 
quant-ph/9703048 (1997).
\bibitem{cleve} R.~Cleve and D.~Gottesman, ``Efficient computations of
encodings for quantum error correction,'' quant-ph/9607030 (1996).
\bibitem{divincenzo-decoder} D.~P.~DiVincenzo, ``Quantum gates and circuits,''
quant-ph/9705009 (1997).
\bibitem{shor-fault-tol} P.~Shor, ``Fault-tolerant quantum computation,''
quant-ph/9605011 (1996).
\bibitem{divincenzo} D.~DiVincenzo and P.~Shor, ``Fault-tolerant error
correction with efficient quantum codes,'' Phys.\ Rev.\ Lett.\ {\bf 77},
3260 (1996).
\bibitem{gottesman-fault-tol} D.~Gottesman, ``A theory of fault-tolerant
quantum computation,'' quant-ph/9702029 (1997).
\bibitem{bennett-teleport} C.~H.~Bennett, G.~Brassard, C.~Crepeau, R.~Josza,
A.~Peres, and W.~K.~Wootters, ``Teleporting an unknown quantum state via dual
classical and Einstein-Podalsky-Rosen channels,'' Phys.\ Rev.\ Lett.\ {\bf 70}, 
1895 (1993).
\bibitem{bennett-EPP} C.~H.~Bennett, G.~Brassard, S.~Popescu, B.~Schumacher,
J.~A.~Smolin, and W.~K.~Wootters, ``Purification of noisy entanglement and
faithful teleportation via noisy channels,'' Phys.\ Rev.\ Lett.\ {\bf 76},
722 (1996).
\bibitem{gottesman-qudit} E.~Knill, R.~Laflamme, and D.~Gottesman, in
preparation.
\bibitem{knill-normalizer} E.~Knill, personal communication.
\bibitem{knill-concatenate2} E.~Knill, R.~Laflamme, and W.~Zurek, ``Accuracy
threshold for quantum computation,'' quant-ph/9610011 (1996); E.~Knill, 
R.~Laflamme, and W.~Zurek, ``Resilient quantum computation: error models
and thresholds,'' quant-ph/9702058 (1997).
\bibitem{evslin} J.~Evslin, S.~Kakade, and J.~P.~Preskill, unpublished.
\bibitem{steane-correction} A.~M.~Steane, ``Active stabilization, quantum
computation and quantum state synthesis,'' Phys.\ Rev.\ Lett.\ {\bf 78},
2252 (1997).
\bibitem{knill-concatenate1} E. Knill and R. Laflamme, ``Concatenated
quantum codes,'' quant-ph/ 9608012 (1996).
\bibitem{zalka} C.~Zalka, ``Threshold estimate for fault tolerant quantum
computing,'' quant-ph/9612028 (1996).
\bibitem{lloyd} S.~Lloyd, ``The capacity of a noisy quantum channel,''
Phys.\ Rev.\ A {\bf 55}, 1613 (1997).
\bibitem{schumacher} B.~Schumacher and M.~A.~Nielsen, ``Quantum data
processing and error correction,'' Phys.\ Rev.\ A {\bf 54}, 2629 (1996).
\bibitem{barnum} H.~Barnum, M.~A.~Nielsen, and B.~Schumacher, ``Information
transmission through a noisy quantum channel,'' quant-ph/9702049 (1997).
\bibitem{ekert} A.~Ekert and C.~Macchiavello, ``Error correction in quantum
communication,'' Phys.\ Rev.\ Lett.\ {\bf 77}, 2585 (1996).
\bibitem{cerf-cleve} N.~J.~Cerf and R.~Cleve, ``Information-theoretic
interpretation of quantum error-correcting codes,'' quant-ph/9702031 (1997).
\bibitem{shor-laflamme-QMW} P.~Shor and R.~Laflamme, ``Quantum analog of
the MacWilliams identities for classical coding theory,'' Phys.\ Rev.\ Lett.\ 
{\bf 78}, 1600 (1997).
\bibitem{rains-enumerators} E.~M.~Rains, ``Quantum weight enumerators,''
quant-ph/9612015 (1996).
\bibitem{rains-poly-invariants} E.~M.~Rains, ``Polynomial invariants of quantum
codes,'' quant-ph/9704042 (1997).
\bibitem{cleve-classical} R.~Cleve, ``Quantum stabilizer codes and classical
linear vodes,'' quant-ph/9612048 (1996).
\bibitem{bennett-erasure} C.~H.~Bennett, D.~P.~DiVincenzo, and J.~A.~Smolin,
``Capacities of quantum erasure channels,'' quant-ph/9701015 (1997).
\bibitem{fuchs-KL} C.~Fuchs and J.~Smolin, unpublished.
\bibitem{shor-smolin} P.~Shor and J.~Smolin, ``Quantum error-correcting codes
need not completely reveal the error syndrome,'' quant-ph/9604006 (1996).
\bibitem{rains-dist2} E.~M.~Rains, ``Quantum codes of minimum distance two,''
quant-ph/ 9704043 (1997).
\bibitem{rains-nonstab} E.~M.~Rains, R.~H.~Hardin, P.~W.~Shor, and
N.~J.~A.~Sloane, ``A nonadditive quantum code,'' quant-ph/9703002 (1997).
\bibitem{lloyd-universal} S.~Lloyd, ``Almost any quantum logic gate is
universal,'' Phys.\ Rev.\ Lett.\ {\bf 75}, 346 (1995).
\bibitem{gates} A.~Barenco, C.~H.~Bennett, R.~Cleve, D.~P.~DiVincenzo,
N.~Margolus, P.~Shor, T.~Sleator, J.~Smolin, and H.~Weinfurter,
``Elementary gates for quantum computation,'' Phys.\ Rev.\ A {\bf 52},
3457 (1995).

\end{thebibliography}
\end{document}